%% file: main.tex
\documentclass[journal=ancac3,manuscript=article,email=true,hyperref=true,keywords=false]{achemso}
\usepackage[utf8]{inputenc}
\usepackage{graphicx}
\usepackage{float}
\usepackage{xcolor}
\usepackage{amsmath}
\usepackage{amssymb}
\usepackage{todonotes}
\usepackage{times}

\usepackage{xr} 
\author{Tian Tian}
\affiliation{Institute for Chemical and Bioengineering, ETH Z{\"{u}}rich,  Vladimir Prelog Weg 1, CH-8093 Z{\"{u}}rich, Switzerland}
\altaffiliation{T. T. and D. S. contributed equally to this work}
\author{Declan Scullion}
\affiliation{School of Mathematics and Physics, Queen's University Belfast, BT7 1NN, United Kingdom}
\altaffiliation{T. T. and D. S. contributed equally to this work}
\author{Dale Hughes}
\affiliation{School of Mathematics and Physics, Queen's University Belfast, BT7 1NN, United Kingdom}
\author{Lu Hua Li}
\affiliation{Institute for Frontier Materials, Deakin University, Victoria, VIC 3216, Australia}
\author{Chih-Jen Shih}
\affiliation{Institute for Chemical and Bioengineering, ETH Z{\"{u}}rich,  Vladimir Prelog Weg 1, CH-8093 Z{\"{u}}rich, Switzerland}
\author{Jonathan Coleman}
\affiliation{School of Physics, Centre for Research on Adaptive Nanostructures and Nanodevices (CRANN) and Advanced Materials and BioEngineering Research (AMBER), Trinity College Dublin, Dublin 2, Ireland}
\author{Manish Chhowalla}
\affiliation{Department of Materials Science \& Metallurgy, University of Cambridge, CB3 0FS, United Kingdom}
\author{Elton J. G. Santos}
\email{e.santos@qub.ac.uk}
\affiliation{School of Mathematics and Physics, Queen's University Belfast, BT7 1NN, United Kingdom}
\date{}
\title{Electronic polarizability as the fundamental variable in the dielectric properties of two-dimensional materials}


\begin{document}

\newpage{}


\begin{abstract}
  The dielectric constant, which defines the polarization of
  the media, is a key quantity in condensed matter. It determines 
  several electronic and optoelectronic properties important for a 
  plethora of modern technologies 
  from computer memory to field effect
  transistors and communication circuits. Moreover, the importance of
  the dielectric constant in 
  describing electromagnetic interactions through screening plays a
  critical role in understanding fundamental molecular
  interactions. Here we show that despite its fundamental
  transcendence, the dielectric constant does not define unequivocally
  the dielectric properties of two-dimensional (2D) materials due to
  the locality of their electrostatic screening. Instead, the
  electronic polarizability correctly captures the dielectric nature
  of a 2D material which is united to other physical quantities in an
  atomically thin layer. 
 %
  %
  We reveal a long-sought universal formalism where electronic, 
  geometrical and dielectric properties are intrinsically correlated  
  through the polarizability opening the door to probe quantities yet  
  not directly measurable including the real covalent thickness of a 
  layer. We unify the concept of dielectric properties in any 
  material dimension finding a global dielectric anisotropy index
  defining their controllability through dimensionality. 
\end{abstract}

\pagebreak{}

\section{Introduction}
\label{sec:introduction}

The dielectric constant $\varepsilon$ (also known as the relative permittivity) 
plays a crucial role in bridging various fundamental material
properties, such as bandgap \cite{Moss_1950_relation,Moss_1985_n_Eg}, 
optical absorption\cite{kittel_2005_introduction} and 
conductivity \cite{Dressel_2001_electrodynamics}   
with elemental interactions. 
The central place of $\varepsilon$ in solid-state physics drives the analysis of several phenomena 
where is common to classify a material accordingly to its ability to screen an 
electric field $\boldsymbol{E}$ in terms of insulators, metals and semiconductors. 
Such definitions determine a broad range of 
condensed matter physics, as well as in related 
fields in chemistry and materials science. 
The ability to compute and measure $\varepsilon$ in 
bulk materials is well established via different 
theoretical \cite{Adler_1962,Hybertsen_1987} and 
experimental techniques \cite{palik_1998handbook} of distinct flavors
where the probe of the dielectric properties 
is made through an external electric field. 
Despite its obvious appeal, however, it is 
still unknown whether such quantity can determine the 
electronic and dielectric properties of 
two-dimensional (2D) materials \cite{Novoselov_2016}.  
The confined nature of such atomically-thin 2D crystals associated
with the attenuated and anisotropic character of the dielectric
screening
\cite{Keldysh_1979_eps_multi,Sharma_1985,Low_2014_screening_BP,Cudazzo_2011_screening_2D,Bechstedt_2012,Cudazzo_2010_screen2D,Nazarov_2015_2D_3D}
has generated long-standing debates whether the dielectric constant
truly represents the dielectric features of such low-dimensional systems. 
%
%
The controversy of values
reported by both theoretical and experimental
approaches can be widely seen throughout the specialized 
literature, see Ref.\cite{Li_2016} for a summary,
where the variation of $\varepsilon$ can be more than one order of
magnitude. As a consequence, several key physical parameters that
scale with $\varepsilon$, such as the exciton binding energy and
Debye screening length, cannot be reliably estimated due to the
discrepancy of the reported magnitudes of $\varepsilon$. 

Here, by using a combination of analytical and numerical models liaised with 
highly-accurate first-principles methods involving high-throughput screening 
techniques, we show that the dielectric constant does not provide a reliable
description of the screening features of a 2D material. The interplay between local 
electrostatic interactions in the monolayer and the volume dependence 
in the definition of $\varepsilon$ makes such quantity questionable. 
We propose however that the electronic polarizability that describes 
the electron dipole in the 2D material as the true descriptor 
of its dielectric nature. 
We overcome several problems intrinsic to thin layers not achievable using conventional 
effective dielectric medium models, such as the real thickness of a monolayer and 
any dependence on the long-range Coulomb potential. 
We unveil universal scaling relations between electronic and dielectric properties through 
the electronic polarizability, such as band gaps, optical spectra and exciton radius, for the 
current library of known 2D materials involving different lattice symmetries, 
atomic elements and chemical and physical properties. 
Moreover, the concept of electronic polarizabilities bridges the
gap between the dielectric properties of 2D and 3D systems through a novel 
dielectric anisotropy index that generalized the concept of dielectric control 
using dimensionality and bandgap. 
Our results open a new avenue for the study of the 
dielectric properties of 2D compounds using techniques yet to be explored.



\section{Results and discussions}
\label{sec:results-discussions}

\subsection{Lattice-dependency of macroscopic dielectric constant}
\label{sec:latt-depend-macr}

\begin{figure}[H]
\centering
\includegraphics[width=0.70\linewidth]{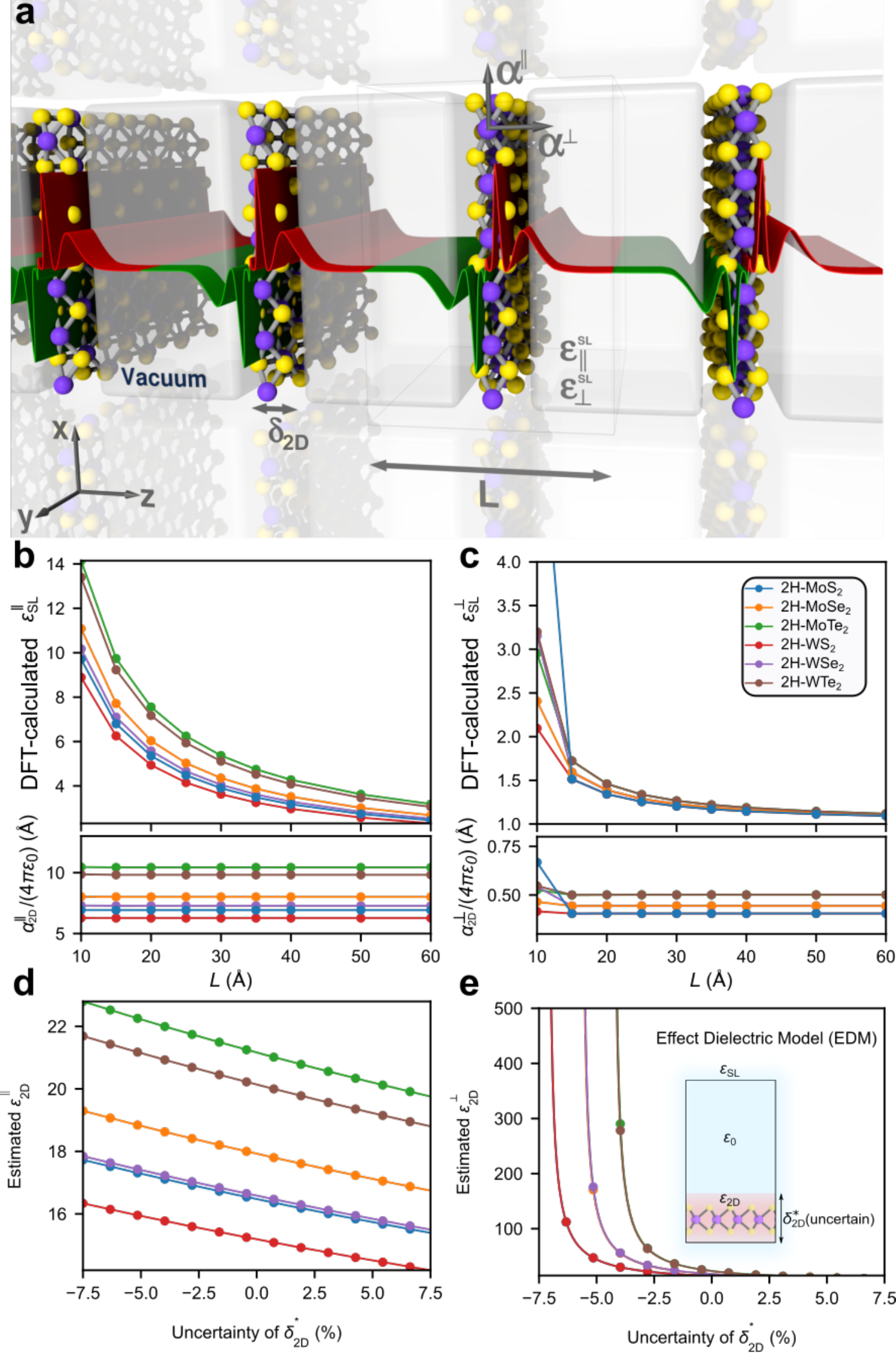}
\caption{\label{fig-1} \textbf{2D polarizability and the breakdown of
    effective dielectric model
    (EDM)} 
  \textbf{a,} 3D illustration of the spatial distribution of the
  charge density change $\Delta \rho(z)$ along the z-direction for
  monolayer 2H-MoS\textsubscript{2} in a periodic superlattice under
  external eletric field of 0.01 V/\AA{}.  The green and red regions
  represent negative and positive induced charges, respectively. The
  macroscopic $\varepsilon_{\mathrm{SL}}^{\parallel}$ and
  $\varepsilon_{\mathrm{SL}}^{\perp}$ are influenced by the lattice
  size $L$, while the 2D polarizabilities
  $\alpha_{\mathrm{2D}}^{\parallel}$ and
  $\alpha_{\mathrm{2D}}^{\perp}$ are invariant to $L$.
  \textbf{b,} $\varepsilon^{\parallel}_{\mathrm{SL}}$ (top) and
  $\alpha_{\mathrm{2D}}^{\parallel}$ (bottom) as functions of $L$ for
  the 2H TMDCs. 
  \textbf{c,} $\varepsilon^{\perp}_{\mathrm{SL}}$ (top)
  and $\alpha_{\mathrm{2D}}^{\perp}$ (bottom) as functions of $L$ for
  the 2H TMDCs. The polarizabilities in \textbf{b} and \textbf{c} are
  constant when $L>$15 \AA{}, compared with the $L$-dependence of
  $\varepsilon_{\mathrm{SL}}$. 
  {\bf d-e,} Estimated $\varepsilon_{\mathrm{2D}}^{\parallel}$ and 
  $\varepsilon_{\mathrm{2D}}^{\perp}$, respectively, using EDM  
  as a function of the uncertainty of the effective layer
  thickness $\delta^{*}_{\mathrm{2D}}$. The inset in {\bf e,} shows schematically the main parameters
  utilized in EMD: the vacuum layer ($\varepsilon^{}_{\mathrm{0}}$), an approximate thickness of the layer   which is given by $\delta^{*}_{\mathrm{2D}}$, and the obtained $\varepsilon^{}_{\mathrm{2D}}$. The length of the box perpendicular to the surface of the layer is given by $L$ (not shown). 
  Overall, there is a large variation and associated errors
  to both components of the dielectric constant for small changes of $\delta^{*}_{\mathrm{2D}}$ in the range of $\pm7.5$\%.  
  %
}
\end{figure}

We first approach the discrepancy of macroscopic dielectric constant
of 2D materials, by showing that the current definition of
$\varepsilon$ used in layered materials is {ill-defined}.  This can be
viewed in a model system as illustrated
in Figure \ref{fig-1}, where an isolated 2D material is placed in
the \textit{xy}-plane of a periodically repeating superlattice (SL)
with a length $L$ along the \textit{z}-direction separating the cell
images. The static macroscopic dielectric tensor from the superlattice
$\varepsilon_{\mathrm{SL}}^{pq}$, is determined through fundamental
electrostatics by the response of the polarization density
$\boldsymbol{P}^{p}$ under small perturbative external field
$\boldsymbol{E}^{q}$, where $p$, $q$ determine their directions,
respectively~\cite{Dressel_2001_electrodynamics}:
\begin{subequations}
  \begin{eqnarray}
      \label{eq:def-eps-1}
    &\varepsilon_{\mathrm{SL}}^{pq} &= \kappa^{pq} +
                                 {\displaystyle \frac{\partial \boldsymbol{P}^{p}}{\varepsilon_{0} \partial \boldsymbol{E}^{q}}} \\
          \label{eq:def-eps-2}
    &\boldsymbol{P}^{p} &=  {\displaystyle \frac{\boldsymbol{u}^{p}}{\Omega}}
                          = {\displaystyle \frac{{\displaystyle
          \int_{\mathrm{SL}} \rho(\boldsymbol{r}) \boldsymbol{r}^{p} d^{3}\boldsymbol{r}}}{AL}}
  \end{eqnarray}
\end{subequations}
where $\kappa$ is the dielectric tensor of the environment,
$\boldsymbol{u}$ is the total dipole moment within the SL, $\rho$ is
the spatial charge density, $\Omega=AL$ is the volume of the
supercell, $A$ is the \textit{xy}-plane area of the SL and
$\varepsilon_{0}$ is vacuum permittivity. Here we limit our study on
the electronic contributions to the macroscopic dielectric constant
where the dipole $\boldsymbol{P}$ results from the response 
of the electron density under an external field.  
Ionic contributions\cite{Sohier_2017} to $\varepsilon_{\mathrm{SL}}$ 
have previously been shown to be negligible\cite{relax-epsilon} and 
are not considered here.  
%
The symmetry of 2D materials leads to inappreciable off-diagonal
elements of the dielectric tensor ($p \neq q$), while the diagonal
elements $\varepsilon_{\mathrm{SL}}^{xx}$,
$\varepsilon_{\mathrm{SL}}^{yy}$ and $\varepsilon_{\mathrm{SL}}^{zz}$
can be different \cite{Sohier_2016}.  
Considering that the 2D material
is placed in vacuum ($\kappa^{pp} = 1$ and $\kappa^{pq} = 0$), we can
distinguish two components of $\varepsilon_{\mathrm{SL}}$, namely the
in-plane ($\varepsilon_{\mathrm{SL}}^{\parallel}$) and out-of-plane
($\varepsilon_{\mathrm{SL}}^{\perp}$) dielectric constants, where
$\varepsilon_{\mathrm{SL}}^{\parallel} =
(\varepsilon_{\mathrm{SL}}^{xx} + \varepsilon_{\mathrm{SL}}^{yy})/2$
and
$\varepsilon_{\mathrm{SL}}^{\perp} = \varepsilon_{\mathrm{SL}}^{zz}$.
The absence of bonding perpendicular to the plane confines the induced
dipole moments along the \textit{z}-direction within a range of $\sim$5--6
\AA{} into the vacuum (Figure \ref{fig-1}{\textbf a} and Supplementary
Figure \ref{SI-fig:rho-profile}).
Under a given external field, the strong confinement of
the induced dipole moment $\boldsymbol{u}$ causes the integral in the 
numerator of Eq.~\ref{eq:def-eps-2} to be converged within few \AA's 
resulting that the dipole moment from the periodic supercell 
images do not mutually interfere. 

Conversely, the increase of
$L$ in the denominator of Eq.~\ref{eq:def-eps-2} dilutes the 
polarization density, and in turn makes both
$\varepsilon^{\parallel}_{\mathrm{SL}}$ and
$\varepsilon^{\perp}_{\mathrm{SL}}$ dropping to
unity when $L$ is infinitely large, which is not physical. 
%
%
%
Despite the simplicity of this argument, any calculation performed
using such definition will intrinsically depend on the magnitude of
$L$, an artificial parameter introduced by the simulation setup. This
dependence can be clearly demonstrated by plotting
$\varepsilon^{\parallel}_{\mathrm{SL}}$ and
$\varepsilon^{\perp}_{\mathrm{SL}}$ calculated from density functional
theory (DFT) (see {\it Theoretical Methods} for details) as a function
of $L$ for P\={6}m2 transition metal dichalcogenides (TMDCs),
2H-MX$_{2}$, where M=Mo, W and X=S, Se, Te (top panels of Figure
\ref{fig-1}{\textbf b} and \ref{fig-1}{\textbf c}, respectively). To
obtain a better description of the electronic band structure, the
calculations of dielectric properties were performed at the level of
Heyd-Scuseria-Ernzerhof (HSE06) hybrid functional
\cite{Heyd_2003,HSE_2006}.  Both components of the dielectric
constant decrease with $L$ as excepted. To rule out the possibility
that the result is affected by the choice of the functional, 
we performed simulations at higher levels of theory
using many-body techniques (G$_{0}$W$_{0}$), which invariably give
alike results (see Supplementary Figure \ref{SI-fig:GW-PBE-alpha}).
The lattice-size dependency also exists for the dielectric function in the 
frequency domain. 

We carried out similar analysis for
frequency-dependent $\varepsilon^{\parallel}_{\mathrm{SL}}(\omega)$
and $\varepsilon^{\perp}_{\mathrm{SL}}(\omega)$ using different 
approaches including Perdew-Burke-Ernzerhof (PBE)
exchange-correlation functional
\cite{Perdew_1996,Ernzerhof99,Paier_2005_PBE}, 
G\textsubscript{0}W\textsubscript{0}\cite{Hedin_1965} and
Bethe-Salpeter equation (G\textsubscript{0}W\textsubscript{0}$-$BSE)\cite{Onida_2002} (see Supplementary Section \ref{SI-ssec:gw} and 
Supplementary Figures \ref{SI-fig:PBE-omega-in}$-$\ref{SI-fig:BSE-omega-out}). 
%
%
Despite the various levels of theory analyzed and the increased 
accuracy of the calculated optical properties due to the inclusion of 
many-body screening and excitonic effects, 
the magnitude of the dielectric function universally decreases
with $L$ over the frequency. The underlying physical reason for such 
dependence can be noticed in the definition of the dielectric 
function versus $\omega$ shown in Eqs. \ref{SI-eq:dft-dielectric}$-$\ref{SI-eq:dft-dielectric-real}, which also depend on the volume of the unit cell. 
These results indicate that any 
quantity that depends on
$\varepsilon^{\parallel}_{\mathrm{SL}}(\omega)$ and
$\varepsilon^{\perp}_{\mathrm{SL}}(\omega)$, such as the optical
absorption ($\mathrm{Im}\{\varepsilon_{\mathrm{SL}}(\omega)\}$), refractive index ($n=\sqrt{\mathrm{Re}\{\varepsilon_{\mathrm{SL}}\}}$) and electron energy loss spectrum (EELS, $\mathrm{Im}\{-1/\varepsilon_{\mathrm{SL}}(\omega)\}$), suffers the same deficiencies for 2D materials.

\subsection{The electronic polarizability of 2D materials}
\label{sec:electr-polar-2d}

To solve the problem described above, we need to find the
$L$-independent alternative of $\varepsilon_{\mathrm{SL}}$, which is
related to both electrostatic and optical properties of a 2D
material~\cite{Matthes_2016}. By multiplying Eq.~\ref{eq:def-eps-2}
with $L$, we obtain the sheet polarization density, that is,
$\boldsymbol{\mu}_{\mathrm{2D}}^{p} =\boldsymbol{u}^{p}/A$, along the
direction $p$. Following the discussion in the previous section,
$\boldsymbol{\mu}_{\mathrm{2D}}^{p}$ becomes independent of the
lattice size when $L$ is large enough, due to the short decay of the 
induced charge density $\Delta \rho$ into the vacuum (see
Supplementary Figure \ref{SI-fig:rho-profile}).
Similar to the molecular polarizability\cite{Israelachvili_2011}, we
utilize the concept of electronic polarizability
$\alpha_{\mathrm{2D}}$, 
which has been used previously to solve
exciton-related problems in 2D materials
\cite{Cudazzo_2011_screening_2D,Olsen_2016_hydrogen,Jiang_2017_Eg_Eb}. 
$\alpha_{\mathrm{2D}}$ is a macroscopic quantity that characterizes
the ability to induce dipole moments in a 2D material, and is
associated with $\boldsymbol{\mu}_{\mathrm{2D}}$ through:
$\boldsymbol{\mu}_{\mathrm{2D}}^{p} = \sum_{q}
\alpha_{\mathrm{2D}}^{pq} \boldsymbol{\overline{E}}_{\mathrm{loc}}^{q}$\cite{T_bik_2004}, where
$\boldsymbol{\overline{E}}_{\mathrm{loc}}$ is the cell-averaged
``local'' electric field acting on the 2D material to induce
polarization\cite{}.
Alike to $\mu_{\mathrm{2D}}$,
$\boldsymbol{\overline{E}}_{\mathrm{loc}}$ is also a macroscopic
quantity that excludes the fields generated by the dipoles of the 2D
sheet from $\boldsymbol{E}$. Note the term ``local'' in
$\boldsymbol{\overline{E}}_{\mathrm{loc}}$ is adapted to resemble the
Lorentz model\cite{Wiser_1963} which has also been used for other
low-dimensional materials (e.g. nanotube\cite{Benedict_1995} and
molecules\cite{T_bik_2004}, and should be distinguished with the
microscopic local field $\boldsymbol{E}_{\mathrm{loc}}(\boldsymbol{r})$ which is spatially
changing.

Such macroscopic treatment of polarizability is valid when the length
of the superlattice is significantly larger than the spatial
distribution of induced charges.  At $L \rightarrow \infty$ limit,
$\boldsymbol{\overline{E}}_{\mathrm{loc}}$ can be solved using
electrostatic boundary conditions of the slab geometry
\cite{Markel_2016,Meyer_2001_dipole_slab}. The continuity of the electric field along the
in-plane direction gives
$\boldsymbol{\overline{E}}^{\parallel}_{\mathrm{loc}}=\boldsymbol{E}^{\parallel}$,
while for the out-of-plane component, the dipole screening yields
$\boldsymbol{\overline{E}}_{\mathrm{loc}}^{\perp}=\boldsymbol{E}^{\perp}+\boldsymbol{\mu}_{\mathrm{2D}}^{\perp}/L$
\cite{Meyer_2001_dipole_slab,T_bik_2004}, where
$\boldsymbol{E}^{\parallel}$ and $\boldsymbol{E}^{\perp}$ are the
external field along the in-plane and out-of-plane directions,
respectively. Combining with Eqs. \ref{eq:def-eps-1} and
\ref{eq:def-eps-2}, $\alpha_{\mathrm{2D}}^{\parallel}$ and
$\alpha_{\mathrm{2D}}^{\perp}$ can be related with
$\varepsilon_{\mathrm{SL}}^{\parallel}$ and
$\varepsilon_{\mathrm{SL}}^{\perp}$, respectively:
\begin{subequations}
\begin{eqnarray}
  \label{eq:alpha-para-def}
  &\varepsilon_{\mathrm{SL}}^{\parallel} &= 1 + \frac{\alpha_{\mathrm{2D}}^{\parallel}}{\varepsilon_{0}L}\\
  \label{eq:alpha-perp-def}
  &\varepsilon_{\mathrm{SL}}^{\perp} &= \left(1 - {\displaystyle \frac{\alpha_{\mathrm{2D}}^{\perp}}{\varepsilon_{\mathrm{0}} L}} \right)^{-1}
\end{eqnarray}
\end{subequations}

%
%
Using these relations, we show that the calculated
$\alpha_{\mathrm{2D}}^{\parallel}$ and $\alpha_{\mathrm{2D}}^{\perp}$
of the selected TMDCs as a function of $L$ in the bottom panels of
Figure \ref{fig-1}b and \ref{fig-1}c, respectively.  In contrast to
$\varepsilon_{\mathrm{SL}}^{\parallel}$ and
$\varepsilon_{\mathrm{SL}}^{\perp}$, we observe that both
$\alpha_{\mathrm{2D}}^{\parallel}$ and $\alpha_{\mathrm{2D}}^{\perp}$
reach convergence when $L \sim$10 \AA $, ~$15 \AA,
respectively. Such results are in good agreement with the spatially
localized induced dipole moment of a 2D material as shown in
Supplementary Figure~\ref{SI-fig:rho-profile}.
%
%
Equations \ref{eq:alpha-para-def}$-$\ref{eq:alpha-perp-def} can 
also be used to remove the dependence on $L$ for
$\varepsilon^{\parallel}_{\mathrm{SL}}(\omega)$
and $\varepsilon^{\perp}_{\mathrm{SL}}(\omega)$, generating
lattice-independent electronic polarizability
$\alpha^{\parallel}_{\mathrm{SL}}(\omega)$ and
$\alpha^{\perp}_{\mathrm{SL}}(\omega)$ in the frequency domain,
respectively (see details in Supplementary Section \ref{SI-ssec:gw}).
These findings indicate that the electronic polarizability $\alpha_{\mathrm{2D}}$ captures the essence
of the dielectric properties of 2D materials. In contrast to the ill-defined macroscopic $\varepsilon_{\mathrm{SL}}$, $\alpha_{\mathrm{2D}}$ has 
a unique definition, and does not suffer from the dependency on the lattice size.
It is worthy mentioning that Eqs.\ref{eq:alpha-para-def}$-$\ref{eq:alpha-perp-def} were 
obtained using purely electrostatic arguments without any assumption 
regarding the medium where the 2D material is immersed or a capacitance 
model where an effective dielectric response can be extracted. 
More details about the choice of the
2D polarizability, comparison with other methods, simulations at the
frequency-dependent domain can be found in Supplementary Section
\ref{SI-ssec:gw}.

\subsection{Comparison with the effective dielectric model (EDM)}
\label{sec:comp-with-effect}

Apart from the 2D electronic polarizability proposed here,  
the effective dielectric model (EDM) is commonly used in
literature to treat the 2D material as a slab with an effective
dielectric tensor $\varepsilon_{\mathrm{2D}}$ and thickness
$\delta^{*}_{\mathrm{2D}}$. Such method can be found in both
experimental and theoretical studies, such as to interpret
ellipsometry data
\cite{graphene-epsilon10,Duesberg14,Chiang13,Kong14}, reflectance /
transmission spectra \cite{Li_2014, Yoffe-Wilson69}, optical
conductance \cite{Matthes_2016} and many-body interactions
\cite{Sohier_2016,Meckbach_2018} of 2D materials. The EDM allows
applying physical concepts of bulk systems directly to their 2D counterparts using
$\varepsilon_{\mathrm{2D}}$. However, there are several drawbacks of
such approach. For instance, the wavevector $q$-dependency of
dielectric screening in 2D
sheets~\cite{Cudazzo_2011_screening_2D,Olsen_2016_hydrogen,Trolle_2017_eps_subst}
is not captured. More severely, here we show that, due to the
uncertainty of $\delta^{*}_{\mathrm{2D}}$, the calculated
$\varepsilon_{\mathrm{2D}}$, in particular its out-of-plane component,
is extremely sensitive to the choice of $\delta^{*}_{\mathrm{2D}}$,
making such model questionable.

The basic assumption of EDM can be
seen in the inset of Figure~\ref{fig-1}\textbf{e}, where the macroscopic
$\varepsilon_{\mathrm{SL}}$ is considered to be composed by ({\it i}) a
2D slab with an effective dielectric constant $\varepsilon_{\mathrm{2D}}$
and a thickness $\delta^{*}_{\mathrm{2D}}$, and ({\it ii}) a vacuum spacing with
distance $L-\delta^{*}_{\mathrm{2D}}$. Using the effective medium theory
(EMT)~\cite{Aspnes_1982,Markel_2016}, the relation between  
$\varepsilon_{\mathrm{SL}}$ and $\varepsilon_{\mathrm{2D}}$ can be
expressed using capacitance-like
equations~\cite{Matthes_2016,Laturia_2018,Tancogne_Dejean_2015}:
\begin{subequations}
  \begin{eqnarray}
    \label{Response:1}
    {\displaystyle \varepsilon_{\mathrm{SL}}^{\parallel}} &= {\displaystyle \frac{\delta^{*}_{\mathrm{2D}}}{L} \varepsilon_{\mathrm{2D}}^{\parallel} + \left(1 - \frac{\delta^{*}_{\mathrm{2D}}}{L} \right)}\\
     \label{eq:emt-2}
    {\displaystyle \frac{1}{\varepsilon_{\mathrm{SL}}^{\perp}}} &= {\displaystyle \frac{\delta^{*}_{\mathrm{2D}}}{L} \frac{1}{\varepsilon_{\mathrm{2D}}^{\perp}} + \left(1 - \frac{\delta^{*}_{\mathrm{2D}}}{L} \right)}
  \end{eqnarray}
\end{subequations}
In principle, both the values of $\varepsilon_{\mathrm{2D}}$ and
$\delta^{*}_{\mathrm{2D}}$ are unknown for a certain 2D material. To
minimize the modeling error, we used non-linear least-square fitting
to extract $\varepsilon_{\mathrm{2D}}$ and $\delta^{*}_{\mathrm{2D}}$ of
selected 2H TMDCs simultaneously from \textit{ab initio}
$\varepsilon_{\mathrm{SL}}$ -- $L$ data in
Figure~\ref{fig-1}\textbf{b} and \ref{fig-1}\textbf{c} (see details in 
Supplementary Figure~\ref{SI-fig:rescale-prb}). The fitted values of
the slab thickness, $\delta_{\mathrm{2D}}^{\parallel, \mathrm{fit}}$
and $\delta_{\mathrm{2D}}^{\perp, \mathrm{fit}}$ from in-plane and
out-of-plane data, respectively, are shown in Supplementary
Table~\ref{SI-tab:delta-L-DFt}.
%
%
Although the uncertainty only corresponds to a few percent of the
interlayer spacing in the bulk structure of these 2D materials, its
influence on the calculated values
$\varepsilon_{\mathrm{2D}}^{\parallel}$ and
$\varepsilon_{\mathrm{2D}}^{\perp}$ is substantial. 
We estimated the dispersion of $\varepsilon_{\mathrm{2D}}^{\parallel}$ and 
$\varepsilon_{\mathrm{2D}}^{\perp}$ considering slightly deviations of 
$\delta^{*}_{\mathrm{2D}}$ from the best fitted value by $\pm{}$7.5\%{} 
(Figures \ref{fig-1}\textbf{d} and \ref{fig-1}\textbf{e}). 
Strikingly, $\varepsilon_{\mathrm{2D}}^{\parallel}$ decays linearly
with $\delta^{*}_{\mathrm{2D}}$, while $\varepsilon_{\mathrm{2D}}^{\perp}$ spans
over more than one order of magnitude. 
The sensitivity of
$\varepsilon_{\mathrm{2D}}^{\perp}$ to $\delta^{*}_{\mathrm{2D}}$
explains the discrepancy in literature for both
isotropic~\cite{Sohier_2016} and highly
anisotropic~\cite{Matthes_2016,Laturia_2018}
$\varepsilon_{\mathrm{2D}}$ tensors on 2D materials extracted 
using EDM. 
As a consequence, the estimated values of
$\varepsilon_{\mathrm{2D}}$, in particular its out-of-plane component,
are highly controversial.

On the contrary, the proposed $\alpha_{\mathrm{2D}}$ approach does not
suffer from such limitations, despite its relatively simple
formalism. The relative uncertainty of $\alpha_{\mathrm{2D}}$ is
generally at the order of 10\textsuperscript{-4} (Supplementary
Figure~\ref{SI-fig:alpha-converg}). In addition, the calculation of
$\alpha_{\mathrm{2D}}$ is technically simpler than 
$\varepsilon_{\mathrm{2D}}$: ({\it i}) $\alpha_{\mathrm{2D}}$ can be
achieved using single-point calculation of macroscopic dielectric
tensor, while $\varepsilon_{\mathrm{2D}}$ requires non-linear fitting
of multiple $\varepsilon_{\mathrm{SL}}$ -- $L$ data points; ({\it ii}) the
values of $\alpha_{\mathrm{2D}}$ typically converge well for $L$
($\sim$20 \AA{}), while $\varepsilon_{\mathrm{2D}}$ suffers from the
uncertainties as described above.



\subsection{Universal scaling laws of $\alpha_{\mathrm{2D}}$}
\label{sec:univ-scal-laws}

For bulk materials, pioneering works from the 1950s had demonstrated
empirical equations between $\varepsilon$ and the bandgap
$E_{\mathrm{g}}$, including the
Moss~\cite{Moss_1950_relation,Moss_1985_n_Eg,Finkenrath_1988} or
Ravindra~\cite{Ravindra_1980_model,Ravindra_1979_eps_Eg} relations.
Such universal relations, if exist in the context of
$\alpha_{\mathrm{2D}}$, would be of high importance for studying and
predicting the screening of 2D materials.  Inspired by the random
phase approximation (RPA) theory \cite{Adler_1962} within the
$\mathbf{k} \cdot \mathbf{p}$
formalism\cite{kittel_2005_introduction,Jiang_2017_Eg_Eb}, we propose the following 
universal relations for $\alpha_{\mathrm{2D}}^{\parallel}$ and $\alpha_{\mathrm{2D}}^{\perp}$,
for 2D materials (see Supplementary Section \ref{SI-sec:theory-1} for details): 
\begin{subequations}
\begin{eqnarray}
\label{eq:2D-Moss-para}
  &E_{\mathrm{g}} &= C^{\parallel}/ \alpha_{\mathrm{2D}}^{\parallel} \\
  \label{eq:2D-Moss-perp}
  &\hat{\delta}_{\mathrm{2D}} & = C^{\perp} \alpha_{\mathrm{2D}}^{\perp} 
\end{eqnarray}
\end{subequations}
where $E_{\mathrm{g}}$ is the fundamental electronic bandgap and
$\hat{\delta}_{\mathrm{2D}}$ is the intrinsic thickness of the 2D
layer, with coefficients
$C^{\parallel} = {\displaystyle \frac{Ne^2}{2 \pi}}$
\cite{Jiang_2017_Eg_Eb}, where $N$ is a pre-factor associated with the
band degeneracy, and $C^{\perp} = {\varepsilon_{0}}^{-1}$. It is worth
noting, unlike the parameter $\delta^{*}_{\mathrm{2D}}$ that is
artificially assigned within the EDM picture (see previous section),
$\hat{\delta}_{\mathrm{2D}}$ can be uniquely defined by
$\alpha_{\mathrm{2D}}^{\parallel}$, a quantity that can be

\begin{figure}[H]
\centering
\includegraphics[width=0.7\linewidth]{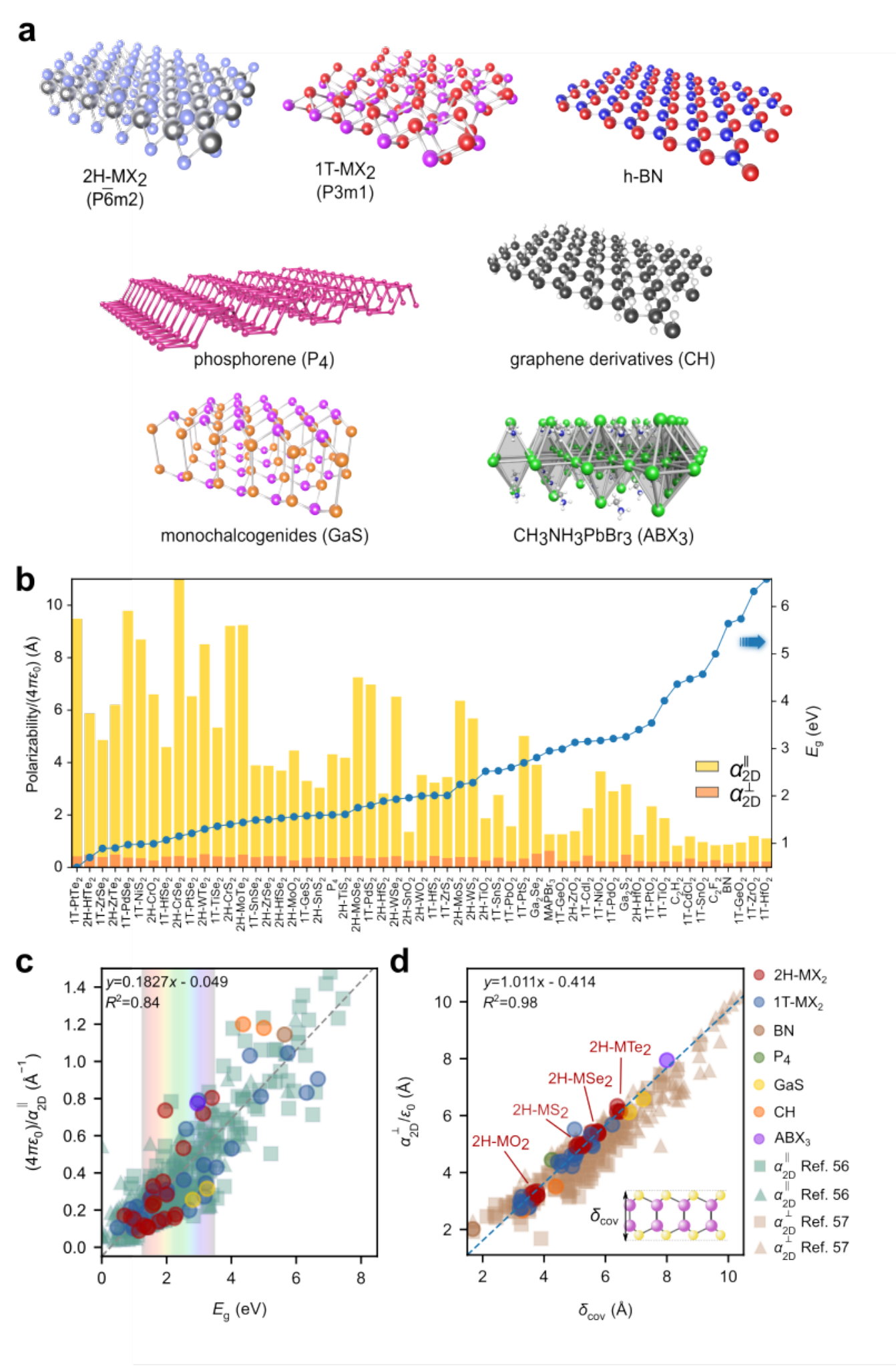}
\caption{\label{fig-3} \textbf{The universal scaling relation of the
    dielectric nature of 2D materials}. 
    \textbf{a,} The structures of
  the 2D materials investigated in this 
  study. 
  \textbf{b,} $\alpha_{\mathrm{2D}}^{\parallel}$,
  $\alpha_{\mathrm{2D}}^{\perp}$ (bar plots) and $E_{\mathrm{g}}$
  (blue dots) for all the 2D materials studied.
  $\alpha_{\mathrm{2D}}^{\parallel}$ is observed to descend with
  increasing $E_{\mathrm{g}}$, while no apparent relation between
  $\alpha_{\mathrm{2D}}^{\perp}$ and $E_{\mathrm{g}}$ is
  observed. HSE06 functional is used to obtain the data. 
  \textbf{c,} $(4\pi \varepsilon_{0})/\alpha_{\mathrm{2D}}^{\parallel}$ (in
  \AA{}$^{-1}$) as a function of $E_{\mathrm{g}}$, showing a linear
  correlation between each other. The energy range of visible light is
  shown in the background. 
  \textbf{d,} $\alpha_{\mathrm{2D}}^{\perp}/(\varepsilon_{0})$ (in \AA{}) as a
  function of $\delta_{\mathrm{cov}}$ (definition schematically shown
  in the inset), showing a perfect linear relation with a slope very
  close to $1$ (i.e.
  $\alpha_{\mathrm{2D}}^{\perp} \approx \varepsilon_{0}
  \delta_{\mathrm{cov}}$ ). The universal scaling relation is also
  revealed using different databank from 
  from Ref.~\citenum{Haastrup_2018} (squares),
  and Ref.~\citenum{Mounet_2018} (triangles) as superimposed on
  \textbf{c} and \textbf{d}. Data corresponding to 2H-MX$_2$ 
  (M=Mo, W; X=O, S, Se, Te) is highlighted in {\bf d}. The very tiny 
  difference in $\alpha_{\mathrm{2D}}^{\perp}/(\varepsilon_{0})$
  between compounds with different metal atoms 
  gives superposed magnitudes not distinguishable in the plot. 
  }
\end{figure}
computationally and experimentally determined. Despite the
simplicity of Eqs. \ref{eq:2D-Moss-para} and \ref{eq:2D-Moss-perp},
they generate direct relationships between the electronic polarizability and
the electronic/structural properties for any 2D material in a new
framework.

Next, we show that these equations are valid for the current
library of known layered materials involving different lattice
symmetry, element composition, optical and electronic properties (Figure \ref{fig-3}{\textbf a}).
A high-throughput screening performed on different families of TMDCs
(MX\(_{\text{2}}\), where M is a metal in groups 4, 6, 10, and X=O, S,
Se, Te) and phases (P\={6}m2, P3m1), metal monochalcogenides
(Ga\textsubscript{2}S\textsubscript{2},
Ga\textsubscript{2}Se\textsubscript{2}), cadmium halides (CdX$_2$,
X=Cl, I), hexagonal boron nitride (BN), graphene derivatives
(fluorographene (C\textsubscript{2}F\textsubscript{2}), graphane
(C\textsubscript{2}H\textsubscript{2})), phosphorene
(P\textsubscript{4}) and thin layer organic-inorganic perovskites
(ABX\textsubscript{3}), shows that our method enables 
full correlation between these disparate variables.
Figure \ref{fig-3}{\textbf b} compares the calculated fundamental
bandgap $E_{\mathrm{g}}$ (blue dots) and 2D electronic
polarizabilities (bar plots) of all the 2D materials investigated,
covering a wide spectrum range from far-infrared to ultraviolet.  Note that 
from dimension analysis, it is more intuitive to express the
polarizability as $\alpha_{\mathrm{2D}}/(4 \pi \varepsilon_{0})$,
which has unit of \AA{}. We find that
$\alpha_{\mathrm{2D}}^{\parallel}$ has a general descending trend when
$E_{\mathrm{g}}$ increases, while no apparent correlation between
$\alpha_{\mathrm{2D}}^{\perp}$ and $E_{\mathrm{g}}$ is observed (see
Supplementary Section \ref{SI-sec:pol-2D-Eg}).  We then examine
Eqs. \ref{eq:2D-Moss-para} and \ref{eq:2D-Moss-perp} using the
polarizabilities by first-principle calculations.  Figure
\ref{fig-3}\textbf{c} shows
$(4 \pi \varepsilon_{0})/\alpha_{\mathrm{2D}}^{\parallel}$ (in
\AA{}$^{-1}$) as a function of $E_{\mathrm{g}}$ (in eV) for the 2D 
materials investigated using HSE06 hybrid functional 
(circular dots) and PBE (triangles and squares).  
A linear
regression coefficient of $R^{2}=0.84$ indicates a strong correlation
between bandgaps and polarizabilities as predicted in
Eq. \ref{eq:2D-Moss-para}.  We also discovered that the linearity
between $(4 \pi \varepsilon_{0})/\alpha_{\mathrm{2D}}^{\parallel}$ and
$E_{\mathrm{g}}$ (measured by the $R^{2}$ value) is higher when the
bandgap is calculated using the HSE06 hybrid functional compared with
that from PBE exchange-correlation functional (see Supplementary
Section \ref{SI-sec:pol-2D-Eg} and Supplementary Figure
\ref{SI-fig:alpha-Eg-diff}). 
This is reasonable as the bandgaps for 2D materials 
obtained at the PBE functional, 
although may be close to experimental reported 
optical transition energies, are an artifact of the simulation due to a 
fortuitous error cancellation\cite{Heine15,Lee_2017}. Thus, the use of a 
time-consuming hybrid functional in our study is justified. 
A detailed benchmark
of Eqs.~\ref{eq:2D-Moss-para} and \ref{eq:2D-Moss-perp} using
different bandgaps, databases, and levels of theory can be seen
in Supplementary Sections~\ref{SI-sec:pol-2D-Eg}$-$\ref{SI-sec:gpaw}.

We further examine the validity of Eq.~\ref{eq:2D-Moss-perp}, that is,
the relation between $\alpha_{\mathrm{2D}}^{\perp}$ and the thickness 
of a 2D material. To test if the quantity $\hat{\delta}_{\mathrm{2D}}$
is physical, we choose the ``covalent'' thickness 
$\delta_{\mathrm{cov}}$ as a comparison.  
$\delta_{\mathrm{cov}}$ is defined as the longest distance
along the \textit{z}-direction between any two atom nuclei plus their covalent
radii:
\begin{equation}
  \label{eq:cov-thick}
  \delta_{\mathrm{cov}} = \mathrm{max}(|z^{i} - z^{j}|
  + r^{i}_{\mathrm{cov}} + r^{j}_{\mathrm{cov}})
\end{equation}
where $i$, $j$ are atomic indices in the 2D material and
$r_{\mathrm{cov}}^{i}$ is the covalent radius of atom $i$ (inset in Figure
\ref{fig-3}{\textbf d}). As shown in Figure~\ref{fig-3}{\textbf
  d}, $\alpha_{\mathrm{2D}}^{\perp}/\varepsilon_{0}$ (or equivalently,
$\hat{\delta}_{\mathrm{2D}}$) is very close to $\delta_{\mathrm{cov}}$
with a good linear correlation of $R^{2}=0.98$. This result indicates
a strong relation between $\alpha_{\mathrm{2D}}^{\perp}$ and the
geometry of the 2D layer, which can be approximated by
$\delta_{\mathrm{cov}}$. Similar to the molecular polarizability which
characterizes the volume of the electron distribution of an isolated molecule
\cite{Israelachvili_2011},
$\alpha_{\mathrm{2D}}^{\perp}/\varepsilon_{0}$ is also naturally
related to the characteristic thickness of the electron density of a 2D
material. Supplementary Section
\ref{SI-ssec:theory-1-perp-fundamental} shows an explanation of this
behavior from fundamental electrostatics and why
$\alpha_{\mathrm{2D}}^{\perp} / \varepsilon_{0}$ is close to
$\delta_{\mathrm{cov}}$. 
The geometric nature of $\alpha_{\mathrm{2D}}^{\perp}$ leads to
several interesting properties. For instance, the points corresponding
to 2H-TMDCs with same chalcogenide element (i.e. 2H-MO$_{2}$,
2H-MS$_{2}$, 2H-MSe2$_{2}$ or 2H-MTe$_{2}$, where M= Mo, W) 
lie very close in Figure
\ref{fig-3}\textbf{b} (detailed values see Supplementary Table
S2). This can be briefly explained by the fact that the difference
between covalent radii of transition metals (e.g. $\sim$8 pm between
Mo and W) is much smaller than that between group 16 elements
(e.g. $\sim$40 pm between O and S).
Our proposed definition of
$\hat{\delta}_{\mathrm{2D}}$ which is based on
Eq. \ref{eq:2D-Moss-perp} will provide insights 
on some long-existing controversies about the experimental thickness
of 2D materials~\cite{Shearer_2016} through a 
measurable quantity, e.g. $\alpha_{\mathrm{2D}}^{\perp}$
\cite{Antoine_1999,Cherniavskaya_2003,Krauss_1999_EFM}.

To rule out the possibility that our conclusion are limited by the
number of materials used at HSE06 level, we further validate
Eqs. \ref{eq:2D-Moss-para} and \ref{eq:2D-Moss-perp} using two
different 2D-material databases based on different codes\cite{Haastrup_2018,Mounet_2018}. 
We extracted the dielectric properties of over 300 compounds
calculated at the PBE level, and superimpose with our results in
Figure \ref{fig-3}{\textbf c} and \ref{fig-3}{\textbf d}. 
The high-throughput datasets also show linear trends for both
$(4\pi\varepsilon_0)/\alpha_{\mathrm{2D}}^{\parallel}$ (in \AA{}) vs
$E_{\mathrm{g}}$ (in eV) ($y=0.190x - 0.0619$, $R^{2}=0.842$) and
$\alpha_{\mathrm{2D}}^{\perp}$ vs $\delta_{\mathrm{cov}}$ (both in
\AA{}, $y=0.904x + 0.0551$, $R^{2} = 0.943$) relations. We notice that the
linear coefficients are similar but not identical to those calculated
at the HSE06 level. The discrepancies may be due to several factors
resulted from different choice of functionals, such as the underestimation
of the bandgap in PBE, and different description of the 
exchange-correlation potentials. 
We note that a more accurate estimation of the
coefficients should be performed with larger datasets and accurate
functionals which requires further work. Nevertheless, 
the validity of the linear trends observed for
$\alpha_{\mathrm{2D}}^{\parallel}$ and $\alpha_{\mathrm{2D}}^{\perp}$ 
is undeniable. We have also
searched for additional relations between the 2D polarizabilities with
other physical quantities, including the effective carrier mass,
quantum capacitance (density of states) and total atomic
polarizabilities with no apparent correlations being found
(Supplementary Section \ref{SI-sec:gpaw-3}).

\subsection{Application in multilayer and bulk systems}
\label{sec:apply-electr-polar}
\begin{figure}[H]
\centering
\includegraphics[width=0.95\linewidth]{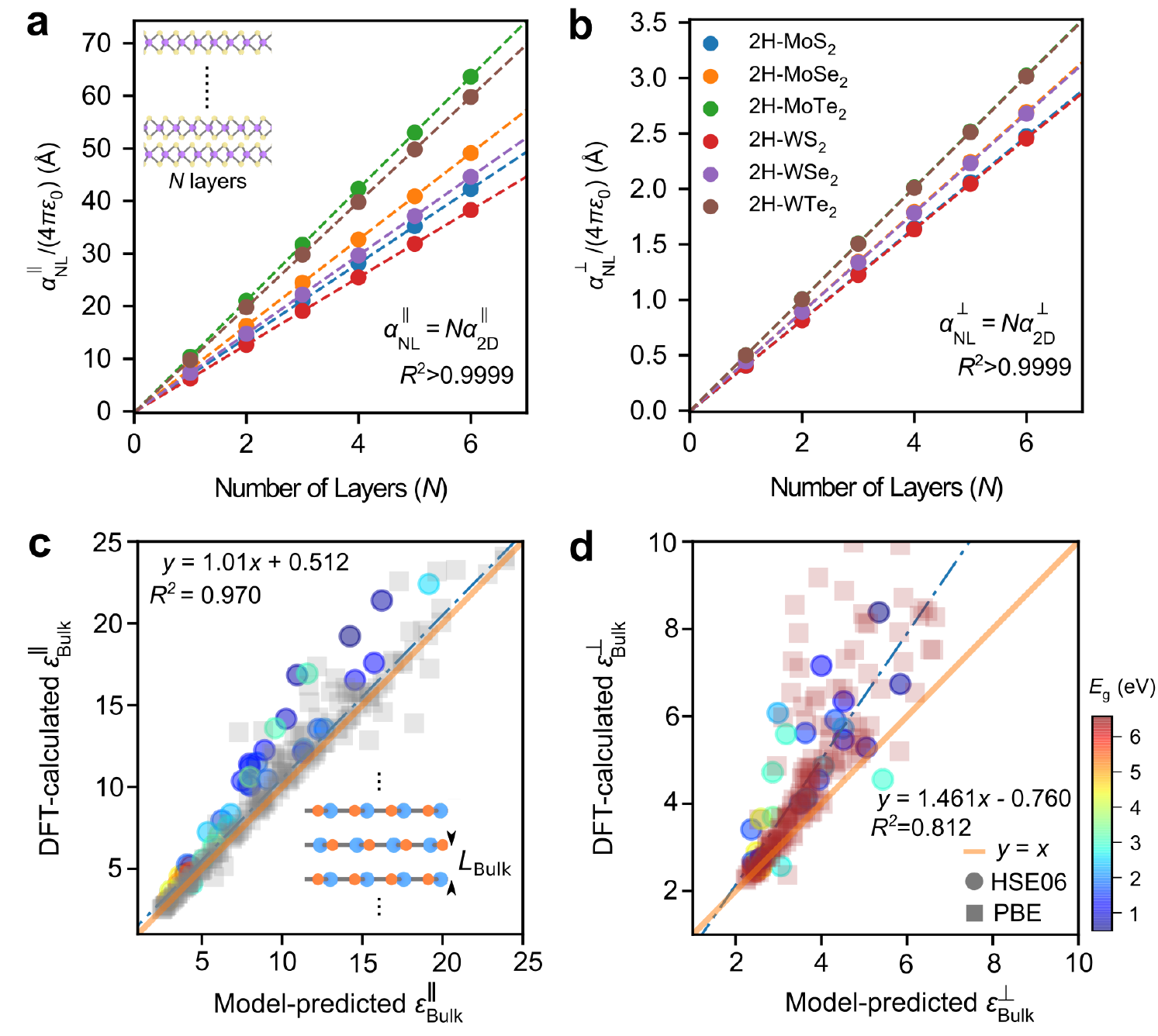}
\caption{\label{fig-4} \textbf{Application of 2D polarizability to
    few-layer and bulk systems}.  \textbf{a-b,} Multilayer
  polarizabilities $\alpha_{\mathrm{NL}}^{\parallel}$ and
  $\alpha_{\mathrm{NL}}^{\perp}$ of selected 2D metal 
  dichacolgenides (2H-MX$_2$, M=Mo, W; X=S, Se, Te)
  as a function of number of layers $N$, respectively.  
  Inset in {\bf a,} shows a 
   scheme of the 2D-3D transition. 
  $\alpha$ in the 2D material is essentially equivalent to
  $\varepsilon$ in its bulk counterpart.  
  Both $\alpha_{\mathrm{NL}}^{\parallel}$
  and $\alpha_{\mathrm{NL}}^{\perp}$ linearly scales with $N$ and the
  electronic polarizability of the monolayer which indicate that
  $\alpha_{\mathrm{2D}}$ is an additive quantity under weak
  interacting regime. 
  {\bf c-d,} DFT calculated $\varepsilon_{\mathrm{Bulk}}^{\parallel}$ and 
  $\varepsilon_{\mathrm{Bulk}}^{\perp}$, respectively, as a function of 
  their predicted values from the 2D polarizability model. A strong correlation 
  is observed in both components with the linear regression slope reaching the 
  unit for $\varepsilon_{\mathrm{Bulk}}^{\parallel}$ but slightly deviating for 
  $\varepsilon_{\mathrm{Bulk}}^{\perp}$ at higher magnitudes. 
  A heat map showing the dependence of $\varepsilon_{\mathrm{Bulk}}^{\parallel,\perp}$
   with the band gaps is included in {\bf d}. The model predicted values for $\varepsilon_{\mathrm{Bulk}}^{\perp}$ are in good aggreement with the DFT calculations when
  $E_{\mathrm{g}}>4$ eV. Inset in {\bf c,} shows the definition 
  of the interlayer distance in bulk $L_{\rm Bulk}$
  utilized to calculate $\varepsilon_{\mathrm{Bulk}}^{\parallel}$ and 
  $\varepsilon_{\mathrm{Bulk}}^{\perp}$ via Eqs. \ref{eq:3D-para}$-$\ref{eq:3D-para}.  
  Calculations at the level of HSE06 and PBE are shown in circles and 
  squares, respectively, in all panels that apply. 
  }
\end{figure}

The concept of electronic polarizability is not limited to monolayer
materials, and can be applied to multilayer and bulk systems as
well. For a 2D-material stack composed of $N$ layers, we can define
the electronic polarizability $\alpha_{\mathrm{NL}}$ similarly to
Eqs. \ref{eq:alpha-para-def}$-$\ref{eq:alpha-perp-def} by replacing 
$\alpha_{\mathrm{2D}}^{\parallel,\perp}$ to $\alpha_{\mathrm{NL}}^{\parallel,\perp}$. 
%
%
To check whether such assumption is valid, 
Figure \ref{fig-4}\textbf{a} and \ref{fig-4}\textbf{b} show
$\alpha_{\mathrm{NL}}^{\parallel}$ and $\alpha_{\mathrm{NL}}^{\perp}$
as functions of $N$ for several TMDCs in 2H-phase, 
respectively. Interestingly, we find that in all cases,
$\alpha_{\mathrm{NL}}$ exhibits nearly ideal linear relation with
$\alpha_{\mathrm{2D}}$, such that
$\alpha_{\mathrm{NL}}^{\parallel}= N \alpha_{\mathrm{2D}}^{\parallel}$
and $\alpha_{\mathrm{NL}}^{\perp}= N
\alpha_{\mathrm{2D}}^{\perp}$. Due to the relatively small
applied electric field (0.01 eV/\AA{}), the interlayer interactions within
the stack are negligible. Under such circumstances, 
$\alpha_{\mathrm{2D}}$ of individual layers is additive, 
which leads to the following general relation: 
\begin{equation}
  \label{eq:alpha-nl}
  \alpha_{\mathrm{NL}}^{p} = \sum_{i=1}^{N} \alpha_{\mathrm{2D, i}}^{p},\quad p=\parallel\ \mathrm{or}\ \perp
\end{equation}
where $\alpha_{\mathrm{2D, i}}$ is the electronic polarizability of
layer $i$, and $p$ is the direction of the polarization. This relation can
be additionally utilized to calculate screening inside 2D heterostructures
\cite{Kumar_2016_jpcc,Andersen_2015_dielec_vdWH}.

In a bulk material with an equilibrium
inter-layer distance $L_{\mathrm{Bulk}}$, we can follow a similar procedure as in multilayer by 
defining the polarizability as $\alpha_{\mathrm{Bulk}}$.  
Inspired by Eqs. \ref{eq:alpha-para-def} and
\ref{eq:alpha-perp-def}, the dielectric constants
$\varepsilon^{\parallel}_{\mathrm{Bulk}}$ and
$\varepsilon^{\perp}_{\mathrm{Bulk}}$ of the bulk layered material can
be reconstructed by $\alpha_{\mathrm{Bulk}}^{\parallel}$ and
$\alpha_{\mathrm{Bulk}}^{\perp}$ as:
\begin{subequations}
\begin{align}
  \label{eq:3D-para}
  \varepsilon^{\parallel}_{\mathrm{Bulk}}
  &= 1 + \frac{\alpha_{\mathrm{Bulk}}^{\parallel}}{\varepsilon_{0} L_{\mathrm{Bulk}}}
  \approx 1 + \frac{\alpha_{\mathrm{2D}}^{\parallel}}{\varepsilon_{0} L_{\mathrm{Bulk}}} \\
  \label{eq:3D-perp}
  \varepsilon^{\perp}_{\mathrm{Bulk}}
  &= \left(1 - \frac{\alpha_{\mathrm{Bulk}}^{\perp}}{\varepsilon_{0} L_{\mathrm{Bulk}}}\right)^{-1}
  \approx \left(1 - \frac{\alpha_{\mathrm{2D}}^{\perp}}{\varepsilon_{0} L_{\mathrm{Bulk}}}\right)^{-1}
\end{align}
\end{subequations}
Here we neglect the effect of the stacking order of the layers and hypothesized 
that the basic building blocks for the dielectric response of the bulk are the polarizability of the 
individual layers subject to vdW and electrostatic interactions. 
The dielectric constant $\varepsilon$ although not well-defined for a
monolayer 2D material becomes applicable when the 2D layers are put
together as shown in the following. 
We compare the values of
$\varepsilon_{\mathrm{bulk}}^{\parallel}$ and
$\varepsilon_{\mathrm{bulk}}^{\perp}$ computed from DFT simulations (\textit{y}-axis)
with those predicted using Eqs \ref{eq:3D-para} and \ref{eq:3D-perp}
(\textit{x}-axis) as shown in Figure \ref{fig-4}{\textbf c} and \ref{fig-4}{\textbf d}. 
Strikingly both HSE06 and PBE datasets give almost identical results which suggest 
a non-method dependent behavior. 
We observe that
$\varepsilon_{\mathrm{bulk}}^{\parallel}$ values calculated by DFT and
predicted by Eq. \ref{eq:3D-para} are in sound agreement with a linear
regression slope of 1.01 and $R^2$ of 0.97. Conversely, 
$\varepsilon_{\mathrm{bulk}}^{\perp}$ values predicted from
Eq. \ref{eq:3D-perp} fairly agree with the DFT-calculated values when
$E_{\mathrm{g}}>4$ eV, while the deviation becomes larger when
$E_{\mathrm{g}}$ reduces. The above results indicate that
$\alpha^{\parallel}_{\mathrm{Bulk}}$ can generally be estimated with 
high accuracy from its 2D counterpart, while $\alpha^{\perp}_{\mathrm{Bulk}}$ differs due
to the interlayer coupling and overlap between induced
dipole\cite{Andersen_2015_dielec_vdWH,Laturia_2018}. 
Nevertheless, as most of the optical response and electronic device properties rely on the 
in-plane dielectric constant for practical applications, the possibility to handily
estimate $\alpha_{\mathrm{2D}}^{\parallel}$ from well established magnitudes of 
$\varepsilon_{\mathrm{bulk}}^{\parallel}$, for instance, from material databases,   
using reverse engineering in Eq.\ref{eq:3D-para}, it is a step forward in the 
design and understanding of the dielectric phenomena in 2D. 


\subsection{Unified geometric representation of $\alpha_{\mathrm{2D}}$}
\label{sec:unif-geom-repr}

\begin{figure}[H]
  \centering
  \includegraphics[width=0.8\linewidth]{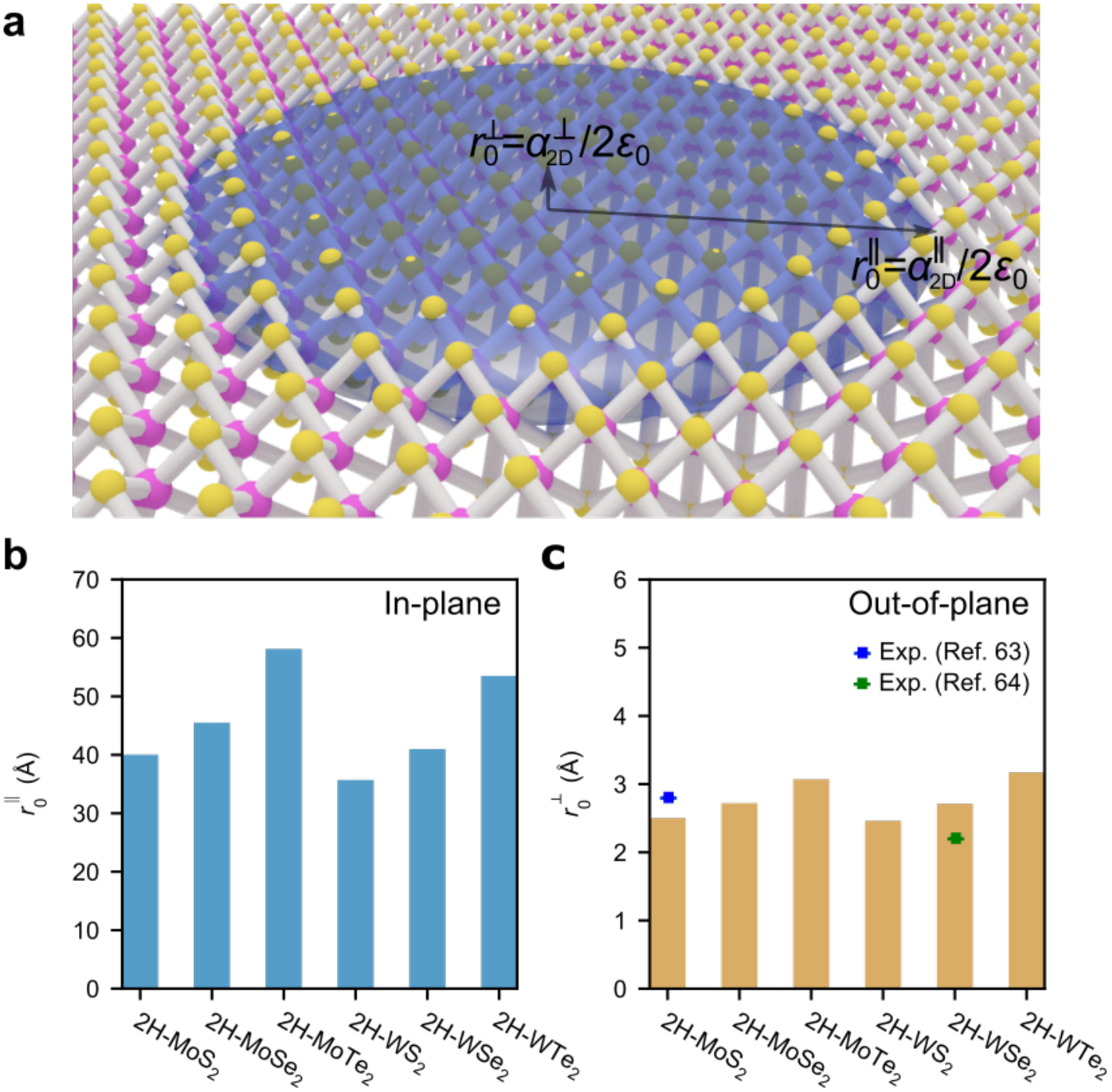}
  \caption{\label{fig-ellip} \textbf{Geometric representation of the
      2D polarizability}. 
      \textbf{a,} Scheme of the polarizability
    ellipsoid of a 2D material, with its in-plane
    ($r_{0}^{\parallel}$) and out-of-plane radii
    ($r_{\mathrm{0}}^{\perp}$) proportional to
    $\alpha_{\mathrm{2D}}^{\parallel}$ and
    $\alpha_{\mathrm{2D}}^{\perp}$, respectively.  
    {\bf b-c,} Calculated magnitudes of 
    $r_{0}^{\parallel}$ and $r_{0}^{\perp}$, respectively, for selected 2D TMDCs. 
    The polarizability ellipsoid is highly anisotropic with screening much
    stronger at in-plane than out-of-plane directions. Comparison with available experimental 
    results\cite{Roch_2018,Verzhbitskiy19} for 2H-MoS$_2$ and 2H-WSe$_2$ is included.} 
\end{figure}

Lastly, we demonstrate that both $\alpha_{\mathrm{2D}}^{\parallel}$
and $\alpha_{\mathrm{2D}}^{\perp}$ can be unified using a geometric
approach. In merit of the unit analysis,
$\alpha_{\mathrm{2D}}^{\parallel}$ and $\alpha_{\mathrm{2D}}^{\perp}$
both have unit of $4\pi\varepsilon_{0} \times$[Length]. In other words,
they represent in- and out-of-plane characteristic lengths,
respectively. 
It is well-known that the in-plane screened
electrostatic potential 
$V(r) = {\displaystyle \frac{e}{4 \alpha_{\mathrm{2D}}^{\parallel}}}
\left[H_{0}({\displaystyle \frac{2\varepsilon_{0}
      r}{\alpha_{\mathrm{2D}}^{\parallel}}}) - Y_{0}( {\displaystyle
    \frac{2
      \varepsilon_{0}r}{\alpha_{\mathrm{2D}}^{\parallel}}})\right]$
from a point charge as a function of distance $r$\cite{Keldysh_1979_eps_multi,Pulci_2014} 
(where $H_{0}$ is the Struve
function and $Y_{0}$ is the Bessel function of second kind) 
is associated with the in-plane screening radius
$r_{0}^{\parallel}=\alpha_{\mathrm{2D}}^{\parallel}/(2
\varepsilon_{0})$, such that $V(r,r/r^{\parallel}_{0} \gg 1)$ reduces
to the simple Coulomb potential in vacuum. Combining with the result
that $\alpha_{\mathrm{2D}}^{\perp}/\varepsilon_{0}$ characterizes the
thickness of a 2D material, we can view the dielectric screening of a
point charge sitting in the middle of a 2D material as an ellipsoid
with the radii of principal axes to be
$r_{0}^{\parallel} = \alpha_{\mathrm{2D}}^{\parallel}/(2
\varepsilon_{0})$ and
$r_{0}^{\perp} = \alpha^{\perp}_{\mathrm{2D}}/(2 \varepsilon_{0})$,
respectively, as illustrated in Figure~\ref{fig-ellip}\textbf{a}.
This is analog to the polarizability ellipsoid picture of molecules
used in spectroscopy \cite{Banwell_1994}. The polarizability ellipsoid
for a 2D material is in general ultra flat, with
$r_{0}^{\parallel} \gg r_{0}^{\perp}$, as demonstrated by layered materials 
of group 6 of 2H-TMDCs (Figure~\ref{fig-ellip}\textbf{b} and
~\ref{fig-ellip}\textbf{c}). 
The picture of the polarizability ellipsoid
provides further insights into the physical nature of
$\alpha_{\mathrm{2D}}$: $r_{0}^{\parallel}$ is close to the exciton
radius that it is confined within the 2D plane~\cite{Pulci_2014}. This radius is
generally larger for a smaller bandgap semiconductor, and can be
converted through the exciton binding energy as proposed in
Refs.\cite{Olsen_2016_hydrogen,Jiang_2017_Eg_Eb}.
%
$r_{0}^{\perp}$ in its turn can be 
indirectly deduced from Stark effect for perpendicular electric fields 
\cite{Pedersen_2016,Klein_2016,Roch_2018}. A comparison with 
available experimental data\cite{Verzhbitskiy19, Roch_2018} gives 
close magnitudes with our predicted values. 

\begin{figure}[H]
  \centering
  \includegraphics[width=1.01\linewidth]{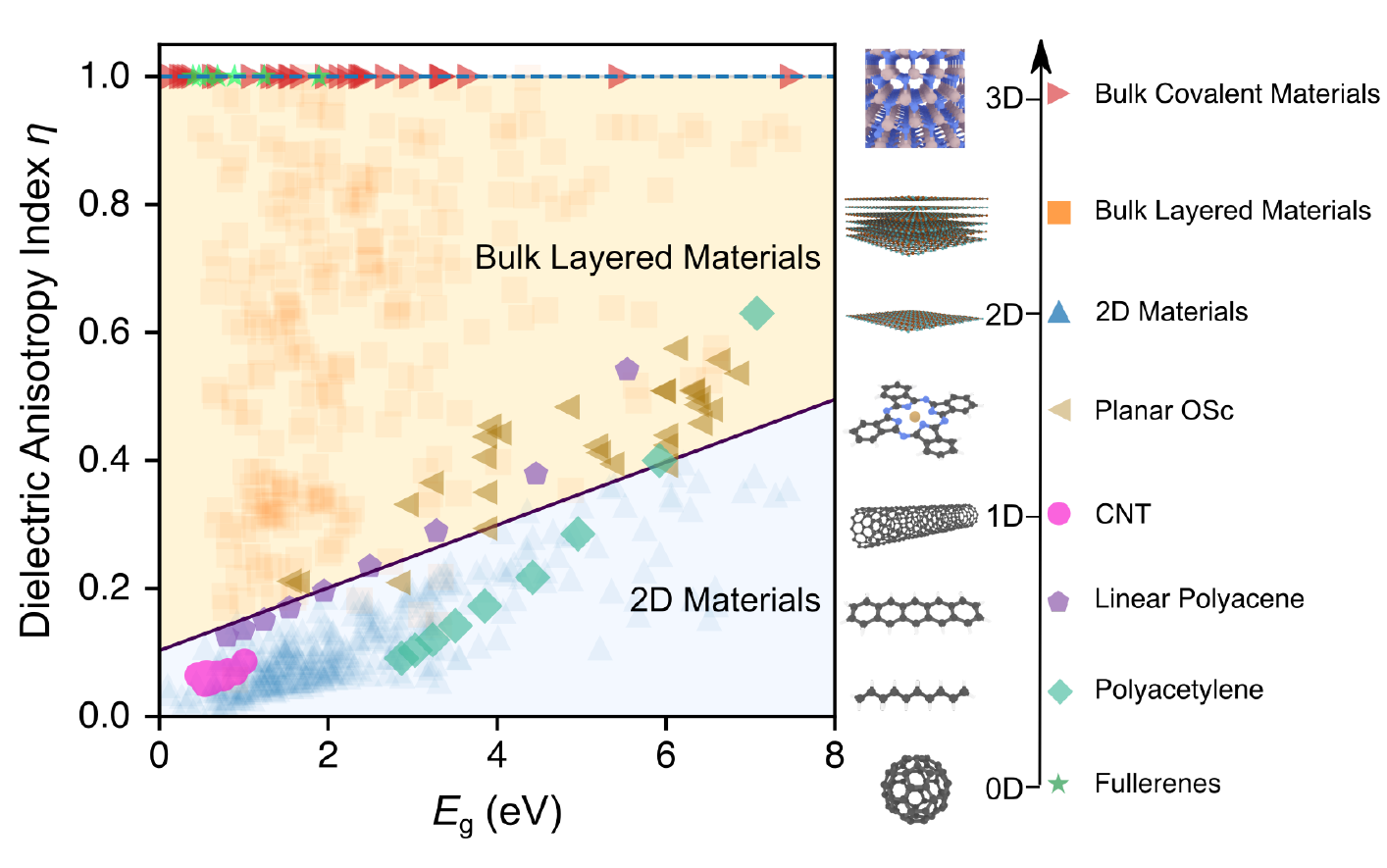}
  \caption{\textbf{Phase diagram of dielectric anisotropy $\eta$ as
      function of bandgap $E_{\mathrm{g}}$}. The
    $\eta$-$E_{\mathrm{g}}$ values of 2D materials (blue triangle) and
    their bulk counterparts (orange square) can be distinguished by
    the line $\eta=0.048(E_{\mathrm{g}}/\mathrm{eV})+0.087$. $\eta-E_{\mathrm{g}}$ values of
    semiconducting materials in other dimensions are also superimposed
    for comparison. Isotropic dielectric property is observed for bulk
    covalent materials (3D, red triangle) and fullerenes (0D, green
    star), while reduced dimensional materials, including planar
    organic semiconductor(OSc, 1D-2D, brown triangle), carbon nanotube
    (CNT, magenta circle) and linear OSc (0D-1D, violet pentagon) are
    scattered along the boundary line. The dimensionality and
    structure of typical materials are shown along the axis on the
    right. Compared with other materials, 2D materials and their bulk
    counterparts provide more flexibility of controlling the
    dielectric anisotropy.}
  \label{fig:aniso}
\end{figure}

Inspired by the polarizability ellipsoid model, we will show that a general 
picture of the dielectric properties in
any dimension can be drawn by studying the dielectric
anisotropy. That is, the dielectric response of a material along 
its different geometrical orientations. 
We define the dielectric anisotropy index $\eta$ as:
\begin{equation}
  \label{eq:anisotropy}
  \begin{aligned}[t]
    \eta =
    \begin{cases}
      {\displaystyle \min_{i \neq j}}
      {\displaystyle
        \left(\frac{\varepsilon^{ii}}{\varepsilon^{jj}}\right)},
      \ \mathrm{Bulk\ Materials}\\
      {\displaystyle \min_{i \neq j}}
      {\displaystyle
        \left(\frac{\alpha_{\mathrm{2D}}^{ii}}{\alpha_{\mathrm{2D}}^{jj}}\right)},
      \ \mathrm{2D\ Materials}\\
    \end{cases}
  \end{aligned}
\end{equation}
$\eta=1$ indicates that the material has isotropic dielectric properties
while $\eta \to 0$ means that the dielectric property is highly
anisotropic. Figure \ref{fig:aniso} shows the phase diagram of $\eta$
as function of $E_{\mathrm{g}}$ for 2D materials and their bulk
counterparts. Interestingly, the 2D materials (blue triangles) can be
clearly distinguished from the bulk layered materials (orange squares)
with the boundary line determined to be
$\eta =0.048 (E_{\mathrm{g}}/ \mathrm{eV})+0.087$. The much lower
$\eta$ values for 2D materials compared with their bulk counterparts
indicates a high dielectric anisotropy, which is responsible for the
unique 2D optoelectronic properties, such as the electrostatic
transparency phenomena\cite{Liluhua_2014,Tian_2016,Li_2018} and the large exciton
binding energies
\cite{Pulci_2014,Tran_2014,Chernikov_2014_EB_MoS2_2D3D,Berkelbach_2013}. From
Eqs. \ref{eq:2D-Moss-para}, \ref{eq:2D-Moss-perp} and
\ref{eq:anisotropy} we can see $\eta$ is roughly proportional to
$E_{\mathrm{g}} \times \delta$, which explains the observation that
$\eta$ for 2D materials increase almost linearly with
$E_{\mathrm{g}}$, since the layer thickness $\delta$ (mostly 3--10
\AA{}) of the 2D materials investigated varies much less than
$E_{\mathrm{g}}$ in the range of 0.1--7 eV (Figure \ref{fig-3}b$-$\ref{fig-3}c). 
Further analysis shows that the dielectric anisotropy
index of any bulk layered material $\eta_{\mathrm{Bulk}}$ obeys
$\eta_{\mathrm{Bulk}} \geq {\displaystyle \frac{4
    \eta_{\mathrm{2D}}}{(\eta_{\mathrm{2D}}+1)^{2}}} \geq
\eta_{\mathrm{2D}}$, where $\eta_{\mathrm{2D}}$ is the anisotropy
index of corresponding 2D layer, which is the basis for the
separation of bulk and 2D regimes in the $\eta-E_{\mathrm{g}}$ phase
diagram (Supplementary Section \ref{SI-sec:aniso}).  
For comparison, we also superimpose the dielectric
anisotropy indices of common semiconducting materials in other
dimensions on
the phase diagram in Figure \ref{fig:aniso}. Bulk covalent 3D (e.g. Si, GaN) and 0D (e.g. fullerenes) semiconductors show isotropic
dielectric properties, scattered along the line $\eta=1$. 
Conversely, reduced dimensionality increases the dielectric anisotropy of
materials such as planar organic semiconductor (OSc) in 1D-2D 
(e.g. CuPc), carbon nanotube (CNT) in 1D, linear OSc in 0D-1D
(e.g. polyacene and polyacetylene). 
Interestingly, most of these
materials also scatter along the boundary line separating the bulk and
2D regimes, indicating that the criteria distinguishing 2D (more 
anisotropic) and bulk materials (more isotropic) from the
$\eta-E_{\mathrm{g}}$ diagram, can also be applied to other
dimensions. From the phase diagram, we can see that 2D and bulk
layered materials, including 2D van der Waals heterostructure
(vdWH)\cite{Novoselov_2016}, provides more flexibility in
controlling the dielectric and electronic properties, compared with
covalent semiconductors (without vdW gaps) in other dimensions.

\section{Conclusion}

Our results show that the 2D electronic polarizability
$\alpha_{\mathrm{2D}}$ is a local variable determining the dielectric
properties of 2D materials.  There exist well-defined relationships
between $\alpha_{\mathrm{2D}}$ and other quantities hidden in the
electronic properties.  According to our analysis, simple scaling
equations involving bandgap and layer thickness can be used to
describe both dielectric and electronic features at the same
footing. A dielectric anisotropy index is found relating any material
dimension with its controllability.  Thus, our results suggest that
the challenge of understanding the dielectric phenomena is in general
a geometrical problem mediated by the bandgap. We believe the
principles presented here will benefit both fundamental understanding
of 2D materials as well as a rational device design and optimization.

\section{Theoretical Methods}
\label{sec:org8457dbb}

Simulations were carried out using plane-wave density functional
theory package VASP \cite{Kresse_1993,Kresse_1996_1,Kresse_1996_2}
using the projector augmented wave (PAW) approach with GW
pseudopotentials \cite{Kresse_1999_pseudopotentials}. Band gaps were
calculated using the Heyd-Scuseria-Ernzerhof hybrid functional (HSE06)
\cite{Heyd_2003,HSE_2006}, with spin orbit coupling (SOC) explicitly
included. The geometries were converged both in cell parameters and
ionic positions, with forces below 0.04 eV/\AA. To ensure the accuracy
of dielectric property of monolayer, a vacuum spacing of $>$ 15 \AA~is
used. A k-point grid of \(7\times7\times1\) was used to relax the
superlattice, with an initial relaxation carried out at the
Perdew-Burke-Ernzerhof
(PBE)\cite{Perdew_1996,Ernzerhof99,Paier_2005_PBE}
exchange-correlation functional level and a subsequent relaxation
carried out at HSE06 level, allowing both cell parameters and ionic
positions to relax each time. In VASP, the tag PREC=High was used,
giving a plane wave kinetic energy cutoff of 30\% greater than the
highest given in the pseudopotentials used in each material. This
guarantees that absolute energies were converged to a few meV and the
stress tensor to within 0.01 kBar.  Calculation of the macroscopic
ion-clamped dielectric tensor were carried out with an
18$\times$18$\times$1 k-grid and electric field strength of 0.001
eV/\AA.  Local field effect corrections are included at the
exchange-correlation potential $V_{\mathrm{xc}}$ at both PBE and HSE06
levels. The materials from Ref.\citenum{Haastrup_2018} for comparison
were choses with the GW bandgap larger than 0.05 eV. Bulk layered
materials were constructed by relaxing the c-axis length of
corresponding monolayer material with the interlayer van der Waals
(vdW) interactions calculated by non-local vdW correlation
functional\cite{Lee_2010_vdFD2}.  The dielectric properties of bulk
layered materials using VASP were calculated at HSE06 level with
18$\times$18$\times$6 k-grid with same parameter as for monolayer,
while the dielectric properties of bulk counterparts of
Ref.~\citenum{Haastrup_2018} are calculated at PBE level with a
k-point density of 10~\AA$^{-1}$. Local field effect corrections are
also used for the dielectric properties of bulk systems.

\section*{Data Availability}
The data that support the findings of this study 
is available within the paper and its Supplementary Information.  

\subsubsection*{Competing interests}
The Authors declare no conflict of interests.

\subsubsection*{Acknowledgments}
C.J.S. and T.T. are grateful for financial support from ETH startup funding. 
L.H.L. thanks the financial support from Australian Research Council (ARC) 
via Discovery Early Career Researcher Award (DE160100796). 
E.J.G.S. acknowledges the use of computational resources from the UK 
Materials and Molecular Modelling Hub for access to THOMAS 
supercluster, which is partially funded by EPSRC (EP/P020194/1); and CIRRUS Tier-2 HPC 
Service (ec019 Cirrus Project) at EPCC (http://www.cirrus.ac.uk) funded 
by the University of Edinburgh and EPSRC (EP/P020267/1). 
The Department for the Economy (USI 097) is acknowledged for funding support.

\subsubsection*{Author Contributions}
E.J.G.S. conceived the idea and supervised the project. 
T.T., D.S., D.H. and E.J.G.S. performed the first-principles simulations 
and data analytics. T.T. developed the analytical model with 
inputs from E.J.G.S. and C.J.S. L.H.L. and J.N.C. 
performed numerical analysis and contributed to the discussions together with M.C.
E.J.G.S. and T.T. co-wrote the manuscript with inputs from all authors. 
All authors contributed to this work, read the manuscript, discussed 
the results, and all agree to the contents of the manuscript.

\section*{Supporting Information}

The Supporting Information contains detailed descriptions and
discussions about dielectric properties of 2D materials, effective
dielectric model, derivations of the 2D polarizability-based model,
dependency of $\alpha_{\mathrm{2D}}$ on the choice of bandgap,
relation between 2D and 3D properties, explanations of the dielectric
anisotropy, as well as raw data sheet from first principles
calculations.

\bibliography{ref}

\label{sec:org34cbe74}
\clearpage

\section*{TOC entry}
\begin{figure}
  \centering
  \includegraphics[width=0.95\linewidth]{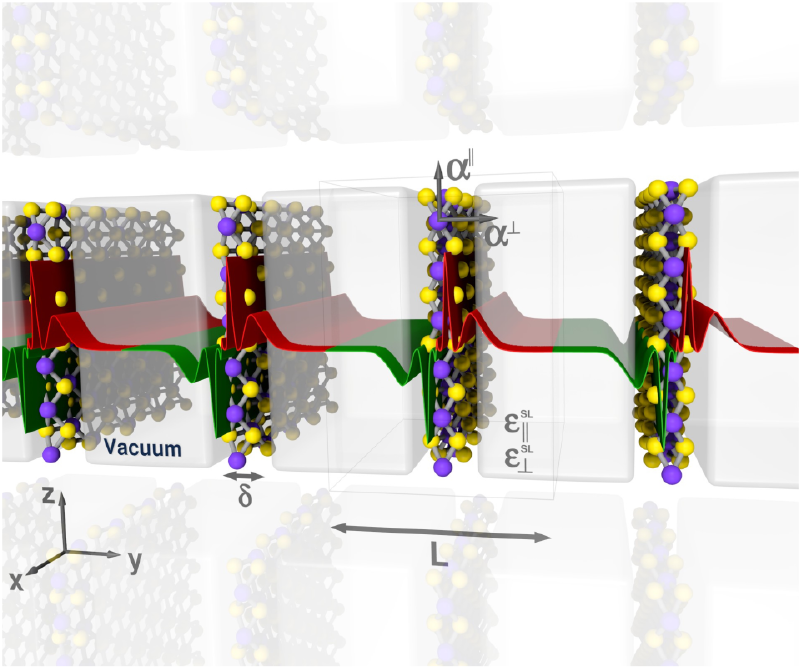}  
\end{figure}

\end{document}


\newpage{}

\section*{Table of Contents}

\begin{itemize}
\item Supplementary Section \ref{sec:polarizability-analysis}: Further analysis on the dielectric properties of 2D materials
  
\item Supplementary Section \ref{sec:2D-3D-rescale}: Hypothetical 2D ``dielectric constant'' rescaled from the 3D dielectric constant

\item Supplementary Section \ref{sec:theory-1}: Polarizability-based theoretical model

\item Supplementary Section \ref{sec:pol-2D-Eg}: Dependence of $\alpha_{\mathrm{2D}}$ on bandgap

\item Supplementary Section \ref{sec:gpaw}: Validation of results using a larger 2D material database

\item Supplementary Section \ref{sec:2D-3D}: More discussion about the
  relation between 2D and 3D properties

  \item Supplementary Section \ref{sec:aniso}: More discussions about the dielectric anisotropy
  
\item Supplementary Section \ref{sec:raw}: Raw data from first
  principles calculations

\end{itemize}

\pagebreak{}

\section{Further analysis on the dielectric properties of 2D materials}
\label{sec:polarizability-analysis}

In this section we provide more analysis on the dielectric properties
of 2D materials calculated using many-body Green function
method (G$_0$W$_0$), including electron-hole interactions at the level
of the Bethe-Salpeter equation (G$_0$W$_0$--BSE), and at the
frequency-dependent regime.

\subsection{Profile of induced dipoles of 2D material}
\label{sec:dipole-plot}
\begin{figure}[htbp]
 \centering
 \includegraphics[width=0.5\linewidth]{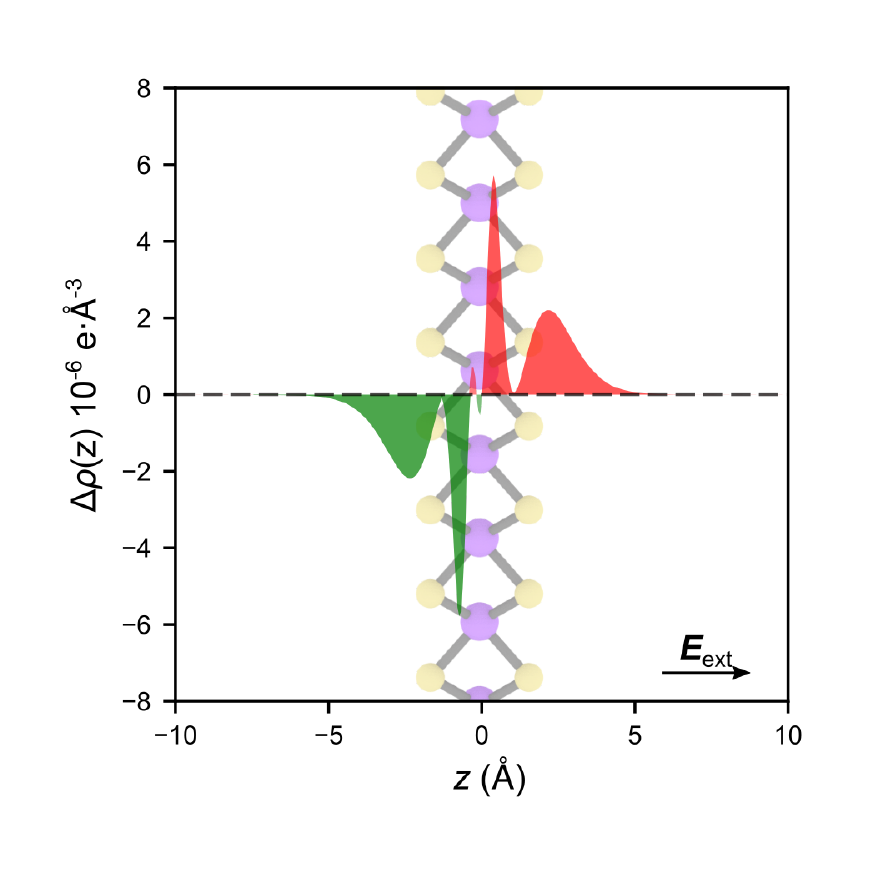}
 \caption{$\Delta \rho$ as a function of $z$ around the
   MoS$_{2}$ layer, corresponding to main text Figure
   \ref{main-fig-1}\textbf{a}. Green and red parts corresponding to negative and
   positive induced charges. The external electric field
   $E_{\mathrm{ext}}$ is 0.01 eV/\AA{}.}
 \label{fig:rho-profile}
\end{figure}

Here we show in detail the
$\Delta {\rho}=\rho(\boldsymbol{E}) - \rho(\boldsymbol{E}=0)$ profile of
the 2H-MoS$_{2}$ slab in main text Figure \ref{main-fig-1}. The
density $\Delta \rho$ is calculated via $\Delta \rho(z) = \frac{1}{S} \int_{S} \Delta \rho (x,y;z) dx dy $, 
where $S$ is the surface of the unit cell perpendicular to a given direction, in this case $z$. 
As can be seen in Figure \ref{fig:rho-profile} the induced charges on
the MoS$_{2}$ layer only extends to a width of $\sim{}$12 \AA{}
centered at the middle of the layer. This corresponds to about 5-6 \AA~
from each side.  When the SL size $L \gg$12 \AA{} as in the first
principle calculations shown in the main text, the induced dipoles from
the periodic images do not interact thus giving the converged
values of $\alpha_{\mathrm{2D}}$.

\subsection{Dielectric properties calculated using many-body Green function method and frequency dependency}
\label{ssec:gw}

Here we will show the results of dielectric properties calculated
using many-body Green function method (G$_{0}$W$_{0}$) and with
electron-hole interactions at the level of the Bethe-Salpeter equation
(G$_{0}$W$_{0}$-BSE). Frequency-dependent dielectric functions,
$\varepsilon_{\mathrm{SL}}(\omega)$, were calculated at the level of
G$_{0}$W$_{0}$ and G$_{0}$W$_{0}$+BSE levels using VASP. For the
calculations on G$_{0}$W$_{0}$, a 12$\times$12$\times$1 $\Gamma$-centered k-grid was
used along with a 800 eV energy cutoff in the plane waves and in
calculation of the response function. 120 bands (4 occupied and 116
unoccupied) were used in the calculation of
$\varepsilon_{\mathrm{SL}}(\omega)$ of monolayer BN, with local-field
effects being included. For the calculation of the
Bethe-Salpeter equation, the Tamm-Dancoff approximation was used with
two occupied and two unoccupied bands being included. 

\begin{figure}[htbp]
  \centering
 \includegraphics[width=0.9\linewidth]{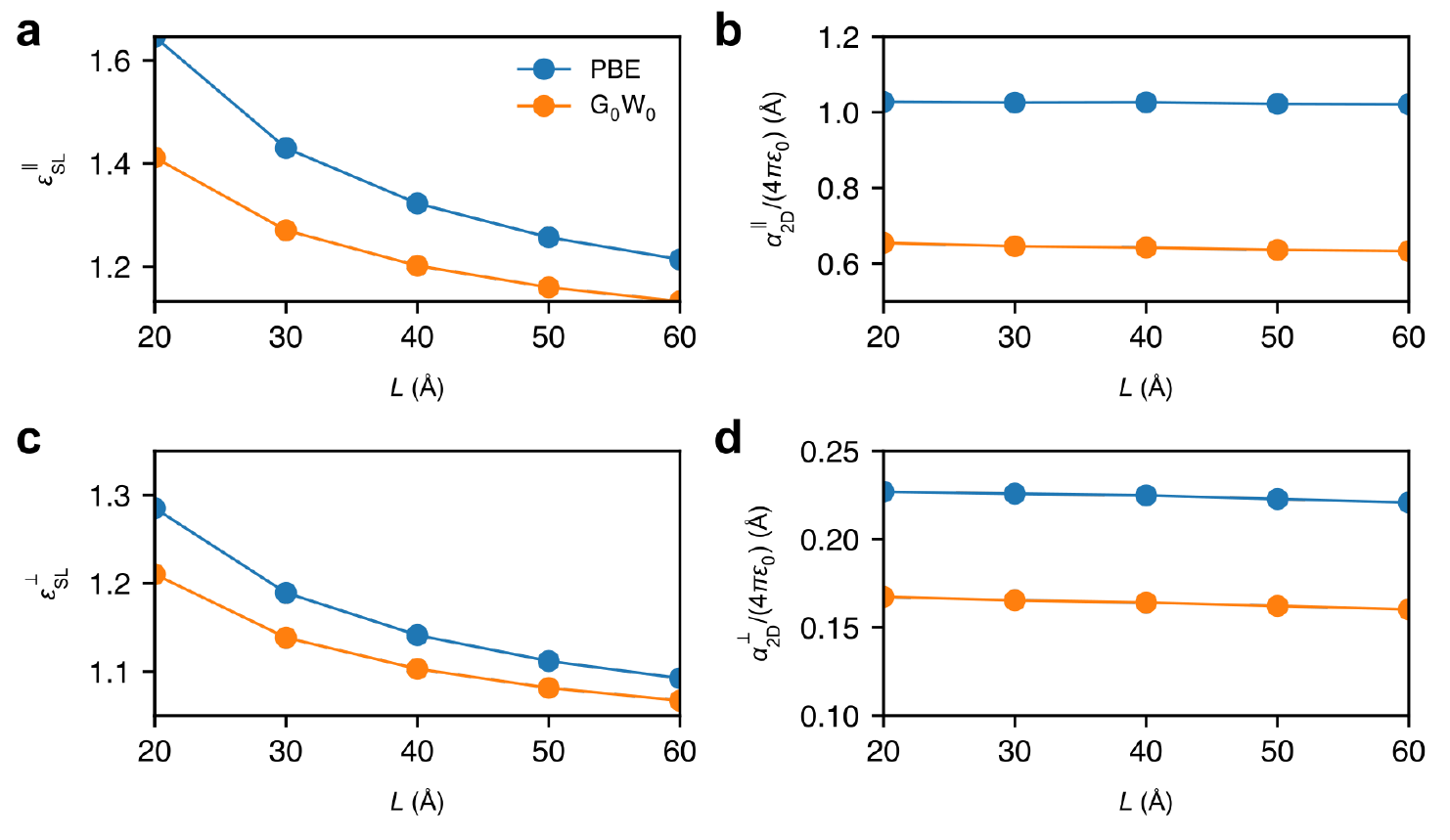}
 \caption{{\bf Variation of 
   $\varepsilon^{\perp}_{\mathrm{SL}}$ as a function of $L$ for
   monolayer BN calculated at the level of PBE and G$_{\rm 0}$W$_{0}$.}
   \textbf{a}-{\bf b}. $\varepsilon_{\mathrm{SL}}^{\parallel}$ and $\alpha_{\mathrm{2D}}^{\parallel}$ as function of $L$, respectively.    
  \textbf{c}-{\bf d}. $\varepsilon_{\mathrm{SL}}^{\perp}$ and 
  $\alpha_{\mathrm{2D}}^{\perp}$ as
   function of $L$. 
   In all cases the values obtained using
   G$_{0}$W$_{0}$ method is smaller than that using PBE method due to
   better estimation of the bandgap.}
  \label{fig:GW-PBE-alpha}
\end{figure}

We first compare the case of a monolayer of BN within a varying
superlattice $L$ calculated using PBE and G$_{0}$W$_{0}$ as shown in
Figure \ref{fig:GW-PBE-alpha}.  Both
$\varepsilon_{\mathrm{SL}}^{\parallel}$ and
$\varepsilon_{\mathrm{SL}}^{\perp}$ do not converge as a function of
$L$ despite of the separation utilized in the simulations (Figure
\ref{fig:GW-PBE-alpha}{\bf a}-{\bf c}). However, corresponding 2D
polarizabilities are almost $L$-independent (Figure
\ref{fig:GW-PBE-alpha}{\bf b}-{\bf d}). It is worth noting that since
the G$_{0}$W$_{0}$ method has better estimation of the electronic bandgap,
$\varepsilon_{\mathrm{SL}}$ and $\alpha_{\mathrm{2D}}$ are smaller
using G$_{0}$W$_{0}$ than in PBE functionals.

Next we will investigate the frequency-dependent dielectric properties of the 2D materials using various methods. The imaginary part of the frequency-dependent dielectric function of a periodic system is calculated using the following relation:

\begin{equation}
\begin{aligned}
  \label{eq:dft-dielectric}
\varepsilon_{\alpha\beta}^{(2)}(\omega)=\frac{4\pi e^2}{\Omega}\lim\limits_{q\rightarrow 0}\frac{1}{q^2}\sum_{v,c,\mathbf{k}}&2\omega_\mathbf{k}\delta(\epsilon_{c\mathbf{k}}-\epsilon_{v\mathbf{k}}-\omega) \\
&\times\braket{u_{c\mathbf{k}+q\mathbf{e}_\alpha}|u_{v\mathbf{k}}}\braket{u_{v\mathbf{k}}|u_{c\mathbf{k}+q\mathbf{e}_\beta}}
\end{aligned}
\end{equation}
Through the Kramers-Kronig transformation the real part of the dielectric function can be obtained as: 

\begin{equation}
\label{eq:dft-dielectric-real}
  \varepsilon_{\alpha\beta}^{(1)}(\omega)=1+\frac{2}{\pi}\int_0^\infty\frac{\varepsilon_{\alpha\beta}^{(2)}(\omega')\omega'}{\omega'^2-\omega^2}d\omega'
\end{equation}

\begin{figure}[htbp]
  \centering
 \includegraphics[width=1.00\linewidth]{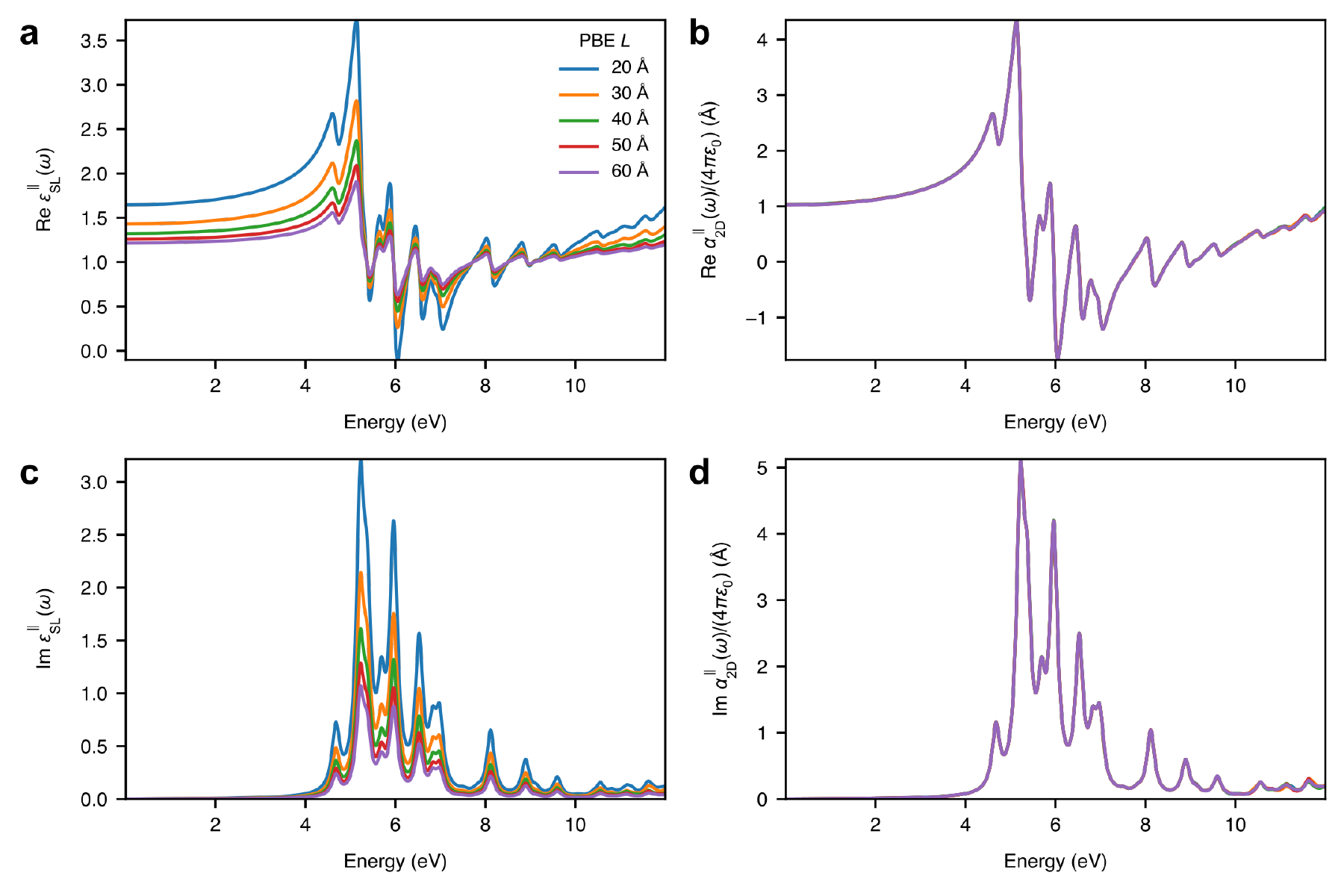}
 \caption{Dependence of in-plane dielectric properties on $L$ for
   monolayer BN calculated using PBE method: \textbf{a} Real part of
   $\varepsilon_{\mathrm{SL}}^{\parallel}$, \textbf{b} Real part of
   $\alpha_{\mathrm{2D}}^{\parallel}$, \textbf{c} Imaginary part of
   $\varepsilon_{\mathrm{SL}}^{\parallel}$, \textbf{d} Imaginary part
   of $\alpha_{\mathrm{2D}}^{\parallel}$.  The same {\bf k}-sampling
   is used in all simulations with the only variable quantity being
   $L$. Clearly, the $L$-dependency of the superlattice dielectric
   function is removed using 2D polarizabilities.}
  \label{fig:PBE-omega-in}
\end{figure}

\begin{figure}[htbp]
  \centering
  \includegraphics[width=1.0\linewidth]{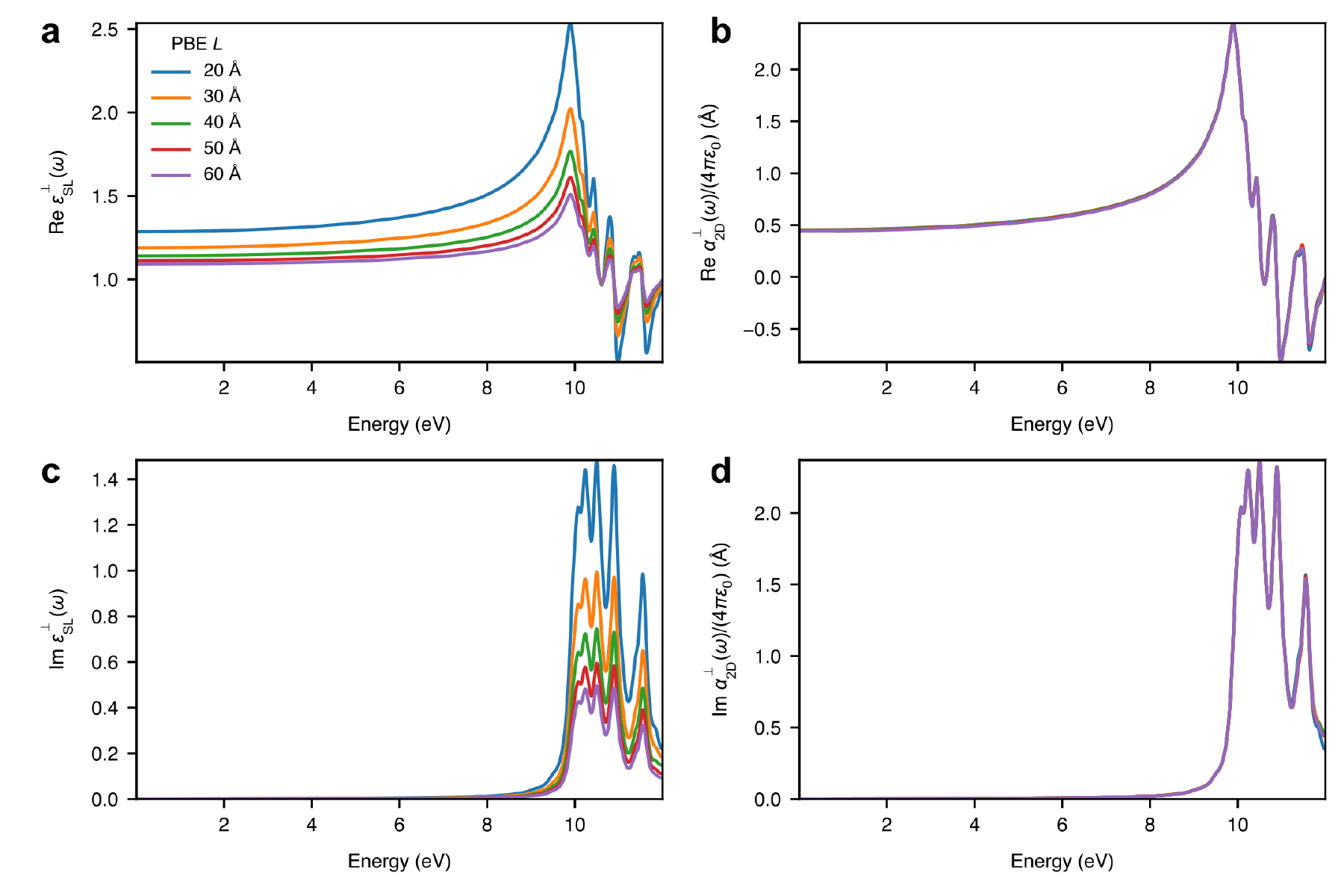}
  \caption{Dependence of out-of-plane dielectric properties on $L$ for
    monolayer BN calculated using PBE method: \textbf{a} Real part of
    $\varepsilon_{\mathrm{SL}}^{\perp}$, \textbf{b} Real part of
    $\alpha_{\mathrm{2D}}^{\perp}$, \textbf{c} Imaginary part of
    $\varepsilon_{\mathrm{SL}}^{\perp}$, \textbf{d} Imaginary part of
    $\alpha_{\mathrm{2D}}^{\perp}$.  The same {\bf k}-sampling is used
    in all simulations with the only variable quantity being
    $L$. Clearly, the $L$-dependency of the superlattice dielectric
    function is removed using 2D polarizabilities.}
  \label{fig:PBE-omega-out}
\end{figure}

\begin{figure}[htbp]
  \centering
 \includegraphics[width=1.0\linewidth]{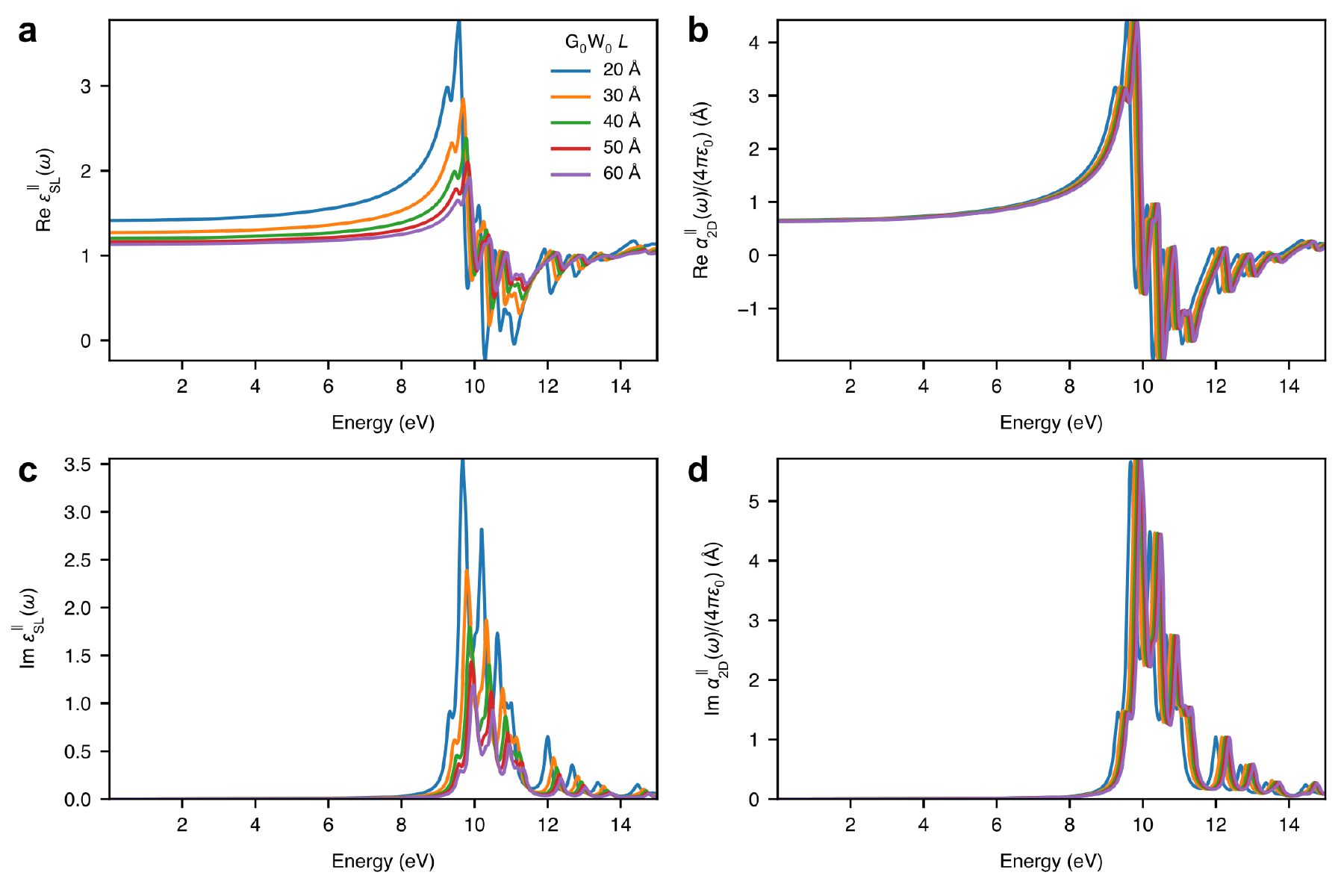}
 \caption{Similar to Figure \ref{fig:PBE-omega-in} but calculated
   using G$_{0}$W$_{0}$ method. A shift of peak position (frequency)
   in the dielectric function spectra is observed, resulted from the
   change of quasi-particle energy in G$_{0}$W$_{0}$ calculations
   invloving the 2D slab.}
  \label{fig:GW-omega-in}
\end{figure}

\begin{figure}[htbp]
  \centering
 \includegraphics[width=1.0\linewidth]{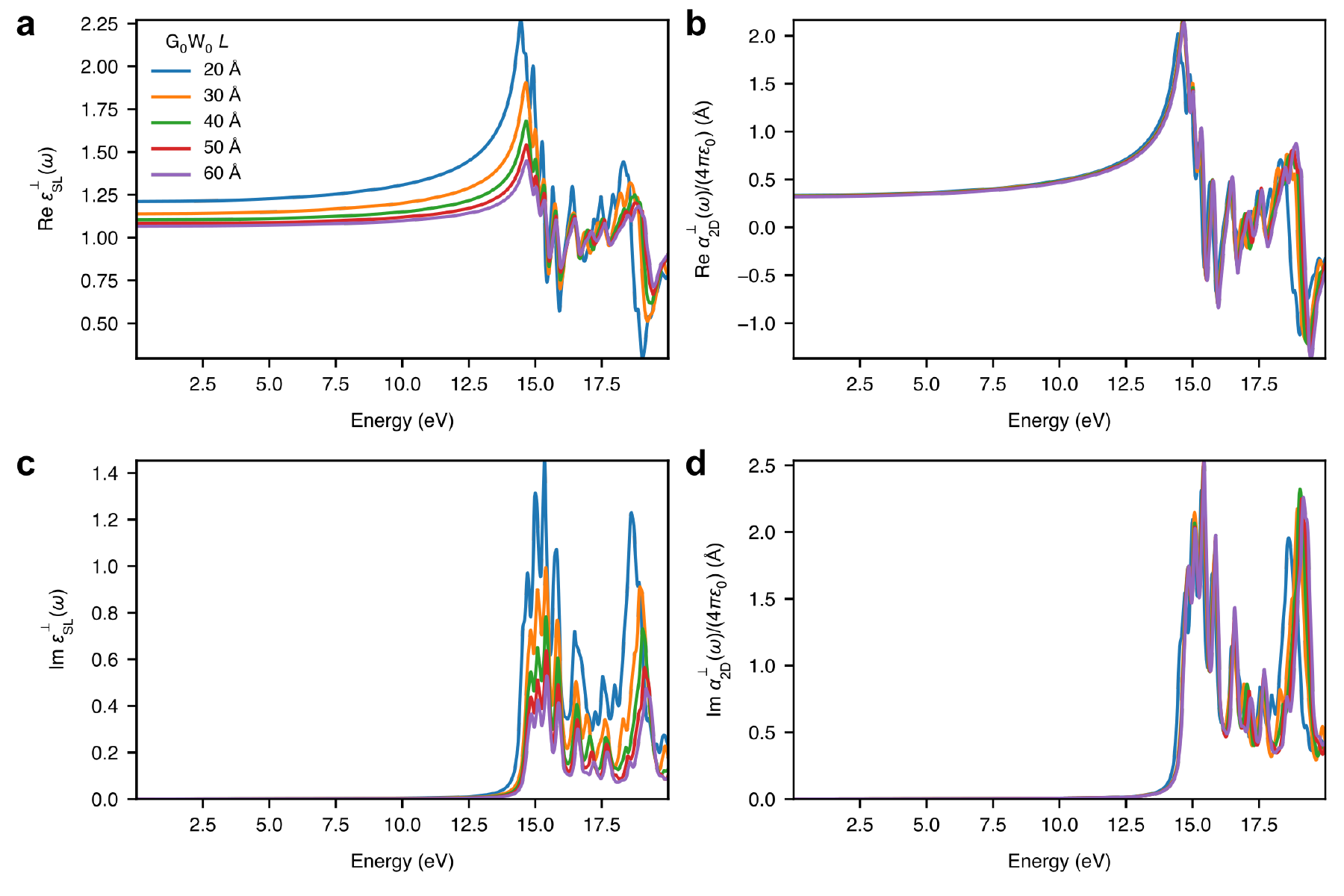}
 \caption{Similar to Figure \ref{fig:PBE-omega-out} but calculated
   using G$_{0}$W$_{0}$ method.  A shift of peak position (frequency)
   in the dielectric function spectra is observed, resulted from the
   change of quasi-particle energy in G$_{0}$W$_{0}$ calculations
   invloving the 2D slab.}
  \label{fig:GW-omega-out}
\end{figure}

We calculate the frequency-dependent dielectric properties of BN with
varying superlattice $L$ at PBE, G$_{0}$W$_{0}$, and G$_{0}$W$_{0}$--BSE. 
Figure \ref{fig:PBE-omega-in} and \ref{fig:PBE-omega-out} show that the
magnitudes of $\varepsilon_{\mathrm{SL}}^{\parallel}(\omega)$ and
$\varepsilon_{\mathrm{SL}}^{\perp}(\omega)$ reduce with increasing $L$
throughout the whole frequency range at PBE level. The $L$-dependency can also be
removed using main text Eqs. \ref{main-eq:alpha-para-def} and
\ref{main-eq:alpha-perp-def}, yielding lattice-independent
polarizabilities throughout the frequency domain (Figure \ref{fig:PBE-omega-in}{\bf b}-{\bf d} and \ref{fig:PBE-omega-out}{\bf b}-{\bf d}). 
Note that when extracting the frequency-dependent 2D polarizabilities
from Eqs. \ref{main-eq:alpha-para-def} and
\ref{main-eq:alpha-perp-def}, the peak position in energy
$\hbar \omega$ from $\varepsilon_{\mathrm{SL}}$ is preserved. This is
explained by the fact that the local extrema from spectra of
$\varepsilon_{\mathrm{SL}}$ is also the local extrema in corresponding
$\alpha_{\mathrm{2D}}$, since when
$\partial \varepsilon_{\mathrm{SL}}(\omega) / \partial \omega = 0 $,
we have
\begin{subequations}
\begin{eqnarray}
  \label{eq:extrema-para}
  \frac{\partial \alpha_{\mathrm{2D}}^{\parallel}(\omega)}{\partial \omega}
  &= \varepsilon_{0}L {\displaystyle \frac{\partial \varepsilon_{\mathrm{SL}}^{\parallel}}{\partial \omega}} = 0   \\
  \label{eq:extrema-perp}
  \frac{\partial \alpha_{\mathrm{2D}}^{\perp}(\omega)}{\partial \omega}
  &= \varepsilon_{0}L {\displaystyle \frac{1}{\varepsilon_{\mathrm{SL}}^{2}}\frac{\partial \varepsilon_{\mathrm{SL}}^{\parallel}}{\partial \omega}} = 0
\end{eqnarray}
\end{subequations}
which indicates that no corrections in energy are present 
when transforming
$\varepsilon_{\mathrm{SL}}(\omega)$ to $\alpha_{\mathrm{2D}}(\omega)$.
Performing the simulations at the level of G$_{0}$W$_{0}$
we observe blue shifts in energy in $\varepsilon_{\mathrm{SL}}(\omega)$ 
with increasing $L$. Figures \ref{fig:GW-omega-in} and
\ref{fig:GW-omega-out} show that not only the magnitudes of the dielectric
functions change with $L$ but also the peak positions. As a result
the obtained polarizabilities also show $L$-dependent peak 
shift (Figures \ref{fig:GW-omega-in} {\bf b}-{\bf d} and
\ref{fig:GW-omega-out}{\bf b}-{\bf d}). This 
can be explained by the long-range nature of the 
Coulomb interactions into the self-energy 
$\Sigma=iGW$. 

The non-interacting Green's function can be constructed as: 

\begin{equation}
    G^{(0)}(\mathbf{r},\mathbf{r}',\omega)=\sum_n\frac{\phi_{\mathrm{n}\mathbf{k}}^{(0)}(\mathbf{r})\phi_{\mathrm{n}\mathbf{k}}^{*(0)}(\mathbf{r'})}{\omega-\epsilon_\mathrm{n}-+i\eta\mathrm{sgn}(\epsilon_\mathrm{n}-\epsilon_\mathrm{F})}
\end{equation}

Here $\epsilon_n(\mathbf{k})$ are the DFT eigenenergies at \textbf{k}, $\epsilon_F$ the Fermi energy, $\omega$ the frequency, $\ket{u_{n\mathbf{k}}}$ the cell periodic Bloch functions, $\phi_{n\mathbf{k}}$ are the one-electron orbitals and $\eta$ is an infintesimal complex shift. It can be seen from equation \ref{eq:dft-dielectric} there is a clear volume dependence on $\Omega$ of the dielectric function from the preceding DFT calculation. Thus, when carrying out a calculation on a slab, the dielectric function will vary with the vacuum spacing used.

Using the dielectric function we can calculate the screened Coulomb interaction:
\begin{equation}
    W=\epsilon^{-1}\nu
\end{equation}
where $\nu$ is the bare Coulomb interaction given by
$e^2$/$|\mathbf{r}-\mathbf{r}'|$. Due to the
1/$|\mathbf{r}-\mathbf{r}'|$ term, images in the non-periodic
direction have a long-range spurious interaction which varies with
vacuum spacing. From the screened Coulomb interaction and
non-interacting Green's function it is possible to calculate the
self-energy of the system:
\begin{equation}\label{self}
\Sigma=iGW
\end{equation}
and the quasi-particle eigenenergies are found using:
\begin{equation}\label{quasi}
E^{\mathrm{QP}}_{\mathrm{n}\mathbf{k}}=\Re\Big[\braket{\phi_{\mathrm{n}\mathbf{k}}|-\frac{1}{2}\Delta+V_{\mathrm{ext}}+V_\mathrm{H}+\Sigma(\epsilon_{\mathrm{n}\mathbf{k}}^{\mathrm{DFT}})|\phi_{\mathrm{n}\mathbf{k}}}\Big]
\end{equation}
Therefore a new set of eigenergies, E$_{n\mathbf{k}}^{QP}$ are found. Using E$_{n\mathbf{k}}^{QP}$ it is then possible to recalculate the dielectric function using the quasi-particle eigenergies as well as the DFT eigenfunctions.

In general terms excitonic effects are not taken into account at the level of PBE or 
G\textsubscript{0}W\textsubscript{0}. 
%
%
%
%
%
%
For the former, this generally leads to an 
underestimation of the electronic bandgap and results in the first optical peak
being lower in energy in comparison to more accurate methods. 
To go beyond
PBE, the G\textsubscript{0}W\textsubscript{0} approximation
replaces the exchange-correlation energy by the self-energy 
to include many-body effects through the interacting Green's function,
$G$, and the screened Coulomb potential, $W$. 
This generally leads to
an overestimation of the first optical peak as electron-hole coupling
is not taken into account. This can be remedied by solving the
Bethe-Salpeter equation using the eigenvalues obtained from
G\textsubscript{0}W\textsubscript{0}
(G\textsubscript{0}W\textsubscript{0} - BSE) which generally gives
good agreement with experiment.

\begin{figure}[htbp]
  \centering
 \includegraphics[width=1.0\linewidth]{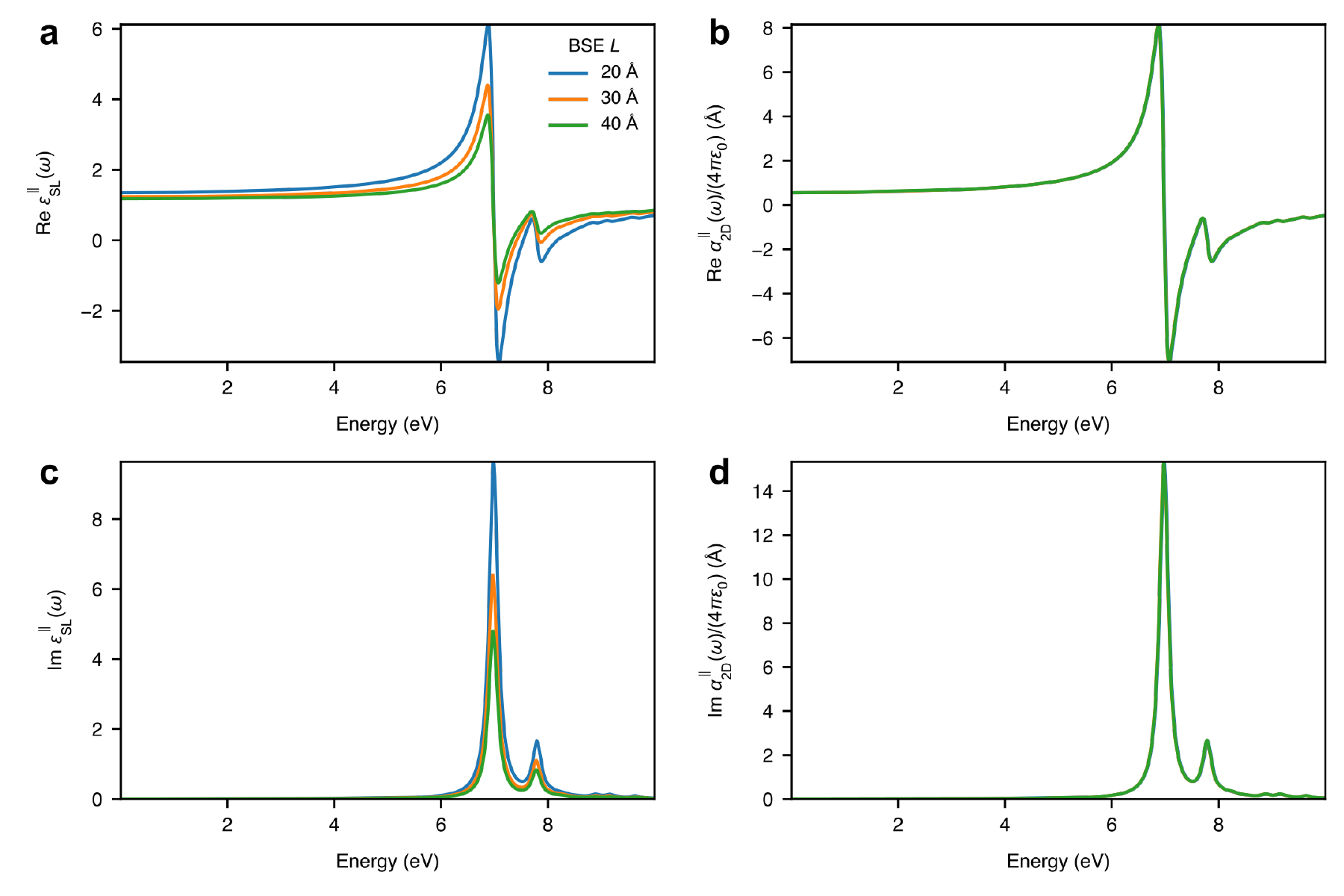}
 \caption{Similar as in Fig. \ref{fig:GW-omega-in} but taking into
   account electron-hole interactions at the level of the
   Bethe-Salpeter equation (G$_{\rm 0}$W$_{0}$ + BSE).  The peak shift
   in energies are reduced in comparison with G$_{\rm 0}$W$_{0}$. As a
   result the polarizabilities becomes almost independent of $L$,
   consistent to results obtained by the PBE method.}
  \label{fig:BSE-omega-in}
\end{figure}

\begin{figure}[htbp]
  \centering
 \includegraphics[width=1.0\linewidth]{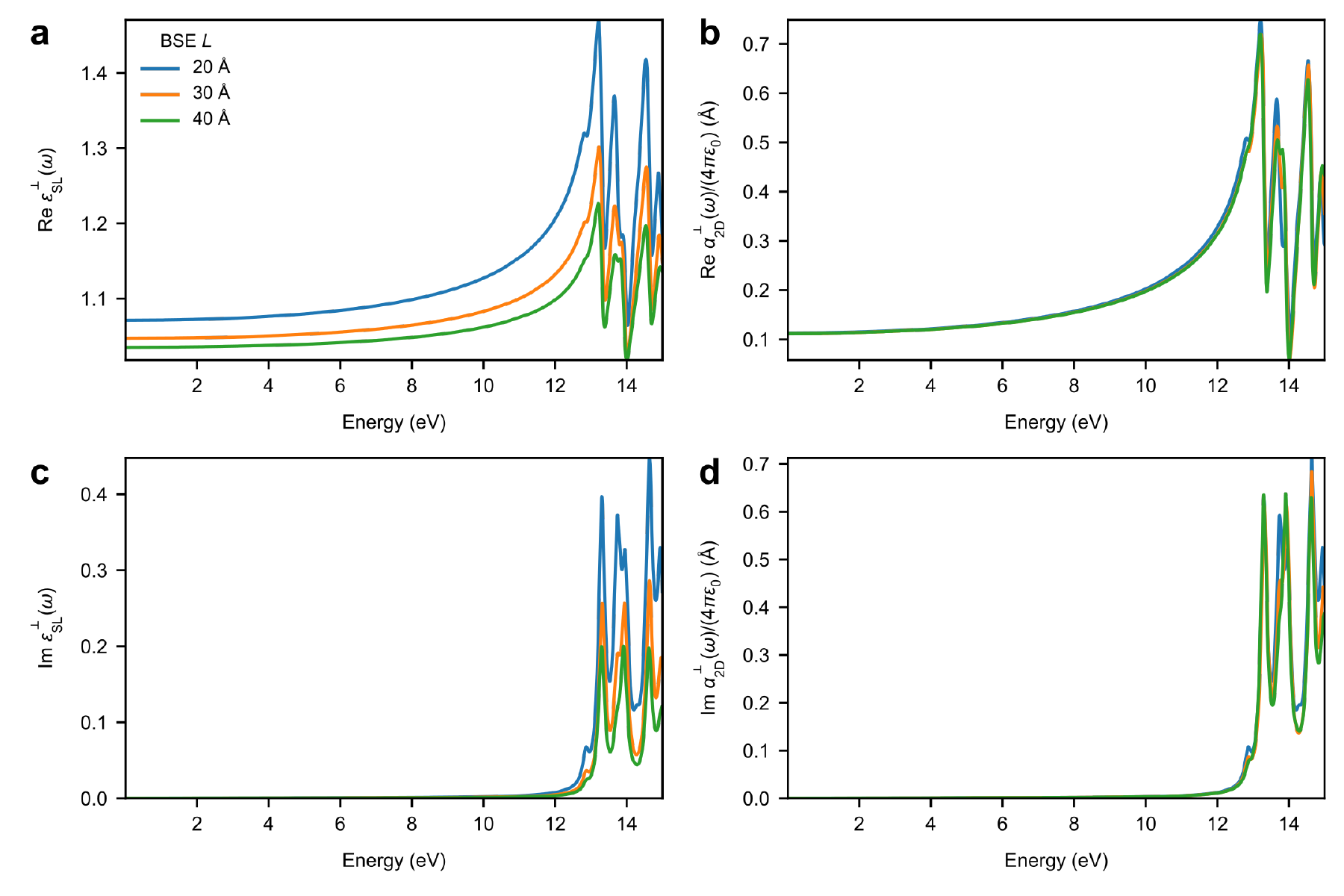}
 \caption{Similar as in Fig. \ref{fig:GW-omega-out} but taking into
   account electron-hole interactions at the level of the
   Bethe-Salpeter equation (G$_{\rm 0}$W$_{0}$ + BSE).  As a result
   the polarizabilities becomes almost independent of $L$, consistent
   to results obtained by the PBE method.}
  \label{fig:BSE-omega-out}
\end{figure}

Due to the volume dependence of $\Sigma$ in equations \ref{self} and
\ref{quasi}, increasing the vacuum spacing leads to a change in the
calculated quasi-particle eigenergies. This, along with the volume
dependence in the calculation of the dielectric function, leads to an
increase in energy of the peak position and a decrease in its
magnitude with increasing vacuum spacing. Such effect is also
discussed in several other studies
\cite{Rozzi_2006,Hueser_2013_2Dvs3D}, when full Coulomb interaction is
used in a supercell. Including excitonic effects on top of a
G$_0$W$_0$ calculation through solving the Bethe--Salpeter equation
correct this energy shift by localizing the exciton within the
slab (Figures \ref{fig:BSE-omega-in} and
\ref{fig:BSE-omega-out}). However, the decrease in the magnitude of
the dielectric function on both components of
$\varepsilon_{\mathrm{SL}}(\omega)$ with increasing vacuum spacing is
still observed.  Using main text Eqs. \ref{main-eq:alpha-para-def} and
\ref{main-eq:alpha-perp-def} we can remove the dependence on $L$ as
plotted in Figures \ref{fig:BSE-omega-in}{\bf b}-{\bf d} and
\ref{fig:BSE-omega-out}{\bf b}-{\bf d} with fewer variations on
$\varepsilon^{\perp}_{\mathrm{SL}}(\omega)$.

Combining the accuray of bandgap estimation, reproducible results of
frequency-dependent dielectric properties and calculation efforts, the
choice of HSE06 hybrid functional used in the main text provides
best trade off between all aspects.

%
%

\begin{figure}[htbp]
  \centering
  \includegraphics[width=0.85\linewidth]{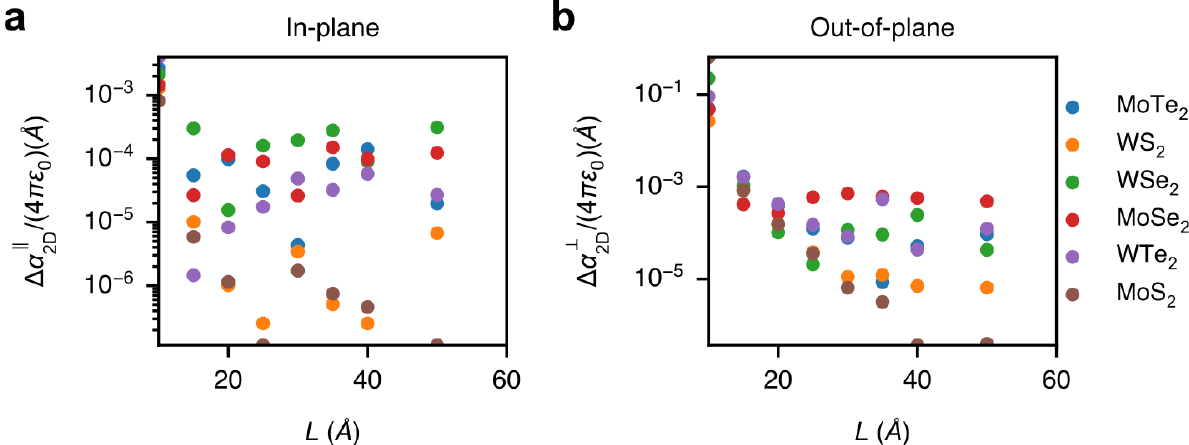}
  \caption{Convergence of electronic polarizabilities
    $\Delta \varepsilon_{\mathrm{2D}}^{\parallel}$ (\textbf{a}) and
    $\Delta \varepsilon_{\mathrm{2D}}^{\perp}$ (\textbf{b}) as
    functions of $L$ for the selected TMDCs, respectively. For both
    in- and out-of-plane polarizabilities, convergence is achieved
    when $L >$15 \AA{}}
  \label{fig:alpha-converg}
\end{figure}

\pagebreak{}
\section{Further Data Concerning the Comparison with EDM}
\label{sec:2D-3D-rescale}

\begin{figure}[htbp]
  \centering
  \includegraphics[width=0.85\linewidth]{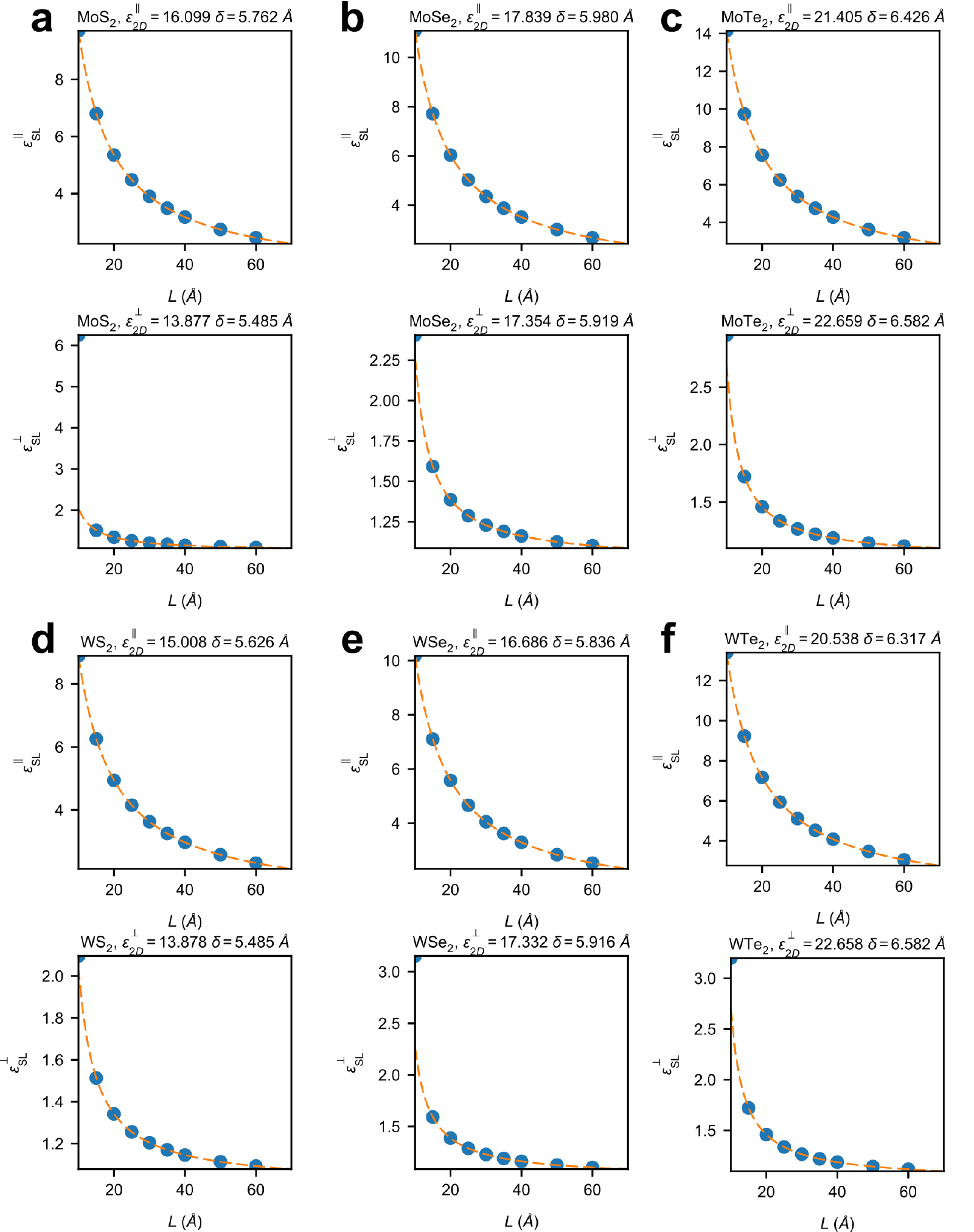}
  \caption{ Calculated (blue dots) and fitted
    (orange broken lines) $\varepsilon_{\mathrm{SL}}$ as function of
    $L$ for the group 6 TMDCs:
    \textbf{a}. MoS$_{2}$. \textbf{b}. MoSe$_{2}$. \textbf{c}.
    MoTe$_{2}$. \textbf{d}. WS$_{2}$. \textbf{e}. WSe$_{2}$. \textbf{f}.
    WTe$_{2}$. The extracted values of $\varepsilon_{\mathrm{2D}}$ and
    $\delta$ are shown in each subfigure.}
  \label{fig:rescale-prb}
\end{figure}

The major issue when using rescale
relations \ref{main-Response:1}$-$\ref{main-eq:emt-2} comes from the determination of $\delta_{\mathrm{2D}}^{*}$. To
eliminate the modeling error caused by the \textit{a priori} selection of this parameter, 
we perform the calculation of
$\varepsilon_{\mathrm{SL}}$ of group 6 TMDCs against different $L$,
and use least-square fitting to extract both
$\varepsilon$ and $\delta$, as shown in
Figure \ref{fig:rescale-prb}.
The values of $\delta_{\mathrm{2D}}^{*}$ extracted
from both $\varepsilon_{\mathrm{SL}}^{\parallel}$ and
$\varepsilon_{\mathrm{SL}}^{\perp}$ are close when $L> 15$
{\AA}. Notably, the $\delta_{\mathrm{2D}}^{*}$ values are generally 10\%
smaller than the interlayer distance in corresponding bulk materials
$L_{\mathrm{\rm Bulk}}$, as shown in Table \ref{tab:delta-L-DFt}. On the
other hand, the extracted $\delta_{\mathrm{2D}}^{*}$ values are closer to
the covalent thickness $\delta_{\rm cov}$ as
described in the main text, with a difference generally smaller than
5\%. Our results indicate that the conventional estimation of the 2D
layer thickness by its bulk interlayer distance
\cite{Matthes_2016,Laturia_2018}, will always lead to
overestimation. On the contrary, the out-of-plane polarizability
$\alpha_{\mathrm{2D}}^{\perp}$ correctly captures the thickness 
of 2D materials.

\begin{table}[htbp]
  \centering
  \begin{tabular}[htbp]{lccccc}
  \hline{}
  Material & $\delta_{\mathrm{2D}}^{\parallel, \mathrm{fit}}$ (\AA) & $\delta_{\mathrm{2D}}^{\perp, \mathrm{fit}}$ ({\AA})& $L_{\mathrm{Bulk}}$ ({\AA}) & $\delta_{\mathrm{cov}}$ ({\AA}) & $\alpha_{\mathrm{2D}}^{\perp}/\varepsilon_{0}$ ({\AA})\\
  \hline{}
  2H-MoS$_{2}$ & 5.76 & 5.49 & 6.15 & 5.22 & 4.98\\
  2H-MoSe$_{2}$ & 5.98 & 5.92 & 6.46 &  5.73 & 5.60\\
  2H-MoTe$_{2}$ & 6.43 & 6.85 & 6.98 & 6.37 & 6.12\\
  2H-WS$_{2}$ & 5.63 & 5.49 & 6.15 & 5.20 & 5.00\\
  2H-WSe$_{2}$ & 5.84 & 5.92 & 6.49 & 5.75 & 5.42\\
  2H-WTe$_{2}$ & 6.32 & 6.58 & 7.06 & 6.38 & 6.33\\
  \hline{}
\end{tabular}

\caption{Fitted effective thickness
  $\delta_{\mathrm{2D}}^{\parallel, \mathrm{fit}}$ and
  $\delta_{\mathrm{2D}}^{\perp, \mathrm{fit}}$ from in- and
  out-of-plane dielectric data, compared with the interlayer distance
  of corresponding bulk material $L_{\mathrm{Bulk}}$, the covalent
  thickness $\delta_{\mathrm{cov}}$, and
  $\alpha_{\mathrm{2D}}^{\perp}/\varepsilon_{0}$ for 2H TMDC
  materials.}
\label{tab:delta-L-DFt}
\end{table}

\begin{figure}[htbp]
  \centering
  \includegraphics[width=0.6\linewidth]{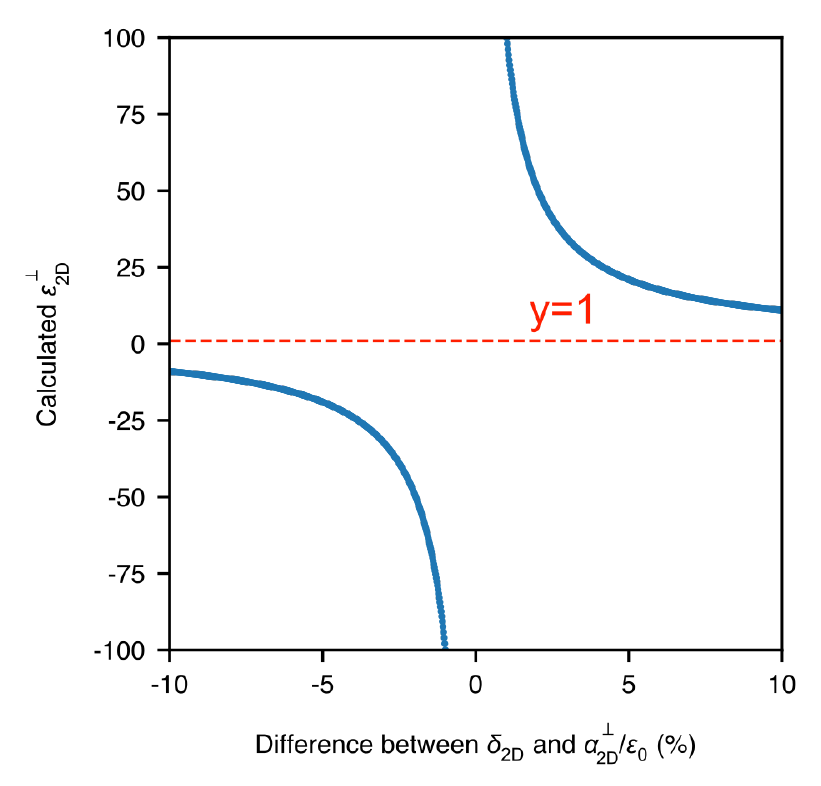}
  \caption{Calculated $\varepsilon_{\mathrm{2D}}^{\perp}$ value as a
    function of the difference between $\delta_{\mathrm{2D}}^{*}$ and
    $\alpha_{\mathrm{2D}}^{\perp}/\varepsilon_{0}$. A small change of
    $\alpha_{\mathrm{2D}}^{\perp}/\varepsilon_{0}$ chosen may lead to divergence of
    the $\varepsilon_{\mathrm{2D}}^{\perp}$ or even negative values,
    which is apparently nonphysical.}
  \label{fig:eps-alpha-error}
\end{figure}

One main drawback of the EDM approach is the
overestimation of the out-of-plane dielectric response. As can be seen
in Figure \ref{fig:rescale-prb}, the extracted
$\varepsilon_{\mathrm{2D}}^{\perp}$ values for the TMDCs studied
are comparable (within a range of 8-13\%) or even larger than
$\varepsilon_{\mathrm{2D}}^{\parallel}$, which does not agree with the
physical picture that electrostatic screening of 2D materials are much
smaller perpendicular to the 2D plane. In fact, combining
Eq. \ref{main-eq:emt-2} and the definition of $\alpha_{\mathrm{2D}}^{\perp}$, we
have:
\begin{equation}
  \label{eq:eps-alpha-perp}
  \frac{\alpha_{\mathrm{2D}}^{\perp}}{\varepsilon_{0}} = \delta^{*}_{\mathrm{2D}}(1 - (\varepsilon_{\mathrm{2D}}^{\perp})^{-1})
\end{equation}
which indicates that the characteristic length
$\alpha_{\mathrm{2D}}^{\perp}/\varepsilon_{0}$, is very close but slightly smaller
than $\delta_{\mathrm{2D}}^{*}$ estimated by the effective medium theory,
if $\varepsilon^{\perp}_{\mathrm{2D}} \gg 1$. Moreover, from
Eq. \ref{eq:eps-alpha-perp}, when $\delta_{\mathrm{2D}}^{*}$ and
$\alpha_{\mathrm{2D}}^{\perp}/\varepsilon_{0}$ are close, slight change of the
$\delta_{\mathrm{2D}}^{*}$ chosen may lead to divergence of
$\varepsilon_{\mathrm{2D}}^{\perp}$, as shown in Figure
\ref{fig:eps-alpha-error}. Therefore cautions must be taken when
treating the dielectric response of the 2D material using effective
medium theory. In comparison, the 2D polarizability does not require
the initial guess of the thickness.  

To conclude, based on theoretical and technical considerations, there are several
advantages of using the electronic polarizability for describing the dielectric
nature of 2D materials, including:
\begin{enumerate}
\item $\alpha_{\mathrm{2D}}$ can be used to describe both the local and macroscopic dielectric properties, while $\epsilon_{\mathrm{2D}}$ is unable. 

\item Calculating $\alpha_{\mathrm{2D}}$ only requires to calculate the dielectric response at single superlattice length at relative small $L$, while $\epsilon_{\mathrm{2D}}$ requires calculation with varied superlattice length. This is a big advantage on the computational side since the larger the supercell the larger the computational cost. 

\item $\alpha_{\mathrm{2D}}$ correctly represents the screening length of 2D material, while $\epsilon_{\mathrm{2D}}$ calculated from EMT does not. 

\item $\alpha_{\mathrm{2D}}$ correctly represents the different degree of screening in/out-of-plane, while $\epsilon_{\mathrm{2D}}$ does not.
  
\item The value of $\epsilon_{\mathrm{2D}}$ hugely depends on the
  choice of the thickness of 2D material, while such information is
  intrinsically embedded in $\alpha_{\mathrm{2D}}$.
\end{enumerate}
%
%


\subsection{Derivation of Eqs.\ref{main-eq:alpha-para-def} and \ref{main-eq:alpha-perp-def}}

To show that Eqs. \ref{main-eq:alpha-para-def} and \ref{main-eq:alpha-perp-def} do not use any arguments based on EMT but rather basic electrostatics, we show the derivation of both equations in the following. 

\subsubsection{Parallel to the surface}

For in-plane electric field, the electrostatic boundary conditions gives the continuity of the applied electric field as  
${E}^{\parallel}_{\mathrm{loc}}={E}^{\parallel}$\cite{Markel_2016}, which resulted in 
${P}^{p}=\frac{1}{L} {\mu}^{p}_{\mathrm{2D}}$. Indeed, ${\mu}^{p}_{\mathrm{2D}}$ can be written in terms of the local electric field ${E}_{\mathrm{loc}}^{p}$ and the 2D polarizability $\alpha_{\mathrm{2D}}$ as:

\begin{equation}
{\mu}^{p}_{\mathrm{2D}} = \alpha^{\parallel}_{\mathrm{2D}} {E}^{\parallel}_{\mathrm{loc}} 
\end{equation}
This gives for the in-plane polarization:  

\begin{equation}
{P}^{p}=\frac{1}{L} \alpha^{\parallel}_{\mathrm{2D}} {E}^{\parallel}_{\mathrm{loc}} = \frac{1}{L} \alpha^{\parallel}_{\mathrm{2D}} {E}^{\parallel} 
\end{equation}
The derivative of this equation relative to the external field ${E}^{\parallel}$ resulted in:  

\begin{equation}
\frac{\partial {P}^{p}}{\partial {E}^{\parallel}} =\frac{1}{L} \alpha^{\parallel}_{\mathrm{2D}} 
\label{parallel}
\end{equation}
Inserting this equation \ref{parallel} into Eq.\ref{main-eq:def-eps-1} with $p=q=\parallel$: 

\begin{equation}
\varepsilon_{\mathrm{SL}}^{\parallel} =  1 + \frac{\alpha_{\mathrm{2D}}^{\parallel}}{\varepsilon_0 L}
\end{equation}

\subsubsection{Perpendicular to the surface}

The same procedure can be used along the out-of-plane component but using ${E}_{\mathrm{loc}}^{\perp}={E}^{\perp}+{\mu}_{\mathrm{2D}}^{\perp}/\varepsilon_0 L$ for the local field. We can write this field as: 

\begin{equation}
{E}_{\mathrm{loc}}^{\perp}={E}^{\perp}+\frac{\alpha^{\perp}_{\mathrm{2D}} {E}_{\mathrm{loc}}^{\perp}}{\varepsilon_0 L}
\end{equation}
If we re-arrange the terms for ${E}_{\mathrm{loc}}^{\perp}$, we can ended up with:  

\begin{equation}
{E}_{\mathrm{loc}}^{\perp}= \frac{{E}^{\perp}}{ 1-\frac{\alpha^{\perp}_{\mathrm{2D}}}{\varepsilon_0 L} }
\end{equation}
We can write the polarization ${P}^{\perp}$ as: 

\begin{equation}
{P}^{\perp} = \frac{\alpha^{\perp}_{\mathrm{2D}}} {L}(\frac{{E}^{\perp}} {1-\alpha^{\perp}_{\mathrm{2D}}/\varepsilon_0 L})
\end{equation} 
Taking the derivative of this equation relative to ${E}^{\perp}$ and inserting in Eq.\ref{main-eq:def-eps-1} with $p=q=\perp$ resulted in: 
\begin{equation}
\varepsilon_{\mathrm{SL}}^{\perp} = 1 +  \frac{\alpha^{\perp}_{\mathrm{2D}}} {\varepsilon_0 L}(\frac{1} {1-\alpha^{\perp}_{\mathrm{2D}}/\varepsilon_0 L})
\end{equation}
Re-arranging the terms, it ended up as: 

\begin{equation}
\varepsilon_{\mathrm{SL}}^{\perp} =\frac{\varepsilon_0 L - \alpha^{\perp}_{\mathrm{2D}} + \alpha^{\perp}_{\mathrm{2D}}}{  \varepsilon_0 L - \alpha^{\perp}_{\mathrm{2D}}} = (1 - \frac{\alpha_{\mathrm{2D}}^{\perp}}{\varepsilon_0 L})^{-1} 
\end{equation}

\section{Polarizability-Based Theoretical Model}
\label{sec:theory-1}
The universal relations for $\alpha_{\mathrm{2D}}^{\parallel}$ and
$\alpha_{\mathrm{2D}}^{\perp}$ revealed by 
Eqs. \ref{main-eq:2D-Moss-para} and \ref{main-eq:2D-Moss-perp} are not
coincidental. Combining recent theoretical findings of the linear
relation between exciton binding energy $E_{\mathrm{b}}$ and
$E_{\mathrm{g}}$ of 2D materials
\cite{Choi_linear_2015,Olsen_2016_hydrogen,Jiang_2017_Eg_Eb}, and the
fact that the $E_{\mathrm{b}}$ is roughly inversely proportional to
$\alpha_{\mathrm{2D}}^{\parallel}$ \cite{Pulci_2014}, it is reasonable
to have a general relation between
$\alpha_{\mathrm{2D}}^{\parallel}$ and $E_{\mathrm{g}}^{-1}$. Moreover,
the bandgap-independent relation of 2D $\alpha_{\mathrm{2D}}^{\perp}$
resembles molecular polarizabilities of conjugate molecules
\cite{Davies_1952}, fullerenes \cite{Sabirov_2014} and carbon
nanotubes \cite{Benedict_1995}, which are also shown to be
geometry-dependent.

In this section we show in detail the polarizability-based theoretical
framework that leads to the 2D Moss-like relations proposed in the
main text.  Due to its highly anisotropic nature, the wave function of
an isolated 2D material $\psi(\mathbf{r})$ can be separated into the
in- and out-of-plane components ($\psi^{\parallel}(\boldsymbol{\rho})$ and $\psi^{\perp}(z)$) 
similar to the treatment of quantum
wells (QW),\cite{davies_physics_1997} such that
$\psi=\psi^{\parallel}\psi^{\perp}$, where $\boldsymbol{\rho}=(x, y)$
is the in-plane coordinate. Using the Bloch theorem, the periodic
$\psi^{\parallel}(\boldsymbol{\rho})$ can be further expressed as
$\psi^{\parallel}(\boldsymbol{\rho})=e^{i\mathbf{k} \cdot
  \boldsymbol{\rho}}u(\boldsymbol{\rho})$, where $\mathbf{k}$ is the
in-plane wave vector and $u(\boldsymbol{\rho})$ is periodic function
in the xy-plane. According to the random phase approximation (RPA)
theory\cite{Adler_1962}, $\varepsilon_{\mathrm{SL}}$ is the
$\mathbf{q} \to 0$ and $\omega \to 0$ limits of the non-interacting
dielectric function $\varepsilon(\mathbf{q}, \omega)$, where
$\mathbf{q}$ is the momentum transfer and $\omega$ is the frequency:
\begin{equation}
  \label{eq:RPA-eps2}
  \varepsilon_{\mathrm{SL}}
  = \lim_{\mathbf{q} \to 0} 1 + \frac{2e^{2}}{\varepsilon_{0} |\mathbf{q}|^{2} \Omega}
  \sum_{\mathrm{k, c, v}}
  \frac{|<\psi_{\mathrm{v}}(\mathbf{k})|e^{-i\mathbf{q}\mathbf{r}}|\psi_{\mathrm{c}}(\mathbf{k+q})>|^{2}}
  {E_{\mathrm{c}}(\mathbf{k+q}) - E_{\mathrm{v}}(\mathbf{k})}
  \left[f(\psi_{\mathrm{c}}) - f(\psi_{\mathrm{v}})\right]
\end{equation}
where $e$ is the unit charge, $c$, $v$ are the conduction and valence
bands, $E$ is the eigenenergy of individual bands, and $f$ is the
Fermi-Dirac distribution function. Taking the
limit that $L\to\infty$, when
$\varepsilon^{\perp}_{\mathrm{SL}} \approx 1$, we have
$1-1/\varepsilon^{\perp}_{\mathrm{SL}} \approx
(\varepsilon_{\mathrm{SL}}^{\perp} - 1)$. Therefore
$\alpha_{\mathrm{2D}}^{\parallel}$ and $\alpha_{\mathrm{2D}}^{\perp}$
at 0 K can be unified by the same equation:
\begin{equation}
  \label{eq:alpha-RPA}
  \alpha_{\rm 2D} = \frac{2e^{2}}{|q|^{2}A} \sum_{\mathrm{k,c,v}}
  \frac{|<\psi_{\mathrm{v}}(\mathbf{k})|e^{-i\mathbf{q}\mathbf{r}}|\psi_{\mathrm{c}}(\mathbf{k+q})>|^{2}}
  {E_{\mathrm{c}}(\mathbf{k+q}) - E_{\mathrm{v}}(\mathbf{k})}
\end{equation}
where the direction is determined by $\mathbf{q}$. Next we will show
that the different behavior of $\psi^{\parallel}$ and $\psi^{\perp}$
give rise to the main text Eqs. \ref{main-eq:2D-Moss-para} and
\ref{main-eq:2D-Moss-perp}.


\subsection{Detailed derivations of main text Eq. \ref{main-eq:2D-Moss-para}}  %
\label{ssec:theory-1-para}

In this section we show how Eq. \ref{main-eq:alpha-para-def}
is derived from Eq. \ref{eq:alpha-RPA}. For the in-plane component
$\alpha_{\mathrm{2D}}^{\parallel}$, $e^{-i\mathbf{qr}}$ is independent
of $z$, therefore the integral in
$|<\psi_{\mathrm{v}}(\mathbf{k})|e^{-i\mathbf{q}\mathbf{r}}|\psi_{\mathrm{c}}(\mathbf{k+q})>|^{2}$
becomes independent of $\psi^{\perp}$, due to the orthogonality
and normalization. The Bloch-wave form of $\psi^{\parallel}$ ensures that
only the cell-function $u(\mathbf{k})$ contributes to the final result
of $\alpha^{\parallel}_{\mathrm{2D}}$ \cite{davies_physics_1997}, such
that:
\begin{equation}
  \label{eq:alpha_para_RPA}
  \alpha_{\mathrm{2D}}^{\parallel} = \frac{2e^{2}}
  {(2 \pi)^{2}} \int d^{2}\mathbf{k} \sum_{\mathrm{c, v}}
  \frac{|<u_{\mathrm{c}}(\mathbf{k})|\nabla|u_{\mathrm{v}}(\mathbf{k})>|^{2}}
  {E_{\mathrm{c}}(\mathbf{k}) - E_{\mathrm{v}}(\mathbf{k})}
\end{equation}
Following the method of $\mathbf{k} \cdot \mathbf{p}$ theory from
Ref.~\citenum{Jiang_2017_Eg_Eb}, the matrix element in the numerator
of Eq. \ref{eq:alpha_para_RPA} is approximated by:

\begin{equation}
\label{eq:matrix-approx}
|<u_{\mathrm{c}}(\mathbf{k})|\nabla|u_{\mathrm{v}}(\mathbf{k})>|^{2}
\approx {\displaystyle \frac{\hbar^{2}}{2 m^{*}}
  \frac{1}{E_{\mathrm{g}} + \frac{\hbar^{2} k^{2}}{2 m^{*}}}}
\end{equation}
plug it into Eq. \ref{eq:alpha_para_RPA} and integrate within the 2D Brillouin zone from $|k|=0$ to $|k|= k_{\mathrm{BZ}}$, where $k_{\mathrm{BZ}}$ is the wavevector at the boundary of the 2D Brillouin Zone, we get:
\begin{equation}
  \label{eq:derivation-2D-Moss-para}
  \begin{aligned}
\alpha_{\mathrm{2D}}^{\parallel} &= N\cdot -\frac{e^{2}}{2 \pi}
\frac{1}{E_{g} + \beta} \Biggr\vert_{\beta=0}^{\beta=\frac{\hbar^{2} k^{2}_{\mathrm{BZ}}}{m^{*}}} \\
&\approx N e^{2}/(2\pi E_{\mathrm{g}}) = C^{\parallel} E_{g}^{-1}
\end{aligned}
\end{equation}
where $N$ is degeneracy of bands associated with $E_{g}$. The
approximation in Eq. \ref{eq:derivation-2D-Moss-para} is due to the
fact that $\frac{\hbar^{2} k^{2}_{\mathrm{BZ}}}{m^{*}} \gg E_{g}$, and
we arrive at Eq. \ref{main-eq:2D-Moss-para}.

The coefficient of $C^{\parallel}$ adapted from
Ref.~\citenum{Jiang_2017_Eg_Eb} $C^{\parallel} = N e^{2}/(2\pi)$
predicts linear correlation between $\alpha^{\parallel}_{\mathrm{2D}}$
and $E_{\mathrm{g}}^{-1}$. We validate this by examining the
DFT-calculated
$(4\pi \varepsilon_{0})/\alpha^{\parallel}_{\mathrm{2D}}$ (measured in
\AA{}$^{-1}$) and $E_{\mathrm{g}}$ (measured in eV) in
Figure \ref{main-fig-3}\textbf{c}. The coefficient $C^{\parallel}$
becomes
$8 \pi^{2} \varepsilon_{0} \AA{} / (e N) \approx 0.436 / N = 0.183$,
corresponding to $N$ between 2 and 3, which is a reasonable result for
the 2D materials studied.


\subsection{Detailed derivation of main text Eq. \ref{main-eq:2D-Moss-perp}}
\label{ssec:theory-1-perp}

For main text Eq. \ref{main-eq:2D-Moss-perp}, treating the in-plane
wave functions as plane wave with form
$\psi^{\parallel}(\rho) \propto e^{i \mathbf{k \rho}}$, the matrix
element of
$<\psi_{\mathrm{v}}(\mathbf{k})|e^{-i\mathbf{qr}}|\psi_{\mathrm{c}}(\mathbf{k+q})>$,
when $\mathbf{q}=(0, 0, q_{z})$, becomes\cite{Hybertsen_1987}:
\begin{equation}
  \begin{aligned}
    \label{eq:matrix-z}
  <\psi_{\mathrm{v}}(\mathbf{k})|e^{-i\mathbf{qr}}|\psi_{\mathrm{c}}(\mathbf{k+q})>
  &= \frac{1}{A} \int dx \int dy
  e^{i(\mathbf{-k \rho} - \mathbf{q \rho} + \mathbf{(k+q) \rho})}
  \int (\psi^{\perp})^{*}_{\mathrm{v}}(\mathbf{k})e^{-iq_{z}z}\psi^{\perp}_{\mathrm{c}}(\mathbf{k+q})\\
  &= <\psi^{\perp}_{\mathrm{v}}(\mathbf{k})|e^{-iq_{z}z}|\psi^{\perp}_{\mathrm{c}}(\mathbf{k+q})>
  \end{aligned}
\end{equation}
Note that the states perpendicular are bound, the integral is
meaningful only when $\mathbf{k=k+q}$ \cite{davies_physics_1997}. By
performing the Taylor expansion of
$e^{-i\mathbf{qr}} \approx 1 - i\mathbf{qr}$, we get:
\begin{equation}
  \begin{aligned}
    \label{eq:matrix-z-2}
    <\psi^{\perp}_{\mathrm{v}}(\mathbf{k})|e^{-iq_{z}z}|\psi^{\perp}_{\mathrm{c}}(\mathbf{k})>
    &\approx <\psi^{\perp}_{\mathrm{v}}(\mathbf{k})|\psi^{\perp}_{\mathrm{c}}(\mathbf{k})> -
    iq_{z} <\psi^{\perp}_{\mathrm{v}}(\mathbf{k})|z|\psi^{\perp}_{\mathrm{c}}(\mathbf{k})>\\
    &= -iq_{z} <\psi^{\perp}_{\mathrm{v}}(\mathbf{k})|z|\psi^{\perp}_{\mathrm{c}}(\mathbf{k})>
   \end{aligned}
\end{equation}
plug this into Eq. \ref{eq:alpha-RPA} and express the summation over
$k_{x}$ and $k_{y}$ in a continuous form within the Brillouin Zone, we
arrive at:
\begin{equation}
\label{eq:alpha_perp_RPA}
\alpha_{\mathrm{2D}}^{\perp} = \frac{2e^{2}}{(2 \pi) ^{2}} \int d^{2}\mathbf{k}
\sum_{\mathrm{c, v}}
\frac{|<\psi_{\mathrm{v}}(\mathbf{k})|z|\psi_{\mathrm{c}}(\mathbf{k})>|^{2}}
{E_{\mathrm{c}}(\mathbf{k}) - E_{\mathrm{v}}(\mathbf{k})}
\end{equation}
The formalism is slightly different from Eq.\ref{eq:alpha_para_RPA}.

The out-of-plane wave function $\psi^{\perp}(z)$ is the solution to the Schr\"{o}dinger equation with
Hamiltonian $\mathcal{H} = -\hbar^{2} \nabla^{2}/2m_{e} + V(z)$, where
$\hbar$ is the reduced Planck constant, $m_{e}$ is electron mass and
$V(z)$ is the confined Coulomb potential along the z-direction
created by the nuclei
\cite{davies_physics_1997,ihn_semiconductor_2009}. Although the exact
form for $\psi^{\perp}$ depends on the exact distribution of $V(z)$,
without loss of generality we can assume the electrons are confined in
a potential well of width $\delta$, which is the typical treatment for
semiconductor QWs \cite{ihn_semiconductor_2009,Fowler_1984,Maize_2011}. 
The allowed bound
states inside the confined region generally have wave vector
$k_{z} \propto n \pi / \delta$. With the total energy
$E_{n}(\mathbf{k}) = {\displaystyle \frac{\hbar^{2} (k_{x}^{2} +
    k_{y}^{2})}{2 m^{\parallel}} + \frac{\hbar^{2} n^{2} \pi^{2}}{2
    m^{\perp} \delta^{2}}}$, where $m^{\parallel}$ and $m^{\perp}$ are
the effective masses parallel and perpendicular to the 2D
plane. Therefore, the denominator of Eq. \ref{eq:alpha_perp_RPA} becomes independent of
$\mathbf{k}$, that
$E_{\mathrm{c}}(\mathbf{k}) - E_{\mathrm{v}}(\mathbf{k}) =
(n_{\mathrm{c}}^{2} - n_{\mathrm{v}}^{2}) {\displaystyle
  \frac{\hbar^{2} \pi^{2}}{2 m^{\perp} \delta^{2}}}$. On the other
hand, the numerator
$<\psi^{\perp}_{\mathrm{v}}(\mathbf{k})|z|\psi^{\perp}_{\mathrm{c}}(\mathbf{k})>$
is proportional to the confinement length $\delta$ which can be seen using the 
particle-in-box solution\cite{davies_physics_1997}. In combination,
the individual terms of the summation in the right hand of Eq. \ref{eq:alpha_perp_RPA} is
independent of neither $E_{\mathrm{g}}$ nor \textbf{k}, proving that
$\alpha_{\mathrm{2D}}^{\perp}$ is independent of the band gap.
In the next section we will
provide a simple explanation for the $\alpha \propto \delta$ relation
from fundamental electrostatics theory.

\subsection{Explanation of main text Eq. \ref{main-eq:2D-Moss-perp}
  from fundamental electrostatics}
\label{ssec:theory-1-perp-fundamental}

\begin{figure}[htbp]
  \centering
  \includegraphics[width=0.99\linewidth]{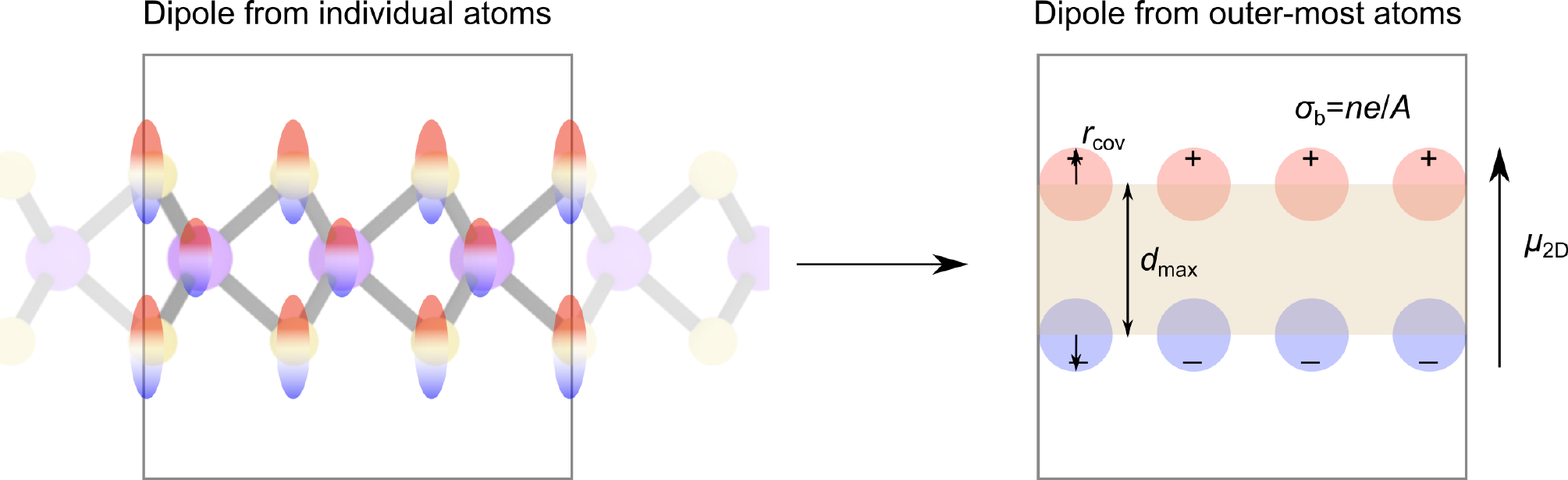}
  \caption{Fundamental electrostatic model for the
    thickness-dependency of $\alpha_{\mathrm{2D}}^{\perp}$, using 2H-MoS2 as an
    example. Left: induced dipoles from individual atoms along the
    z-direction. The positive and negative induced charges from inner
    atoms cancel out. Right: simplified model for the thickness
    dependency of $\alpha_{\mathrm{2D}}^{\perp}$, where the surface dipole density
    $\boldsymbol{\mu}$ comes only from the outer-most atoms.}
  \label{fig:classic-model}
\end{figure}

The dependency of $\alpha_{\mathrm{2D}}^{\perp}$ on the thickness
$\delta$ of a 2D material, can also be regarded using fundamental
electrostatic model. Consider the smallest repeating unit of the 2D
material with xy-plane area $A$, under small perturbation field $E$
along the z-direction.  Note that the surface bound charge
$\sigma_{\mathrm{b}}=n e /A$, where $n$ is the number of unit charges
contributes to the bound charges, comes only from the dipoles of the
outer-most atoms, since the induced charges from inner atoms are
cancel out (see Figure \ref{fig:classic-model}).
From the definition of
$\alpha_{\mathrm{2D}}^{\perp}$, we have:
\begin{equation}
  \label{eq:alpha-classic}
  \alpha_{\mathrm{2D}}^{\perp} = \frac{\boldsymbol{u}_{z}}{E_{\mathrm{loc}} A}
  = \frac{(d_{\mathrm{max}} + r_{\mathrm{cov}}^{i} + r_{\mathrm{cov}}^{j}) \sigma_{\mathrm{b}}}{E}
\end{equation}
where $r_{\mathrm{cov}}^{i}$ and $r_{\mathrm{cov}}^{j}$ are the
covalent radii of the outer-most atoms, the characteristic length of
the dipole extension in z-direction, respectively, and $d_{\mathrm{max}}$ is the
z-distance between the nuclei of such atoms.  The field $E$
counterbalances the field from the surface bound charges and equals
$E = \sigma_{\mathrm{b}}/\varepsilon_{0}$. Therefore we have:
\begin{equation}
  \label{eq:alpha-classic-2}
  \alpha_{\mathrm{2D}}^{\perp} = (d_{\mathrm{max}} + r_{\mathrm{cov}}^{i} + r_{\mathrm{cov}}^{j})\varepsilon_{0}
                = \delta_{\mathrm{cov}} \varepsilon_{0}
\end{equation}
which explains the linear relation seen in main text Figure
\ref{main-fig-3}d. We can see that such simple model nicely captures
the thickness feature of $\alpha_{\mathrm{2D}}^{\perp}$, and
reproduces the right coefficient between $\delta_{\mathrm{cov}}$ and
$\alpha_{\mathrm{2D}}^{\perp}$.

\section{Dependence of $\alpha_{\mathrm{2D}}$ on bandgap}
\label{sec:pol-2D-Eg}

\begin{figure}[htbp]
  \centering
  \includegraphics[width=0.99\linewidth]{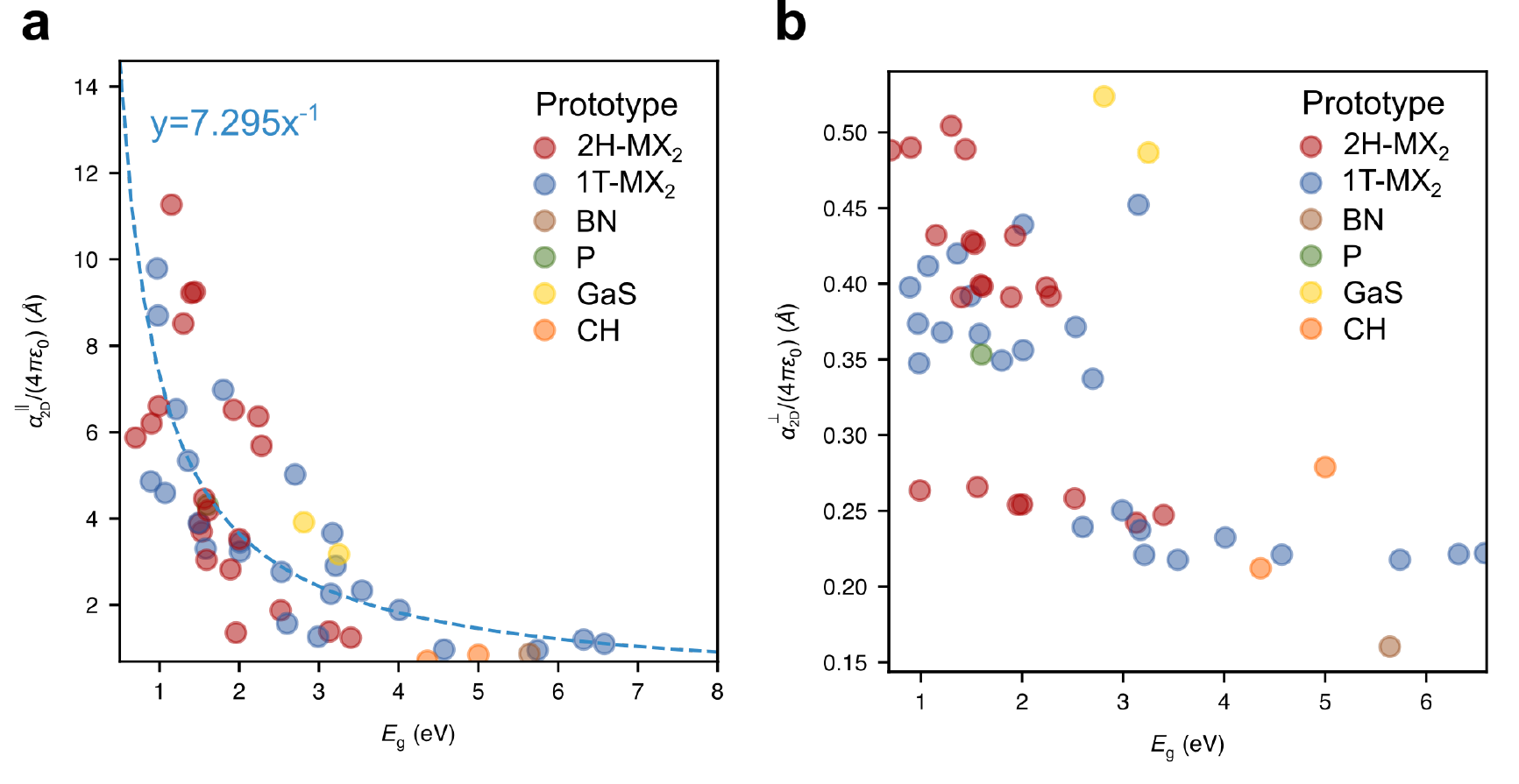}
  \caption{$E_{\mathrm{g}}$-dependence of \textbf{a,} $\alpha_{\mathrm{2D}}^{\parallel}$ and
    \textbf{b,} $\alpha_{\mathrm{2D}}^{\perp}$ for the 2D materials investigated here using HSE06.}
  \label{fig:SI-raw-HSE}
\end{figure}

In this section we further look into the bandgap dependency of the 2D
polarizability. Figure \ref{fig:SI-raw-HSE} shows the raw data of
$\alpha_{\mathrm{2D}}^{\parallel}$ and $\alpha_{\mathrm{2D}}^{\perp}$
as functions of $E_{\mathrm{g}}$ of the 2D materials investigated. We
observe that $\alpha_{\mathrm{2D}}^{\parallel}$ can be approximated by
a reciprocal function of $E_{\mathrm{g}}$, that
$\alpha_{\mathrm{2D}}^{\parallel}\sim{}
7.295(E_{\mathrm{g}})^{-1}$. On the other hand, the plot of
$\alpha_{\mathrm{2D}}^{\perp}$ against $E_{\mathrm{g}}$ shows no
apparent correlation.

\begin{figure}[htbp]
  \centering
  \includegraphics[width=0.6\linewidth]{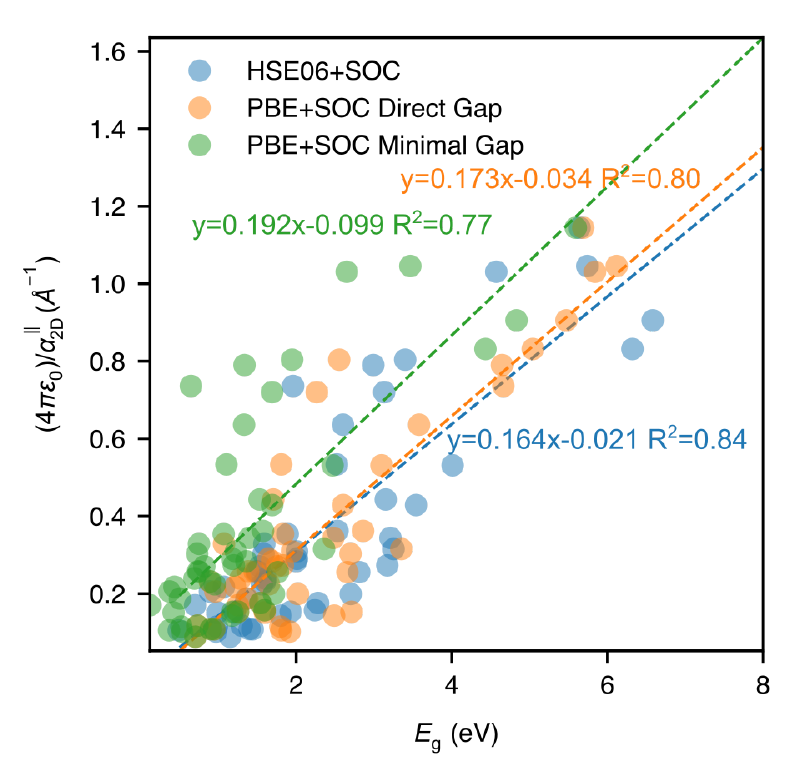}
  \caption{Relation between $1/\alpha_{\mathrm{2D}}^{\parallel}$ and various choices
    of $E_{\mathrm{g}}$: minimal gap from HSE06 (blue), minimal gap
    from PBE (orange) and direct gap from PBE (green). The linear
    regression results are shown as broken lines.}
  \label{fig:alpha-Eg-diff}
\end{figure}

We also investigate the relation of 2D polarizabilities with
difference choices of \textit{ab initio} bandgaps. It is widely
accepted that the PBE exchange correlation, tends to underestimate the
bangap \cite{Heyd_2005,Kumar_2016_PRB,Kumar_2016_jpcc}.  Indeed,
changing the choice of $E_{\mathrm{g}}$ yields different regression
relation with $1/\alpha_{\mathrm{2D}}^{\parallel}$, as shown in Figure
\ref{fig:alpha-Eg-diff}. We see that due to the underestimation of PBE
bandgap, the slope of linear regression is larger than that from
HSE-bandgap. We also observe that the $1/\alpha-E_{\mathrm{g}}$
relation is better presented by using the minimal HSE bandgap than the
minimal PBE bandgap, due to higher regression $R^{2}$ coefficient of
the former. We note that the higher $R^{2}$ coefficient observed using
the direct PBE bandgap than the minimal PBE bandgap may be solely
caused by the fact that the direct bandgap of 2D materials on the PBE
level is closer to the HSE bandgap. From the random phase
approximation theory of dielectric response, the polarizability is
contributed by all possible transition between valence and conduction
bands, with the minimal bandgap as the least possible transition. In
this sense, $\alpha_{\mathrm{2D}}^{\parallel}$ is mostly like to be associated with
the minimal, not direct bandgap, as also observed in the original Moss
relation. We also examine the validity of such statement based on the
analysis of a different database\cite{Haastrup_2018} 
as will be discussed in the following sections.

\section{Using a different dataset of 2D materials}
\label{sec:gpaw}
\subsection{Validation of the universal description of 2D polarizabilities}
\label{sec:gpaw-1}

\begin{figure}[htbp]
  \centering
  \includegraphics[width=1.05\linewidth]{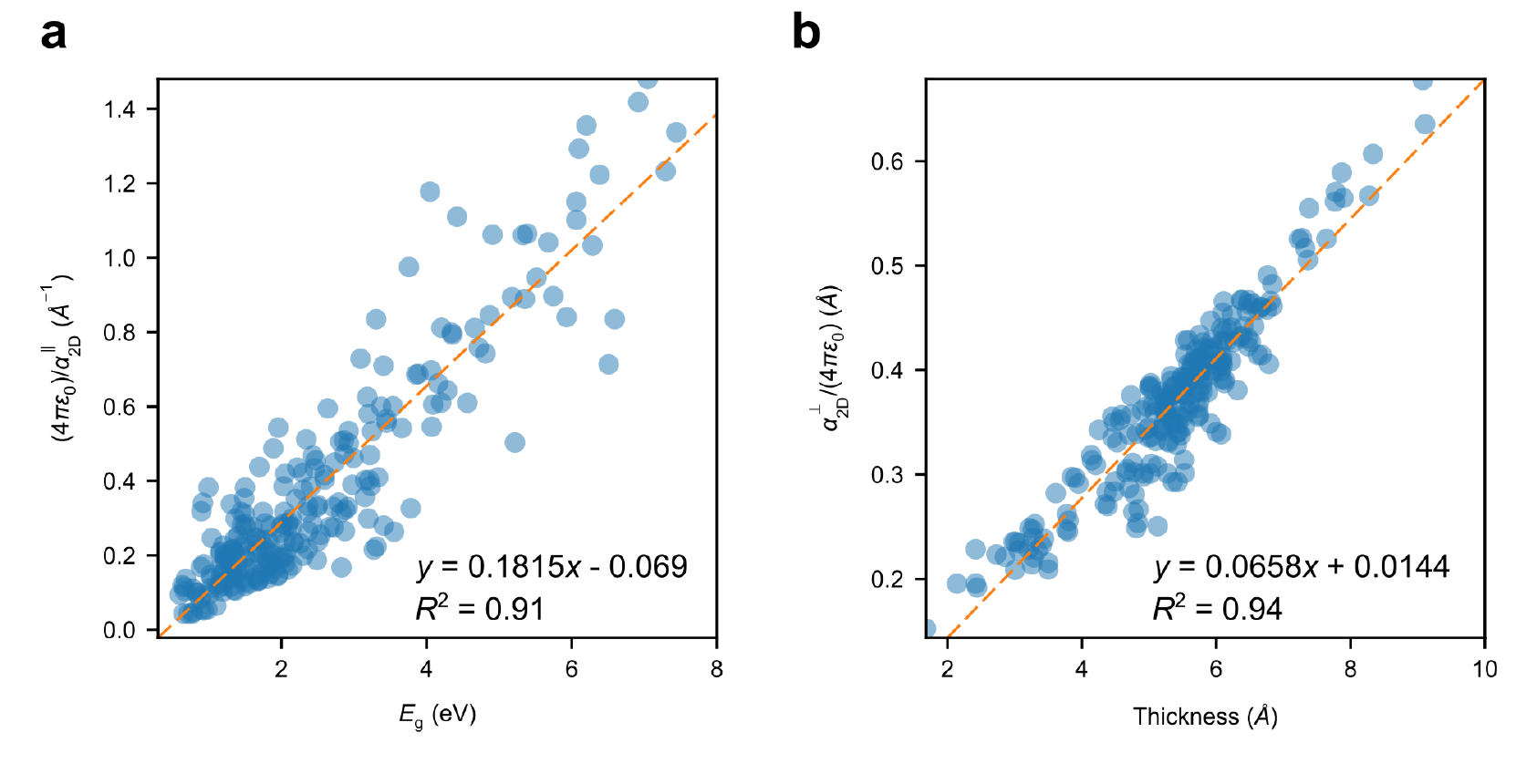}
  \caption{Validation of the linear relation between \textbf{a,}
    $1/\alpha_{\mathrm{2D}}^{\parallel}-E_{\mathrm{g}}(\mathrm{HSE})$
    and \textbf{b,}
    $\alpha_{\mathrm{2D}}^{\perp}-\delta_{\mathrm{cov}}$ from 
    Ref.~\citenum{Haastrup_2018} corresponding to main text Figure \ref{main-fig-3}{\bf c}
    and \ref{main-fig-3}{\bf d}.}
  \label{fig:gpaw-alpha-relation}
\end{figure}

Due to time-consuming simulations and significant increment in 
memory overload, high-accurate calculations at hybrid HSE06 level is limited 
to about 55 compounds. 
It is desirable to validate our proposed
relations on an even larger scale database. We select over 248
semiconducting 2D materials from Ref.~\citenum{Haastrup_2018} 
with a GW bandgap
larger than 0.05 eV and extracted the 2D polarizabilities calculated
on the PBE level. The proposed linear relations between
$1/\alpha_{\mathrm{2D}}^{\parallel}-E_{\mathrm{g}}(\mathrm{HSE})$ and
$\alpha_{\mathrm{2D}}^{\perp}-\delta_{\mathrm{cov}}$ are also valid,
as shown in Figure \ref{fig:gpaw-alpha-relation}. Excellent linear
correlation is observed in both cases with the $R^{2}$ coefficient
larger than 0.9 which indicates the existence of a universal description 
of 2D dielectric nature through the proposed relations with the 2D
polarizabilities. We note that the slope of the linear regression is
slightly different from the one proposed from the dielectric
response at the HSE06 level. 

\subsection{Choice of bandgap}
\label{sec:gpaw-2}

\begin{figure}[htbp]
  \centering
  \includegraphics[width=1.05\linewidth]{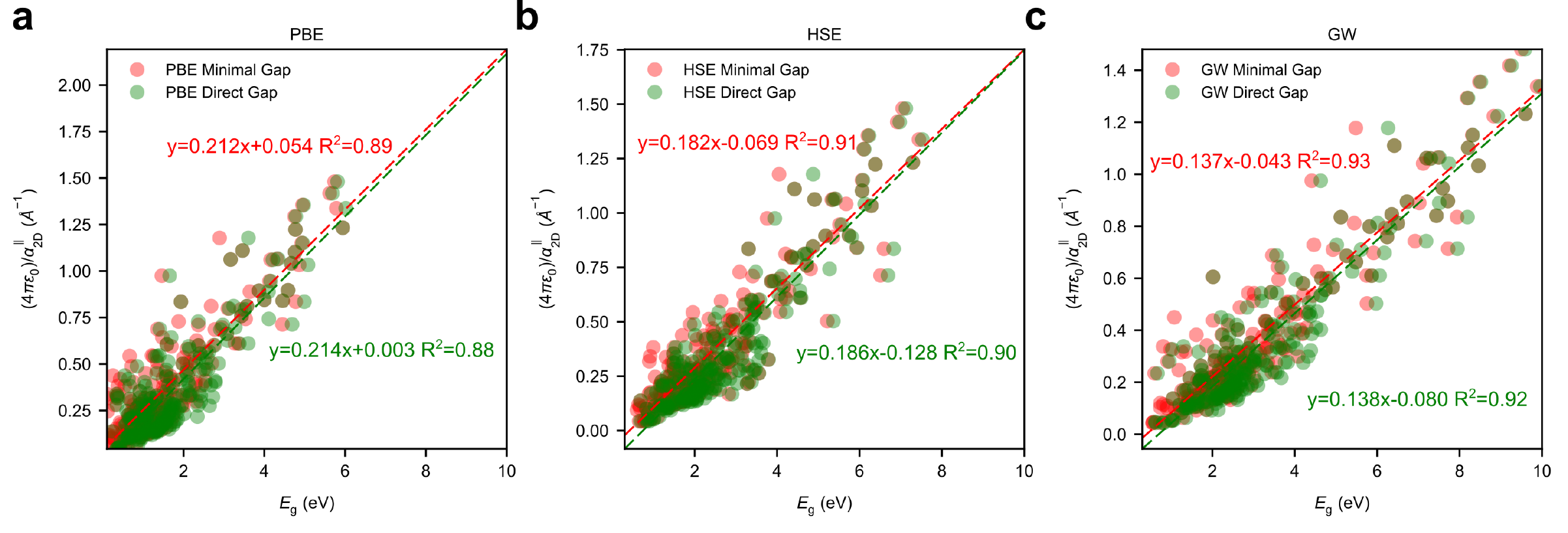}
  \caption{$\alpha_{\mathrm{2D}}^{\parallel}$ as function of minimal and direct
    $E_{\mathrm{g}}$ calculated on different theoretical levels:
    \textbf{a,} PBE, \textbf{b,} HSE and \textbf{c,} GW of Ref.~\citenum{Haastrup_2018}.}
  \label{fig:SI-gpaw-alpha-Eg-all}
\end{figure}

Next we investigate the influence of choice of $E_{\mathrm{g}}$ on the
regression of $\alpha_{\mathrm{2D}}^{\parallel}-E_{\mathrm{g}}$ relation. Figure
\ref{fig:SI-gpaw-alpha-Eg-all} shows $1/\alpha_{\mathrm{2D}}^{\parallel}$ from 
Ref.~\citenum{Haastrup_2018} as a function of minimal and direct bandgap calculated
on PBE, HSE06 and GW levels. We observe,
although the regression $R^{2}$ coefficient in all cases are around
0.9, the $\alpha_{\mathrm{2D}}^{\parallel}-E_{\mathrm{g}}$ is better described using
the HSE and GW bandgaps than the PBE bandgaps. On the other hand,
using indirect or minimal bandgaps on the same level gives almost
identical regression slope. The observations are in good agreement
with our calculations on the HSE level discussed in Section
\ref{sec:pol-2D-Eg}. In combination with the physical contribution of
$E_{\mathrm{g}}$ to the dielectric screening, we conclude that the
minimal bandgap should be used for quantitative prediction of the
in-plane 2D polarizability. The prediction is greatly improved when
more accurate theory level for bandgap is used (for instance, HSE and GW).

\begin{figure}[htbp]
  \centering
  \includegraphics[width=0.70\linewidth]{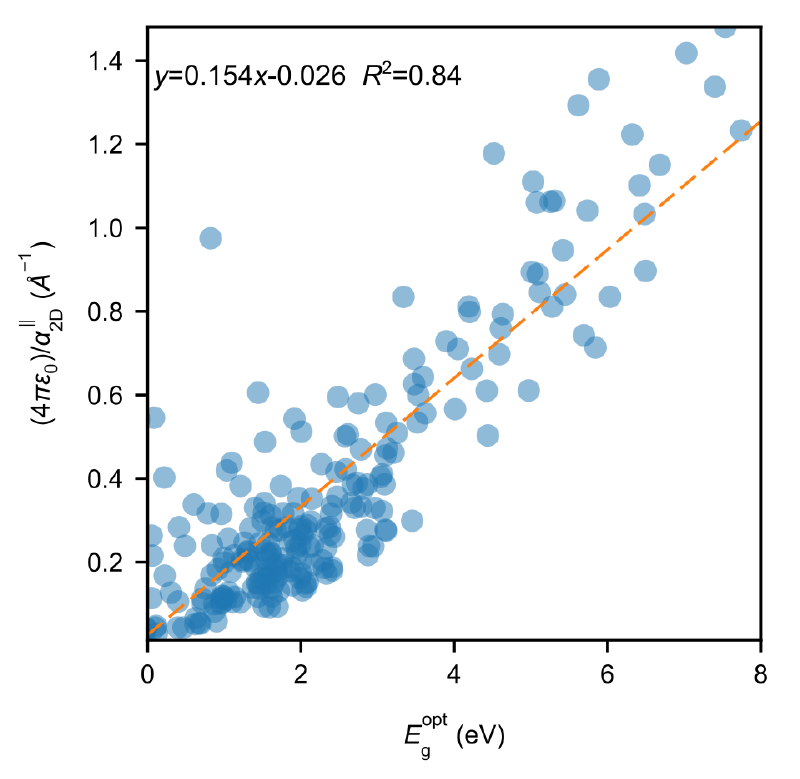}
  \caption{$(4\pi\varepsilon_{0})/\alpha_{\mathrm{2D}}^{\parallel}$ as a function of
    $E_{\mathrm{g}}^{\mathrm{opt}}$ from Ref.~\citenum{Haastrup_2018}.}
  \label{fig:opt}
\end{figure}

For 2D materials, the exciton effect plays an important role in
determining the experimentally accessible bandgap
\cite{Arnaud_2006_exc_hBN,Pulci_2014,Ramasubramaniam_2012,Chernikov_2014_EB_MoS2_2D3D}. The experimentally observed optical bandgap
$E_{\mathrm{g}}^{\mathrm{opt}}$, is usually lower than the direct
bandgap from band structure $E_{\mathrm{g}}^{\mathrm{dir}}$ by the
exciton binding energy $E_{\mathrm{b}}$, which is at the 10$^{-1}$ to
10$^{1}$ eV for different 2D materials due to the attenuated
dielectric screening. Next we examine the relation between
$\alpha_{\mathrm{2D}}^{\parallel}$ and $E_{\mathrm{g}}^{\mathrm{opt}}$
from Ref.~\citenum{Haastrup_2018} with
$E_{\mathrm{g}}^{\mathrm{opt}}=E_{\mathrm{g}}^{\mathrm{dir,QP}}-E_{\mathrm{b}}^{\mathrm{BSE}}$,
where $E_{\mathrm{g}}^{\mathrm{dir,QP}}$ is the direct quasi-particle
bandgap calculated using G$_0$W$_0$ method and
$E_{\mathrm{b}}^{\mathrm{BSE}}$ is the exciton binding energy
calculated using the Bethe-Salpethe equation. Figure \ref{fig:opt}
shows $(4\pi\varepsilon_{0})/\alpha_{\mathrm{2D}}^{\parallel}$ as a
function of $E_{\mathrm{g}}^{\mathrm{opt}}$, with a linear regression
slope of 0.154 and $R^{2}$ of 0.84, similar to the relation between
$(4\pi \varepsilon_{0})/\alpha_{\mathrm{2D}}^{\parallel}$ and
$E_{\mathrm{g}}$ (from HSE06 level, see Figures \ref{fig:SI-raw-HSE}
and \ref{fig:SI-gpaw-alpha-Eg-all}). The roughly linear correlation
between $(4\pi \varepsilon_{0})/\alpha_{\mathrm{2D}}^{\parallel}$ and
$E_{\mathrm{g}}^{\mathrm{opt}}$ is not coincidental: in fact,
theoretical analysis shows that the binding energy $E_{\mathrm{b}}$ is
proportional to the direct bandgap $E_{\mathrm{g}}$
\cite{Jiang_2017_Eg_Eb}, taking into account that
$(\alpha_{\mathrm{2D}}^{\parallel})^{-1} \propto E_{\mathrm{g}}$, we
rationalize that
$(\alpha_{\mathrm{2D}}^{\parallel})^{-1} \propto
E_{\mathrm{g}}^{\mathrm{opt}}=E_{\mathrm{g}}^{\mathrm{dir}}-E_{\mathrm{b}}$. The
slightly smaller linearity than the 2D Moss-like relation is caused
from multiple approximations used. Nevertheless we show that
$(4\pi\varepsilon_{0})/\alpha_{\mathrm{2D}}^{\parallel}$ can be
equivalently predicted using the experimentally accessible optical
bandgap.

\subsection{Relation between 2D polarizabilities and other physical quantities}
\label{sec:gpaw-3}

\begin{figure}[htbp]
  \centering
  \includegraphics[width=0.99\linewidth]{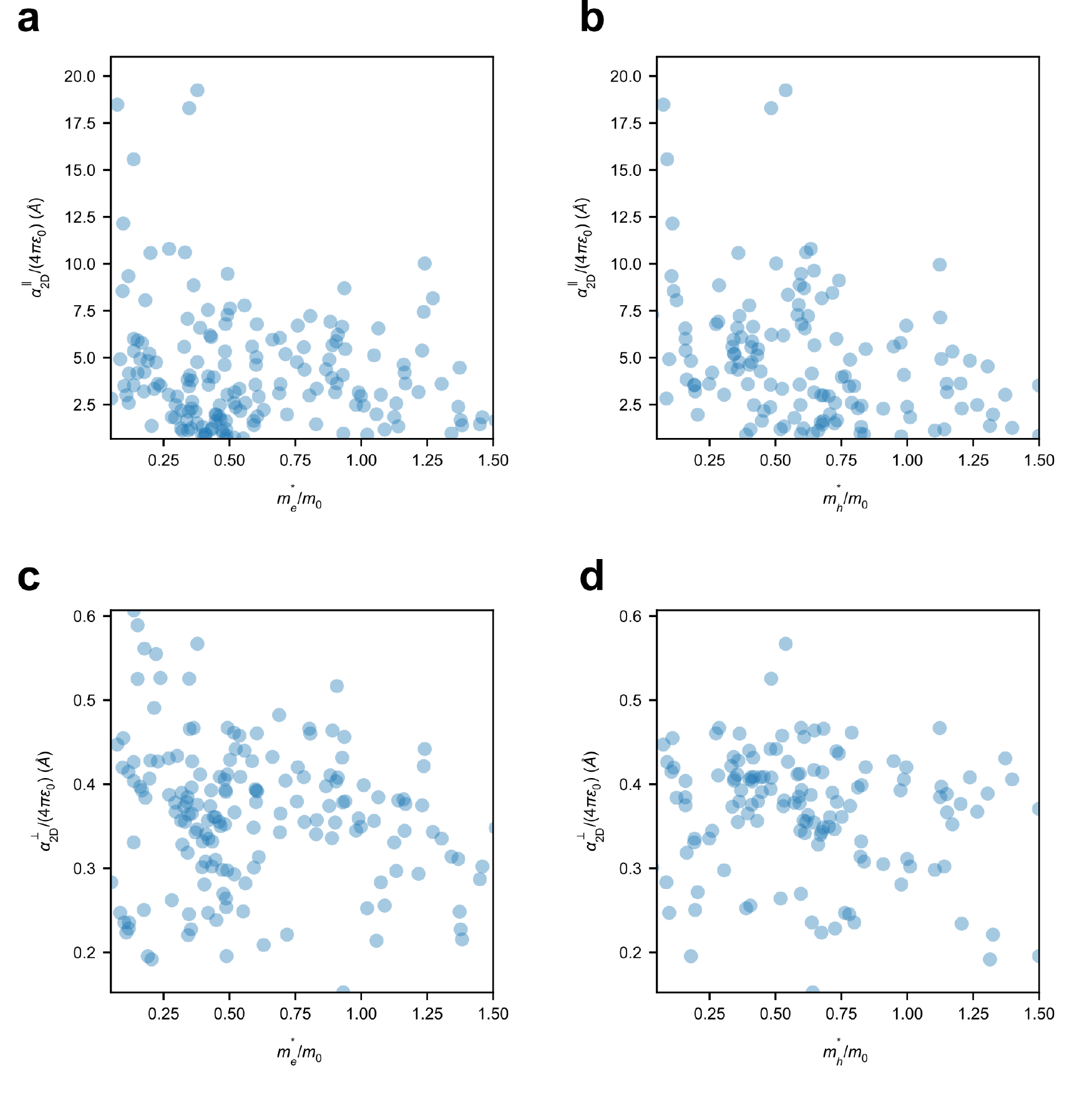}
  \caption{Relation between 2D polarizabilities and the effective
    carrier mass from Ref.~\citenum{Haastrup_2018}. 
    \textbf{a}. $\alpha_{\mathrm{2D}}^{\parallel}$ as a function of the
    electron mass $m_{e}^{*}$.  \textbf{b}. $\alpha_{\mathrm{2D}}^{\parallel}$ as a
    function of the hole mass
    $m_{h}^{*}$. \textbf{c}. $\alpha_{\mathrm{2D}}^{\perp}$ as a function of the
    electron mass $m_{e}^{*}$.  \textbf{d}. $\alpha_{\mathrm{2D}}^{\perp}$ as a
    function of the hole mass $m_{h}^{*}$. No apparent correlation
    between the 2D polarizabilities and the effective carrier masses
    is observed.}
  \label{fig:gpaw-2D-quantities-1}
\end{figure}

\begin{figure}[htbp]
  \centering
  \includegraphics[width=0.95\linewidth]{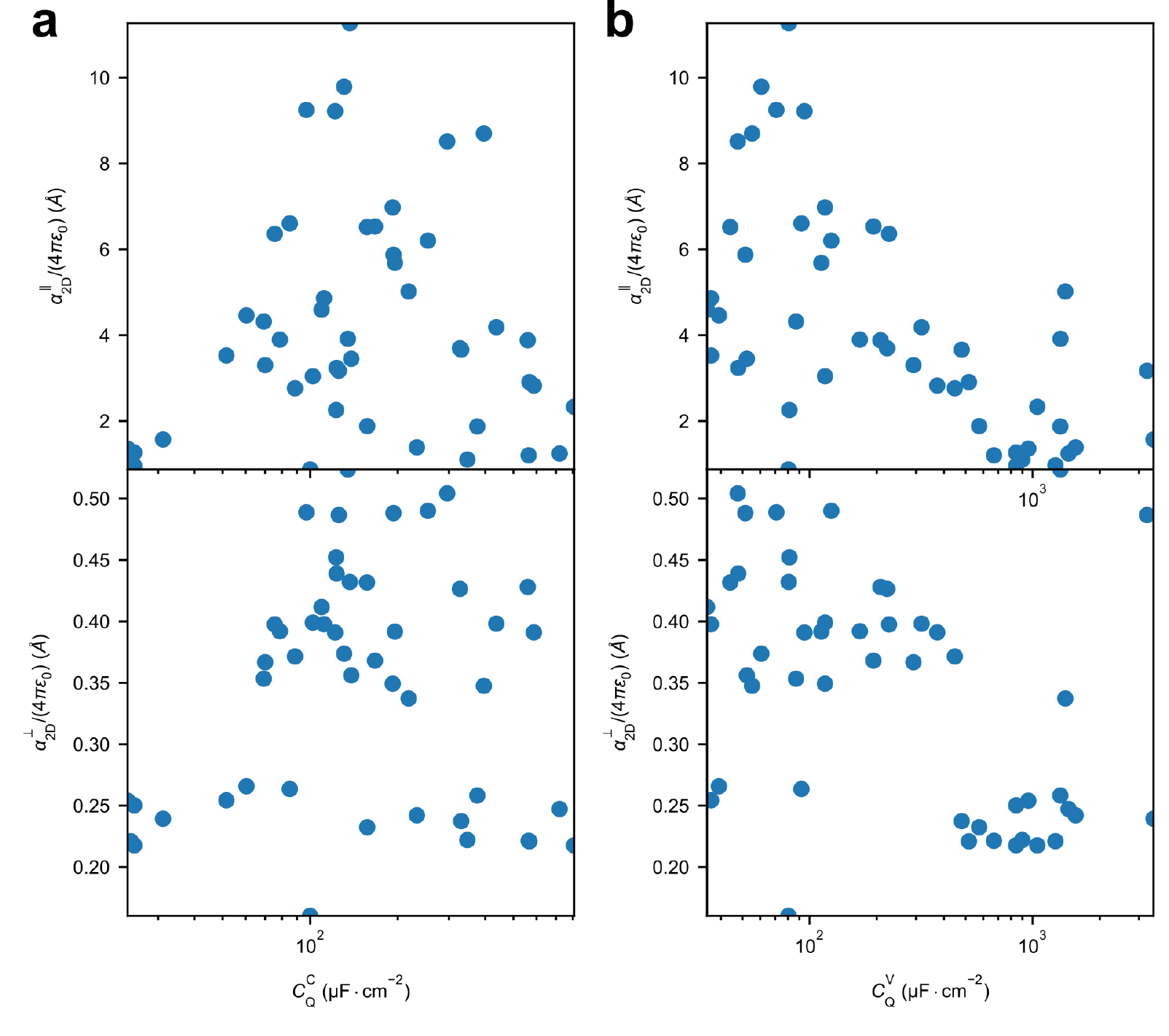}
  \caption{Relation between the 2D polarizabilities with the quantum
    capacitance. \textbf{a} $\alpha_{\mathrm{2D}}^{\parallel}$ (top) and
    $\alpha_{\mathrm{2D}}^{\perp}$ (bottom) as functions of the quantum capacitance
    of the conduction band edge,
    $C_{\mathrm{Q}}^{\mathrm{C}}$. \textbf{b} $\alpha_{\mathrm{2D}}^{\parallel}$
    (top) and $\alpha_{\mathrm{2D}}^{\perp}$ (bottom) as functions of the quantum
    capacitance of the valence band edge,
    $C_{\mathrm{Q}}^{\mathrm{V}}$. Similar to the case of effective
    carrier mass, no apparent correlation between 2D polarizabilities
    and the quantum capacitance can be found.}
  \label{fig:gpaw-2D-quantities-2}
\end{figure}

\begin{figure}[htbp]
  \centering
  \includegraphics[width=1.07\linewidth]{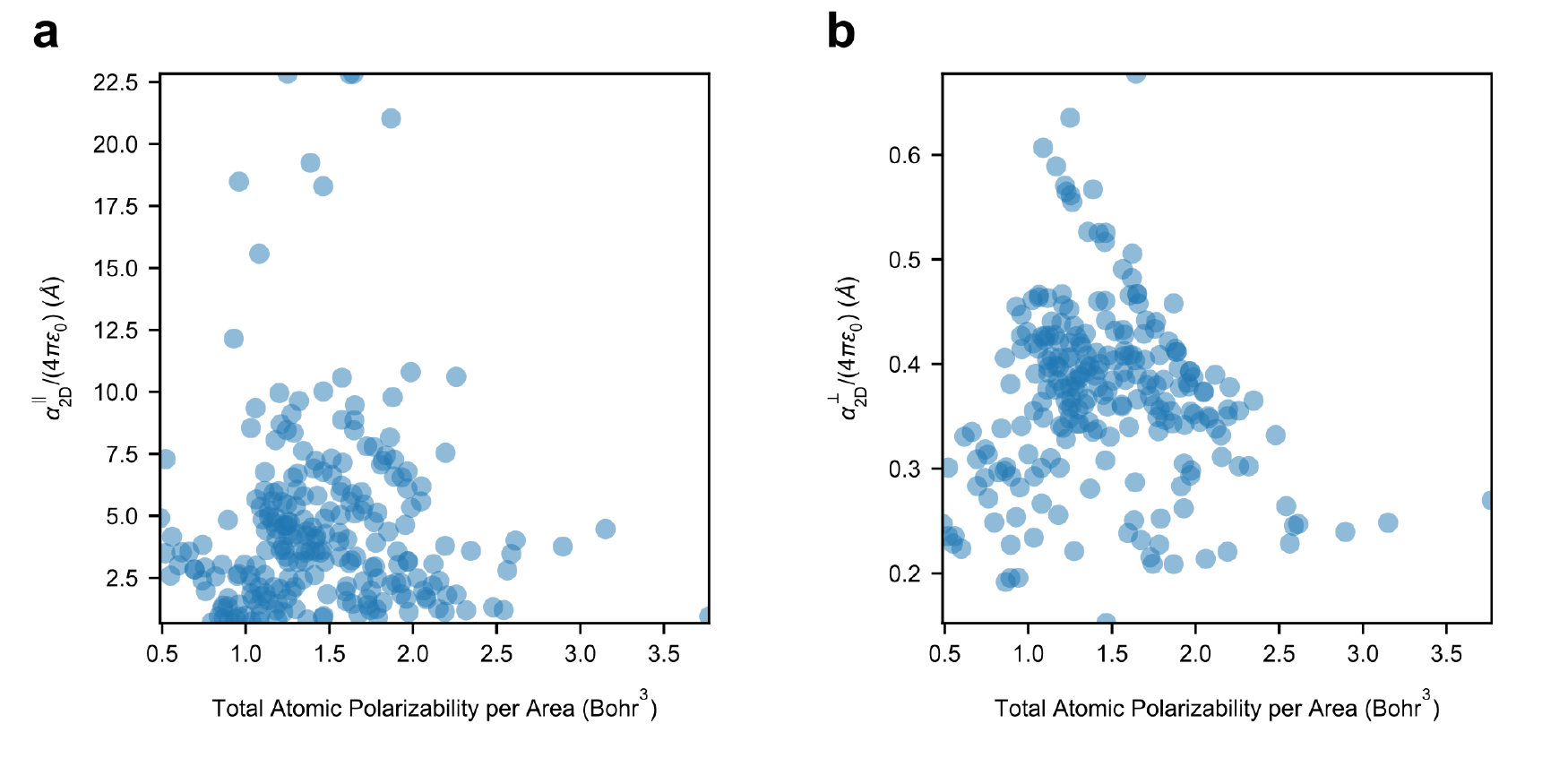}
  \caption{Relation between the 2D polarizabilities
    (\textbf{a}. $\alpha_{\mathrm{2D}}^{\parallel}$ and
    \textbf{b}. $\alpha_{\mathrm{2D}}^{\perp}$) with the total atomic polarizability per area.}
  \label{fig:gpaw-2D-quantities-3}
\end{figure}

The relatively large size of Ref.~\citenum{Haastrup_2018} database allows us
to examine the relation between 2D polarizabilities and other physical
quantities. We choose the following quantities for comparison,
corresponding to Figures \ref{fig:gpaw-2D-quantities-1} to
\ref{fig:gpaw-2D-quantities-3}:
\begin{enumerate}
\item The effective carrier mass for electron $m_{e}^{*}$ and hole $m_{h}^{*}$
  
\item The quantum capacitance at the conduction band edge
  $C_{\mathrm{Q}}^{\mathrm{C}}$ and valence band edge
  ($C_{\mathrm{Q}}^{\mathrm{V}}$).

\item The total atomic polarizabilities per area $\alpha_{\mathrm{2D}}^{\mathrm{sum}}$.
\end{enumerate}

The quantum capacitance
$C_{\mathrm{Q}}(E)$ at certain energy level $E$ is calculated using
the relation $C_{\mathrm{Q}}(E)=\mathrm{DOS}(E)e^{2}$, where
$\mathrm{DOS}(E)$ is the density of states at the conduction or valence band edge (averaged by cell
area). The DOS value is calculated at the energy level with a charge
cutoff such that
$|n_{\mathrm{2D}}(E)| = 5 \times 10^{13}\ \mathrm{cm}^{-2}$, calculated by
the relation of accumulated charge $n_{\mathrm{2D}}(E)$ at CB or VB:
\begin{equation}
  \label{eq:CQ-method}
  |n_{\mathrm{2D}}(E)| = \left|\int_{E_{\mathrm{BE}}}^{E} \mathrm{DOS}(E') dE' \right|
\end{equation}
where $E_{\mathrm{BE}}$ is the energy of the CB or VB band edge.

The total polarizability $\alpha_{\mathrm{2D}}^{\mathrm{sum}}$ is calculated by the
summation of the atomic polarizabilities $\alpha_{\mathrm{2D}}^{\mathrm{atom}}$
\cite{Gould_2016_jctc} of individual atoms per area $A$, such that:
\begin{equation}
  \label{eq:atom-polar}
  \alpha_{\mathrm{2D}}^{\mathrm{sum}} = \frac{\sum_{i} \alpha^{\mathrm{atom}}_{i}}{A}
\end{equation}

From Figures \ref{fig:gpaw-2D-quantities-1} to
\ref{fig:gpaw-2D-quantities-3} we can see that none of the above
quantities have apparent relation with the 2D polarizabilities, as
compared with the bandgap and covalent thickness proposed in the main
text. 

\section{More discussion about the relation between 2D and 3D properties}
\label{sec:2D-3D}

\subsection{Comparing 2D and 3D Moss relations}
\label{ssec:theory-2D}
The 2D Moss-like relation $\alpha^{\parallel} \propto E_{g}^{-1}$ is
similar to the 3D Moss relation $\varepsilon \propto E_{g}^{-1/2}$,
with a different power law. Such difference in the power law can
indeed be explained by modern theory of dielectric properties. From
the 2D material to a bulk covalent semiconductor, the wave function
becomes periodic in all directions. Considering only one pair of
valence-conduction transition and uniform effective mass $m^{*}$,
extending the approach Eq. \ref{eq:alpha_perp_RPA} to the bulk
material, and using the Bloch presentation for wave functions in all
dimensions, we get \cite{Jiang_2017_Eg_Eb}:
\begin{equation}
  \begin{aligned}
    \label{eq:eps-bulk}
    \varepsilon_{\mathrm{bulk}} - 1 &\propto \int d^{3}\mathbf{k}
    {\displaystyle \frac{1}{(E_{\mathrm{g}} + {\displaystyle
          \frac{\hbar^{2} k^{2}}{m^{*}}})^{2}}}\\
    &= \int_{0}^{k_{\mathrm{BZ}}} {\displaystyle
      \frac{4 \pi k^{2}}{(E_{\mathrm{g}} + {\displaystyle \frac{\hbar^{2}
            k^{2}}{m^{*}}})^{2}}} dk
  \end{aligned}
\end{equation}
where $k_{\mathrm{BZ}}$ is the boundary for the Brillouin Zone. The
last step in Eq. \ref{eq:eps-bulk} assumes the integral within the
Brillouin Zone is equivalent to integral inside a sphere of
k-space. Let $\hbar^{2}/(2 m^{*})=\beta$, the integral becomes:
\begin{equation}
  \begin{aligned}
    \label{eq:integral-BZ-bulk}
    \varepsilon_{\mathrm{Bulk}} &\propto {\displaystyle \frac{2 \pi
        \mathrm{arctan}(\sqrt{\beta k^{2}/E_{\mathrm{g}}})}{\sqrt{E_{\mathrm{g}} \beta^{3}}}
        - \frac{2\pi k}{\beta(\beta +E_{\mathrm{g}}k^{2})}
      } \bigg\rvert_{0}^{k_{\mathrm{BZ}}}\\
      &\propto 1/\sqrt{E_{\mathrm{g}}}
  \end{aligned}
\end{equation}
when $\varepsilon_{\mathrm{Bulk}} \gg 1$. since generally
$\hbar^{2}k_{\mathrm{BZ}}^{2}/(2m^{*}) \gg
E_{\mathrm{g}}$ \cite{Finkenrath_1988}. The final result $\varepsilon_{\mathrm{Bulk}} \propto E_{\mathrm{g}}^{-1/2}$ recovers the original Moss relation for bulk semiconductors.

\subsection{Static 2D polarizability and 2D plasma frequency}
\label{ssec:omega-p}

\begin{figure}[htbp]
  \centering
  \includegraphics[width=0.85\linewidth]{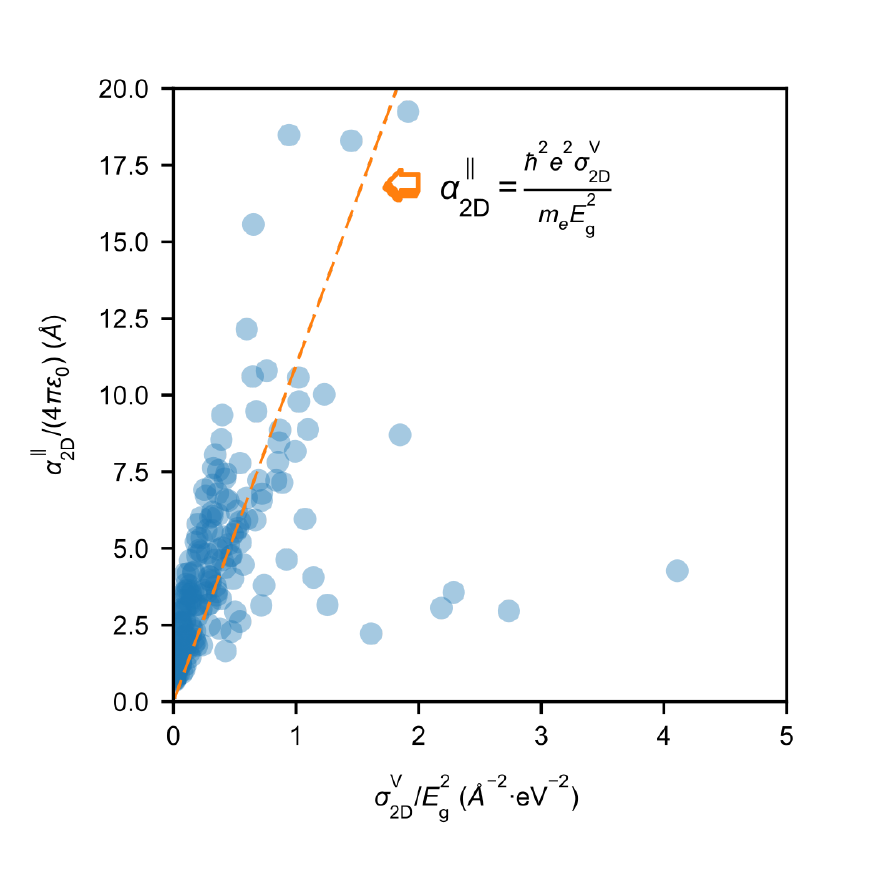}
  \caption{Calculated $\alpha_{\mathrm{2D}}^{\parallel}$ as a function of
    $\sigma_{\mathrm{2D}}^{\mathrm{v}}/E_{\mathrm{g}}^{2}$ using data from Ref.~\citenum{Haastrup_2018}. The broken line shows the theoretical prediction
    from single-oscillator model.}
  \label{fig:plasma}
\end{figure}

A common approach for describing the bulk dielectric function of bulk
semiconductors is via the Lorentz oscillator model, where the
dielectric function is dominated by the plasma frequencies
$\omega_{\mathrm{3D}}^{\mathrm{p}}$ and bandgap $E_{\mathrm{g}}$ of
individual oscillators \cite{ketterson_physics_2016}. At zero optical
frequency and the static limit, the dielectric constant for single
oscillator is:
\begin{equation}
  \label{eq:eps-plas-3D}
  \varepsilon_{\mathrm{3D}} = 1 +
  \frac{\hbar^{2} (\omega_{\mathrm{3D}}^{\mathrm{p}})^{2}}{E_{\mathrm{g}}^{2}}
\end{equation}
where
$\omega_{\mathrm{3D}}^{\mathrm{p}} = {\displaystyle \sqrt{\frac{e^{2}
      n_{\mathrm{3D}}}{\varepsilon_{0} m_{e}}}}$, where
$n_{\mathrm{3D}}$ is the 3D number density of valence
electrons. Combine Eq. \ref{eq:eps-plas-3D} with main text
Eq. \ref{main-eq:alpha-para-def}, we get:
\begin{equation}
  \begin{aligned}
  \label{eq:alpha-plas}
  \alpha_{\mathrm{2D}}^{\parallel} &= \frac{e^{2} n_{\mathrm{3D}} L}{m_{e} E_{\mathrm{g}}^{2}} \\
  &= \frac{e^{2} n_{\mathrm{2D}}}{m_{e} E_{\mathrm{g}}^{2}} \\
  &= \varepsilon_{0} \frac{\hbar^{2}
    (\omega_{\mathrm{2D}}^{\mathrm{p}})^{2}}{E_{\mathrm{g}}^{2}}
\end{aligned}
\end{equation}
where $n_{\mathrm{2D}} =n_{\mathrm{3D}} L$ is the 2D number density of
valence electrons and
$\omega_{\mathrm{2D}}^{p}=\omega_{\mathrm{3D}}^{p}\sqrt{L}$ is the 2D
plasma frequency at static limit \cite{Nazarov_2015_2D_3D}, as
discussed in the main text. Apparently $n_{\mathrm{2D}}$ and
$\omega_{\mathrm{2D}}^{p}$ defines the superlattice-independent 2D
quantity $\alpha_{\mathrm{2D}}^{\parallel}$, while its 3D counterpart
$\varepsilon_{\mathrm{3D}}$ is dependent on $L$. By defining the 2D
valence charge density
$\sigma_{\mathrm{2D}}^{\mathrm{v}}=n_{\mathrm{2D}}e$, we have also
calculated $\alpha_{\mathrm{2D}}^{\parallel}$ as a function of
$\sigma_{\mathrm{2D}}^{\mathrm{v}}/E_{\mathrm{g}}^{2}$ using Ref.~\citenum{Haastrup_2018} database, as shown in Figure \ref{fig:plasma}.
It can be seen that, a large number of materials are close to the
theoretical value of
$\alpha_{\mathrm{2D}}^{\parallel} = \frac{e^{2} n_{\mathrm{3D}} L}{m_{e}
  E_{\mathrm{g}}^{2}}$ (broken line). However there are also many
violations to this simple relation, making such model not suitable for
quantitative prediction of the 2D dielectric nature, due to the
oversimplification of single Lorentz oscillator. Nevertheless, this
example shows excellently how the quantities in both dimensions are
related to each other.

\subsection{The relation between 2D and 3D physical quantities}
\begin{figure}[htbp]
\centering
\includegraphics[width=0.85\linewidth]{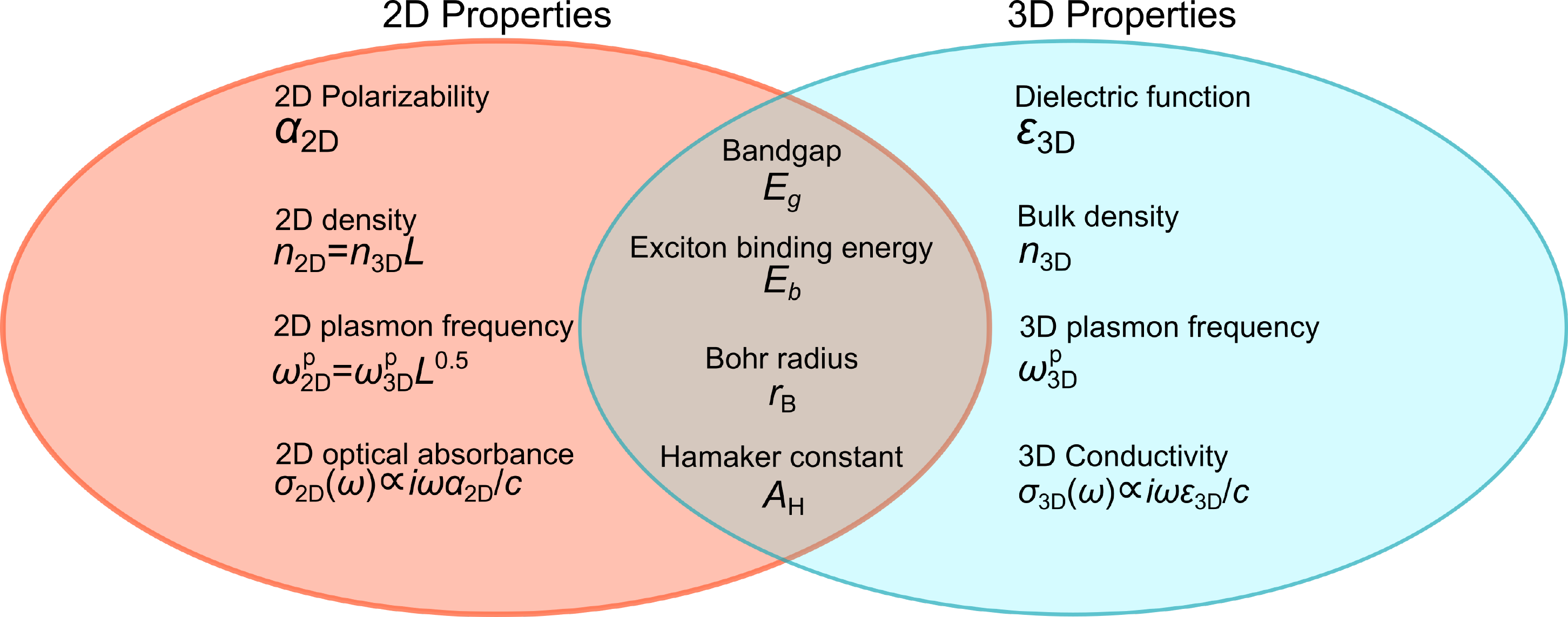}
\caption{\label{fig-2D-3D} Dielectric-related physical quantities in
  both 2D (red circle) and 3D (cyan circle) systems. The
  dimension-dependent quantities can be related with $\alpha_{\mathrm{2D}}$ and
  $\varepsilon$, respectively. The intersection between the circles
  present the quantities are well-defined in both dimensions, but may have a
  different scaling relation with others quantities.}
\end{figure}

As schematically shown in Figure \ref{fig-2D-3D}, the physical
quantities related to the dielectric properties can be categorized
into (i) strictly 2D (microscopic), (ii) strictly 3D (macroscopic) and
(iii) valid both 2D and 3D.  $\alpha_{\mathrm{2D}}$ and $\varepsilon$ are the
starting point for the strictly 2D and 3D quantities, which require
distinct definitions when dimensionality changes. Such quantities
include (but not limited to):
\begin{enumerate}
\item The densities $n_{\mathrm{2D}}$ and $n_{\mathrm{3D}}$ for
  charge, polarization, electronic states, etc.
  
\item The plasma frequencies $\omega^{\mathrm{p}}_{\mathrm{2D}}$ and
  $\omega^{\mathrm{p}}_{\mathrm{2D}}$\cite{Nazarov_2015_2D_3D} (see
  Section \ref{ssec:omega-p} and Figure \ref{fig:plasma}).

\item The optical conductivity $\sigma_{\mathrm{2D}}$ and
  $\sigma_{\mathrm{3D}}$\cite{Bechstedt_2012,Matthes_2016}.
\end{enumerate}
These quantities have distinct units in both dimensions, and related
by $L$ (for density and optical conductivity) or $\sqrt{L}$ (for
plasma frequency), which requires prudent interpretation of
theoretical and experimental results. For instance, the experimentally
observed ``dielectric constant'' of monolayer 2D materials
\cite{Ning_2015,Li_2014,Yao_2014,Wu_2015} would be questionable
without considering the effect of mixed medium. Instead, the 2D slab
polarizability, either transformed from the vacuum-containing
macroscopic dielectric constant, or predicted from the bandgap and
geometry as proposed here, will be a better descriptor for the true 2D
dielectric nature. There are also
dimension-independent quantities that are valid for both 2D and 3D
systems, for instance the bandgap $E_{\mathrm{g}}$, exciton binding
energy $E_{\mathrm{b}}$, Bohr radius $r_{\mathrm{B}}$ of the exciton
as well as the Hamaker constant of van der Waals interaction
$A_{\mathrm{H}}$. All these quantities are well-defined and can be
measured in both dimensions, while their relation with the dielectric
property varies with dimensionality. The well-known examples are the
different Wannier-Mott laws for exciton binding energy
\cite{Olsen_2016_hydrogen}, the dielectric-bandgap relation
proposed here, and the distinct power laws for van der Waals
interactions \cite{Gobre_2013}. To get a accuracy description of
dielectric-related properties of the 2D materials and their
heterostructures, one has to distinguish between the 2D and 3D
properties, and choose a suitable relation with the
dimension-dependent and dimension-independent quantities.

\section{More discussions about the dielectric anisotropy}
\label{sec:aniso}

\subsection{Choice of materials for main text Figure \ref{main-fig:aniso}}
\label{ssec:aniso-materials}
The dielectric anisotropy $\eta$ proposed in the main text is also
applied to other dimensions. Similar to the case of 2D and bulk
layered materials, $\epsilon$ is used to compare the anisotropy when
the material is periodic in all dimensions (bulk covalent materials),
while the polarizability $\alpha$ is used for reduced dimensional
materials.

The following types materials are chosen for comparison:
\begin{itemize}
\item Bulk covalent materials.  \\
  The list of materials and bandgap are chosen according to
  Ref. \citenum{sze_appendix_2006}, including IV-IV, III-V, II-VI, and
  IV-VI semiconductors. All these materials have isotropic dielectric
  properties.
\item Planar OSc \\
  The planar OScs include metal phthalocyanines, disk-like polycyclic
  aromatic hydrocarbon (PAHs), and benzene derivatives. The
  dimensionality of these materials are close to 2D materials due to
  their planar shape. The bandgap values (mostly at B3LYP density
  functional with 6-31G** basis sets) are extracted from the NIST
  Computational Chemistry Comparison and Benchmark Database
  (http://cccdb.nist.gov, Release 19, April 2018) and the
  polarizability values are obtained from
  Refs. \citenum{Miller_1990,Ramprasad_2006}.
\item Carbon nanotubes (CNT) \\
  Like 2D materials, CNTs are periodic along the 1D directions, and
  should be treated in a similar way to get the polarizability
  proportional to [Length]$^{2}$ \cite{Benedict_1995}. Semiconducting
  zigzag and armchair CNTs are considered, with their electronic
  properties obtained from Refs
  \citenum{Benedict_1995,Matsuda_2010}. and dielectric properties
  obtained from
  Refs. \citenum{Benedict_1995,Jensen_2000,Brothers_2008}.
\item Linear OSc \\
  We choose the linear polyacenes (linear PAHs from benzene to
  nonacene) and zigzag polyacetylene (1--9 repeating units) as model
  systems of linear OScs. The bandgaps are obtained from
  \citenum{Salzner_1998} and the polarizabilities are obtained from
  \citenum{Hinchliffe_2005}.
\item Fullerenes \\
  The bandgap of fullerenes (C$_{n}$ where $n=60*m^{2}$ where
  $m=1\sim{}7$) are taken from \citenum{Lin_1994} and the
  polarizabilities are taken from \citenum{Martin_2008}. All these
  materials have isotropic polarizability due to the high symmetry.
\end{itemize}

\subsection{Explanation for the separation between 2D and 3D regimes in main text Figure \ref{main-fig:aniso}}
\label{sssec:aniso-2}
In this section we give an analytical explanation for the separation
between the dielectric anisotropy indices of 2D and their bulk
counterparts. From main text Eqs. \ref{main-eq:alpha-para-def},
\ref{main-eq:alpha-perp-def} and \ref{main-eq:anisotropy},
$\eta_{\mathrm{Bulk}}$ of a bulk layered material is expressed as:
\begin{equation}
  \label{eq:eta-bulk}
  \begin{aligned}[t]
    \eta_{\mathrm{Bulk}} &= \frac{\varepsilon_{\mathrm{Bulk}}^{\perp}}
    {\varepsilon_{\mathrm{Bulk}}^{\parallel}}\\
    &= \frac{1}{\left(1 + {\displaystyle
          \frac{\alpha_{\mathrm{Bulk}}^{\parallel}}{\varepsilon_{0}L_{\mathrm{Bulk}}}}\right)
      \left(1 - {\displaystyle \frac{\alpha_{\mathrm{Bulk}}^{\perp}}{\varepsilon_{0}L_{\mathrm{Bulk}}}} \right)}\\
    &= \frac{1}{ \left[1-\left( {\displaystyle
            \frac{\alpha_{\mathrm{Bulk}}^{\perp}}{\alpha_{\mathrm{Bulk}}^{\parallel}}}\right)
        \left(
          {\displaystyle \frac{\alpha_{\mathrm{Bulk}}^{\parallel}}{\varepsilon_{0}
            L_{\mathrm{Bulk}}}} \right) \right]
      \left[ 1 + \left(
          {\displaystyle \frac{\alpha_{\mathrm{Bulk}}^{\parallel}}{\varepsilon_{0}
            L_{\mathrm{Bulk}}}} \right)\right]
    }
  \end{aligned}
\end{equation}
Name take the fact that
$\alpha_{\mathrm{Bulk}}^{\parallel} \approx
\alpha^{\parallel}_{\mathrm{2D}}$ (main text Figure
\ref{main-fig-4}b), we name
$\alpha_{\mathrm{Bulk}}^{\parallel} /\epsilon_{0}L_{\mathrm{Bulk}}
\approx \alpha_{\mathrm{2D}}^{\parallel}/
\epsilon_{0}L_{\mathrm{Bulk}}$ as $\gamma$. Furthermore we name
$\alpha_{\mathrm{Bulk}}^{\perp}/\alpha_{\mathrm{Bulk}}^{\parallel}$ as
$\hat{\eta_{\mathrm{2D}}}$, that
$\hat{\eta}_{\mathrm{2D}} =
\alpha_{\mathrm{Bulk}}^{\perp}/\alpha_{\mathrm{Bulk}}^{\parallel} \geq
\alpha_{\mathrm{2D}}^{\perp}/\alpha_{\mathrm{2D}}^{\parallel}=\eta_{\mathrm{2D}}$,
and Eq. \ref{eq:eta-bulk} is reduced to:
\begin{equation}
  \label{eq:eta-bulk-2}
  \eta_{\mathrm{Bulk}} = \frac{1}{(1 - \hat{\eta}_{\mathrm{2D}} \gamma) (1 + \gamma)}
\end{equation}
The minimal value for $\eta_{\mathrm{Bulk}}$ when $\gamma>0$ is obtained by solving:
\begin{equation}
  \label{eq:eta-bulk-min}
  \frac{\partial \eta_{\mathrm{Bulk}}}{\partial \gamma}
  = \frac{2 \hat{\eta}_{\mathrm{2D}} \gamma + \hat{\eta}_{\mathrm{2D}} - 1}
  {(\gamma+1)^{2}(1 - \hat{\eta}_{\mathrm{2D}} \gamma)^{2}} = 0
\end{equation}
which gives that
$\eta_{\mathrm{Bulk}} \geq \frac{4
  \hat{\eta}_{\mathrm{2D}}}{(\hat{\eta}_{\mathrm{2D}} + 1)^{2}}$,
where the minimal value is taken at
$\gamma = \frac{1}{2}(\frac{1}{\hat{\eta}_{\mathrm{2D}}} - 1)$. Since
$\frac{4 \hat{\eta}_{\mathrm{2D}}}{(\hat{\eta}_{\mathrm{2D}} +
  1)^{2}}$ monotonically increases when
$0 < \hat{\eta}_{\mathrm{2D}} < 1$, we get the comparison between the
dielectric anisotropy indices between 2D materials and their bulk
counterparts:
\begin{equation}
  \label{eq:aniso-final}
\eta_{\mathrm{Bulk}} \geq \frac{4
  \eta_{\mathrm{2D}}}{(\eta_{\mathrm{2D}} + 1)^{2}} \geq
\eta_{\mathrm{2D}} 
\end{equation}
Since $\gamma$ is actually $2r_{0}^{\parallel}/L_{\mathrm{Bulk}}$, the
ratio between the in-plane screening length $r_{0}^{\parallel}$ and
inter-plane distance $L_{\mathrm{Bulk}}$, and in general
$r_{0}^{\parallel} \gg L_{\mathrm{Bulk}}$, we can conclude that the
case when $\eta_{\mathrm{Bulk}} = \eta_{\mathrm{2D}}$ only happens
when $\eta_{\mathrm{2D}}$ is much smaller than 1. Therefore the
separation between the 2D and 3D regimes in main text Figure
\ref{main-fig:aniso} is explained.


\section{Raw data from first principles calculations}
\label{sec:raw}
\subsection{Quantities from first principles calculation}

Table S2 shows the parameters and results from the
first principles calculations for the 2D materials studied. The 2D
screening lengthes $r_{0}^{\parallel}$ and $r_{0}^{\perp}$ can be
obtained by multiplying $2 \pi$ to the columns
$\alpha_{\mathrm{2D}}^{\parallel}/(4\pi \varepsilon_{0})$ and
$\alpha_{\mathrm{2D}}^{\perp}/(4\pi \varepsilon_{0})$, respectively.

\input{raw_data}

\clearpage{}
\section*{}
\label{sec:ref}
\bibliography{ref}

%% file: raw_data.tex

\begin{center}
  \footnotesize
  \setlongtables
  \begin{tabularx}{1.1\linewidth}{lXXXXXXXXX}
      \caption{Raw data of the materials calculated in this study.}\\
    \hline
    Material & L (\AA) & HSE06 $E_{\mathrm{g}}^{\mathrm{min}}$ (eV) & PBE $E_{\mathrm{g}}^{\mathrm{min}}$ (eV) & PBE $E_{\mathrm{g}}^{\mathrm{direct}}$ (eV) & $\varepsilon_{\mathrm{SL}}^{\mathrm{xx}}$ & $\varepsilon_{\mathrm{SL}}^{\mathrm{yy}}$ & $\varepsilon_{\mathrm{SL}}^{\mathrm{zz}}$ & $\alpha_{\mathrm{2D}}^{\parallel}/(4\pi \varepsilon_{0})$ (\AA) & $\alpha_{\mathrm{2D}}^{\perp}/(4\pi \varepsilon_{0})$ (\AA)\\
    \hline
    \endhead
    1T-TiO$_{2}$ & 26.668  & 4.010  & 3.096  & 2.467  & 1.887  & 1.887  & 1.123  & 1.882  & 0.232 \\
    2H-TiO$_{2}$ & 27.648  & 2.520  & 1.808  & 1.103  & 1.852  & 1.852  & 1.133  & 1.875  & 0.258 \\
    1T-TiSe$_{2}$ & 33.049  & 1.360  & 1.372  & 0.505  & 3.029  & 3.029  & 1.190  & 5.336  & 0.420 \\
    1T-ZrO$_{2}$ & 26.561  & 6.320  & 5.039  & 4.431  & 1.569  & 1.569  & 1.117  & 1.203  & 0.221 \\
    1T-ZrS$_{2}$ & 32.622  & 2.010  & 1.643  & 1.180  & 2.329  & 2.329  & 1.159  & 3.450  & 0.356 \\
    1T-ZrSe$_{2}$ & 34.056  & 0.890  & 0.961  & 0.371  & 2.794  & 2.794  & 1.172  & 4.862  & 0.398 \\
    2H-ZrO$_{2}$ & 28.188  & 3.130  & 2.264  & 1.690  & 1.619  & 1.619  & 1.121  & 1.389  & 0.242 \\
    2H-ZrSe$_{2}$ & 33.692  & 1.500  & 1.382  & 0.738  & 2.448  & 2.448  & 1.190  & 3.882  & 0.428 \\
    2H-ZrTe$_{2}$ & 35.904  & 0.900  & 1.216  & 0.284  & 3.171  & 3.171  & 1.207  & 6.203  & 0.490 \\
    1T-HfO$_{2}$ & 26.636  & 6.580  & 5.471  & 4.830  & 1.521  & 1.521  & 1.117  & 1.104  & 0.222 \\
    1T-HfS$_{2}$ & 32.558  & 2.010  & 1.949  & 1.224  & 2.250  & 2.250  & 1.204  & 3.239  & 0.439 \\
    1T-HfSe$_{2}$ & 33.916  & 1.070  & 1.215  & 0.435  & 2.702  & 2.702  & 1.180  & 4.594  & 0.412 \\
    2H-HfO$_{2}$ & 28.167  & 3.400  & 2.552  & 1.948  & 1.555  & 1.555  & 1.124  & 1.244  & 0.247 \\
    2H-HfS$_{2}$ & 32.678  & 1.890  & 1.831  & 1.068  & 2.087  & 2.087  & 1.177  & 2.827  & 0.391 \\
    2H-HfSe$_{2}$ & 33.419  & 1.530  & 1.754  & 0.819  & 2.390  & 2.390  & 1.191  & 3.697  & 0.426 \\
    2H-HfTe$_{2}$ & 35.629  & 0.700  & 1.251  & 0.121  & 3.072  & 3.072  & 1.208  & 5.875  & 0.488 \\
    1T-GeO$_{2}$ & 26.526  & 5.740  & 6.118  & 3.466  & 1.453  & 1.453  & 1.115  & 0.956  & 0.218 \\
    1T-GeS$_{2}$ & 31.883  & 1.580  & 2.697  & 0.726  & 2.302  & 2.302  & 1.169  & 3.303  & 0.367 \\
    1T-GeO$_{2}$ & 27.908  & 2.990  & 4.643  & 1.335  & 1.570  & 1.570  & 1.127  & 1.266  & 0.250 \\
    1T-SnO$_{2}$ & 27.147  & 4.570  & 5.840  & 2.649  & 1.449  & 1.449  & 1.114  & 0.970  & 0.221 \\
    1T-SnS$_{2}$ & 32.793  & 2.530  & 2.859  & 1.574  & 2.059  & 2.059  & 1.166  & 2.764  & 0.372 \\
    1T-SnSe$_{2}$ & 34.077  & 1.490  & 1.466  & 0.751  & 2.437  & 2.437  & 1.169  & 3.897  & 0.392 \\
    2H-SnO$_{2}$ & 28.938  & 1.960  & 4.661  & 0.647  & 1.590  & 1.590  & 1.124  & 1.359  & 0.254 \\
    2H-SnS$_{2}$ & 32.873  & 1.590  & 1.072  & 0.750  & 2.164  & 2.164  & 1.180  & 3.045  & 0.399 \\
    1T-PbO$_{2}$ & 27.862  & 2.600  & 3.578  & 1.330  & 1.709  & 1.709  & 1.121  & 1.572  & 0.239 \\
    BN & 29.995  & 5.640  & 5.688  & 5.592  & 1.366  & 1.366  & 1.072  & 0.874  & 0.160 \\
    C$_{2}$F$_{2}$ & 31.998  & 5.000  & 3.173  & 3.173  & 1.318  & 1.348  & 1.123  & 0.810  & 0.279 \\
    P$_{4}$ & 27.097  & 1.600  & 0.888  & 0.895  & 2.894  & 3.115  & 1.196  & 4.084  & 0.353 \\
    C$_{2}$H$_{2}$ & 31.015  & 4.360  & 3.468  & 3.468  & 1.288  & 1.288  & 1.094  & 0.711  & 0.212\\ 
    1T-NiO$_{2}$ & 26.112  & 3.170  & 1.828  & 1.198  & 2.763  & 2.763  & 1.129  & 3.663  & 0.237\\ 
    1T-PdO$_{2}$ & 26.712  & 3.210  & 2.475  & 1.397  & 2.368  & 2.368  & 1.116  & 2.908  & 0.221 \\
    1T-PdS$_{2}$ & 30.361  & 1.800  & 2.487  & 1.178  & 3.888  & 3.888  & 1.169  & 6.978  & 0.349 \\
    1T-PtO$_{2}$ & 26.316  & 3.540  & 2.602  & 1.691  & 2.114  & 2.114  & 1.116  & 2.333  & 0.218 \\
    1T-PtS$_{2}$ & 30.239  & 2.700  & 2.022  & 1.714  & 3.086  & 3.086  & 1.163  & 5.020  & 0.337 \\
    1T-PdSe$_{2}$ & 31.080  & 0.970  & 1.917  & 0.534  & 4.958  & 4.958  & 1.178  & 9.789  & 0.374 \\
    1T-NiS$_{2}$ & 29.616  & 0.980  & 1.797  & 0.523  & 4.691  & 4.691  & 1.173  & 8.699  & 0.348 \\
    1T-PtSe$_{2}$ & 31.058  & 1.210  & 2.710  & 1.180  & 3.643  & 3.643  & 1.175  & 6.532  & 0.368 \\
    Ga$_{2}$Se$_{2}$ & 30.000  & 2.810  & 2.657  & 1.764  & 2.640  & 2.640  & 1.281  & 3.915  & 0.524 \\
    Ga$_{2}$S$_{2}$ & 30.000  & 3.250  & 3.351  & 2.358  & 2.329  & 2.329  & 1.256  & 3.173  & 0.487 \\
    CdCl$_{2}$ & 31.085 & 4.480 & 3.172 & 3.172 & 1.48 & 1.48 & 1.157 & 1.187 & 0.336 \\
    CdI$_{2}$ & 35.281  & 3.150  & 1.706  & 1.528  & 1.804  & 1.804  & 1.192  & 2.257  & 0.452 \\
    2H-MoS$_{2}$ & 32.296  & 2.240  & 1.594  & 1.594  & 3.475  & 3.475  & 1.183  & 6.361  & 0.398 \\
    2H-MoSe$_{2}$ & 40.854 & 1.752 & 1.449 &   1.449 & 3.231 &  3.231 & 1.154 & 7.253 & 0.433 \\
    2H-WS$_{2}$ & 32.271  & 2.280  & 1.540  & 1.540  & 3.214  & 3.214  & 1.180  & 5.686  & 0.392 \\
    2H-WSe$_{2}$ & 32.965  & 1.930  & 1.253  & 1.253  & 3.485  & 3.485  & 1.197  & 6.519  & 0.432 \\
    2H-WO$_{2}$ & 29.183  & 2.000  & 1.693  & 1.359  & 2.519  & 2.519  & 1.123  & 3.528  & 0.254 \\
    2H-MoO$_{2}$ & 29.231  & 1.560  & 1.648  & 0.952  & 2.918  & 2.918  & 1.129  & 4.462  & 0.266 \\
    2H-MoTe$_{2}$ & 34.061  & 1.440  & 0.946  & 0.946  & 4.412  & 4.412  & 1.220  & 9.248  & 0.489 \\
    2H-WTe$_{2}$ & 33.883  & 1.300  & 0.731  & 0.731  & 4.158  & 4.158  & 1.230  & 8.515  & 0.504 \\
    2H-CrS$_{2}$ & 31.759  & 1.400  & 0.902  & 0.902  & 4.647  & 4.647  & 1.183  & 9.217  & 0.391 \\
    2H-CrSe$_{2}$ & 32.446  & 1.150  & 0.704  & 0.704  & 5.364  & 5.364  & 1.201  & 11.268  & 0.432 \\
    2H-CrO$_{2}$ & 28.027  & 0.990  & 1.596  & 0.424  & 3.961  & 3.961  & 1.134  & 6.604  & 0.264 \\
    2H-TiS$_{2}$ & 32.199  & 1.610  & 1.284  & 0.692  & 2.634  & 2.634  & 1.184  & 4.187  & 0.398 \\
    1T-PtTe$_{2}$ & 32.005  & 0.490  & 1.809  & 0.366  & 4.726  & 4.726  & 1.200  & 9.490  & 0.424 \\
    MAPbBr$_{3}$ & 23.018 & 3.163 & 2.444 & 2.444 & 1.608 & 1.778 & 1.527 & 1.265 & 0.632 \\
    \hline
  \end{tabularx}

\end{center}

%% file: main.bbl
\providecommand{\latin}[1]{#1}
\makeatletter
\providecommand{\doi}
  {\begingroup\let\do\@makeother\dospecials
  \catcode`\{=1 \catcode`\}=2 \doi@aux}
\providecommand{\doi@aux}[1]{\endgroup\texttt{#1}}
\makeatother
\providecommand*\mcitethebibliography{\thebibliography}
\csname @ifundefined\endcsname{endmcitethebibliography}
  {\let\endmcitethebibliography\endthebibliography}{}
\begin{mcitethebibliography}{77}
\providecommand*\natexlab[1]{#1}
\providecommand*\mciteSetBstSublistMode[1]{}
\providecommand*\mciteSetBstMaxWidthForm[2]{}
\providecommand*\mciteBstWouldAddEndPuncttrue
  {\def\EndOfBibitem{\unskip.}}
\providecommand*\mciteBstWouldAddEndPunctfalse
  {\let\EndOfBibitem\relax}
\providecommand*\mciteSetBstMidEndSepPunct[3]{}
\providecommand*\mciteSetBstSublistLabelBeginEnd[3]{}
\providecommand*\EndOfBibitem{}
\mciteSetBstSublistMode{f}
\mciteSetBstMaxWidthForm{subitem}{(\alph{mcitesubitemcount})}
\mciteSetBstSublistLabelBeginEnd
  {\mcitemaxwidthsubitemform\space}
  {\relax}
  {\relax}

\bibitem[Moss(1950)]{Moss_1950_relation}
Moss,~T.~S. A {Relationship} between the {Refractive} {Index} and the
  {Infra}-{Red} {Threshold} of {Sensitivity} for {Photoconductors}. \emph{Proc.
  Phys. Soc. B} \textbf{1950}, \emph{63}, 167\relax
\mciteBstWouldAddEndPuncttrue
\mciteSetBstMidEndSepPunct{\mcitedefaultmidpunct}
{\mcitedefaultendpunct}{\mcitedefaultseppunct}\relax
\EndOfBibitem
\bibitem[Moss(1985)]{Moss_1985_n_Eg}
Moss,~T.~S. Relations between the Refractive Index and Energy Gap of
  Semiconductors. \emph{Phys. Status Solidi B} \textbf{1985}, \emph{131},
  415--427\relax
\mciteBstWouldAddEndPuncttrue
\mciteSetBstMidEndSepPunct{\mcitedefaultmidpunct}
{\mcitedefaultendpunct}{\mcitedefaultseppunct}\relax
\EndOfBibitem
\bibitem[Kittel(2005)]{kittel_2005_introduction}
Kittel,~C. \emph{Introduction to solid state physics}; Wiley, 2005\relax
\mciteBstWouldAddEndPuncttrue
\mciteSetBstMidEndSepPunct{\mcitedefaultmidpunct}
{\mcitedefaultendpunct}{\mcitedefaultseppunct}\relax
\EndOfBibitem
\bibitem[Dressel and Gruner(2001)Dressel, and
  Gruner]{Dressel_2001_electrodynamics}
Dressel,~M.; Gruner,~G. \emph{Electrodynamics of solids: optical properties of
  electrons in matter}; Cambridge University Press, 2001\relax
\mciteBstWouldAddEndPuncttrue
\mciteSetBstMidEndSepPunct{\mcitedefaultmidpunct}
{\mcitedefaultendpunct}{\mcitedefaultseppunct}\relax
\EndOfBibitem
\bibitem[Adler(1962)]{Adler_1962}
Adler,~S.~L. Quantum Theory of the Dielectric Constant in Real Solids.
  \emph{Phys. Rev.} \textbf{1962}, \emph{126}, 413--420\relax
\mciteBstWouldAddEndPuncttrue
\mciteSetBstMidEndSepPunct{\mcitedefaultmidpunct}
{\mcitedefaultendpunct}{\mcitedefaultseppunct}\relax
\EndOfBibitem
\bibitem[Hybertsen and Louie(1987)Hybertsen, and Louie]{Hybertsen_1987}
Hybertsen,~M.~S.; Louie,~S.~G. Ab initiostatic dielectric matrices from the
  density-functional approach. I. Formulation and application to semiconductors
  and insulators. \emph{Phys. Rev. B} \textbf{1987}, \emph{35},
  5585--5601\relax
\mciteBstWouldAddEndPuncttrue
\mciteSetBstMidEndSepPunct{\mcitedefaultmidpunct}
{\mcitedefaultendpunct}{\mcitedefaultseppunct}\relax
\EndOfBibitem
\bibitem[Palik(1998)]{palik_1998handbook}
Palik,~E.~D., Ed. \emph{Handbook of optical constants of solids}; Academic
  Press, 1998\relax
\mciteBstWouldAddEndPuncttrue
\mciteSetBstMidEndSepPunct{\mcitedefaultmidpunct}
{\mcitedefaultendpunct}{\mcitedefaultseppunct}\relax
\EndOfBibitem
\bibitem[Novoselov \latin{et~al.}(2016)Novoselov, Mishchenko, Carvalho, and
  Castro~Neto]{Novoselov_2016}
Novoselov,~K.~S.; Mishchenko,~A.; Carvalho,~A.; Castro~Neto,~A.~H. 2D materials
  and van der Waals heterostructures. \emph{Science} \textbf{2016}, \emph{353},
  aac9439\relax
\mciteBstWouldAddEndPuncttrue
\mciteSetBstMidEndSepPunct{\mcitedefaultmidpunct}
{\mcitedefaultendpunct}{\mcitedefaultseppunct}\relax
\EndOfBibitem
\bibitem[Keldysh(1979)]{Keldysh_1979_eps_multi}
Keldysh,~L.~V. Coulomb interaction in thin semiconductor and semimetal films.
  \emph{Sov. Phys. JETP} \textbf{1979}, \emph{29}, 658\relax
\mciteBstWouldAddEndPuncttrue
\mciteSetBstMidEndSepPunct{\mcitedefaultmidpunct}
{\mcitedefaultendpunct}{\mcitedefaultseppunct}\relax
\EndOfBibitem
\bibitem[Sharma(1985)]{Sharma_1985}
Sharma,~A. Dielectric function of a semiconductor slab. \emph{J. Phys. C: Solid
  State Phys.} \textbf{1985}, \emph{18}, L153--L156\relax
\mciteBstWouldAddEndPuncttrue
\mciteSetBstMidEndSepPunct{\mcitedefaultmidpunct}
{\mcitedefaultendpunct}{\mcitedefaultseppunct}\relax
\EndOfBibitem
\bibitem[Low \latin{et~al.}(2014)Low, Rold{\'a}n, Wang, Xia, Avouris, Moreno,
  and Guinea]{Low_2014_screening_BP}
Low,~T.; Rold{\'a}n,~R.; Wang,~H.; Xia,~F.; Avouris,~P.; Moreno,~L.~M.;
  Guinea,~F. Plasmons and {Screening} in {Monolayer} and {Multilayer} {Black}
  {Phosphorus}. \emph{Phys. Rev. Lett.} \textbf{2014}, \emph{113}, 106802\relax
\mciteBstWouldAddEndPuncttrue
\mciteSetBstMidEndSepPunct{\mcitedefaultmidpunct}
{\mcitedefaultendpunct}{\mcitedefaultseppunct}\relax
\EndOfBibitem
\bibitem[Cudazzo \latin{et~al.}(2011)Cudazzo, Tokatly, and
  Rubio]{Cudazzo_2011_screening_2D}
Cudazzo,~P.; Tokatly,~I.~V.; Rubio,~A. Dielectric screening in two-dimensional
  insulators: Implications for excitonic and impurity states in graphane.
  \emph{Phys. Rev. B} \textbf{2011}, \emph{84}, 085406\relax
\mciteBstWouldAddEndPuncttrue
\mciteSetBstMidEndSepPunct{\mcitedefaultmidpunct}
{\mcitedefaultendpunct}{\mcitedefaultseppunct}\relax
\EndOfBibitem
\bibitem[Bechstedt \latin{et~al.}(2012)Bechstedt, Matthes, Gori, and
  Pulci]{Bechstedt_2012}
Bechstedt,~F.; Matthes,~L.; Gori,~P.; Pulci,~O. Infrared absorbance of silicene
  and germanene. \emph{Appl. Phys. Lett.} \textbf{2012}, \emph{100},
  261906\relax
\mciteBstWouldAddEndPuncttrue
\mciteSetBstMidEndSepPunct{\mcitedefaultmidpunct}
{\mcitedefaultendpunct}{\mcitedefaultseppunct}\relax
\EndOfBibitem
\bibitem[Cudazzo \latin{et~al.}(2010)Cudazzo, Attaccalite, Tokatly, and
  Rubio]{Cudazzo_2010_screen2D}
Cudazzo,~P.; Attaccalite,~C.; Tokatly,~I.~V.; Rubio,~A. Strong Charge-Transfer
  Excitonic Effects and the Bose-Einstein Exciton Condensate in Graphane.
  \emph{Phys. Rev. Lett.} \textbf{2010}, \emph{104}, 226804\relax
\mciteBstWouldAddEndPuncttrue
\mciteSetBstMidEndSepPunct{\mcitedefaultmidpunct}
{\mcitedefaultendpunct}{\mcitedefaultseppunct}\relax
\EndOfBibitem
\bibitem[Nazarov(2015)]{Nazarov_2015_2D_3D}
Nazarov,~V.~U. Electronic excitations in quasi-2D crystals: what theoretical
  quantities are relevant to experiment? \emph{New J. Phys.} \textbf{2015},
  \emph{17}, 073018\relax
\mciteBstWouldAddEndPuncttrue
\mciteSetBstMidEndSepPunct{\mcitedefaultmidpunct}
{\mcitedefaultendpunct}{\mcitedefaultseppunct}\relax
\EndOfBibitem
\bibitem[Li \latin{et~al.}(2016)Li, Tsukagoshi, Orgiu, and
  Samor{\`\i}]{Li_2016}
Li,~S.-L.; Tsukagoshi,~K.; Orgiu,~E.; Samor{\`\i},~P. Charge transport and
  mobility engineering in two-dimensional transition metal chalcogenide
  semiconductors. \emph{Chem. Soc. Rev.} \textbf{2016}, \emph{45},
  118--151\relax
\mciteBstWouldAddEndPuncttrue
\mciteSetBstMidEndSepPunct{\mcitedefaultmidpunct}
{\mcitedefaultendpunct}{\mcitedefaultseppunct}\relax
\EndOfBibitem
\bibitem[Sohier \latin{et~al.}(2017)Sohier, Gibertini, Calandra, Mauri, and
  Marzari]{Sohier_2017}
Sohier,~T.; Gibertini,~M.; Calandra,~M.; Mauri,~F.; Marzari,~N. Breakdown of
  Optical Phonons' Splitting in Two-Dimensional Materials. \emph{Nano Lett.}
  \textbf{2017}, \emph{17}, 3758--3763\relax
\mciteBstWouldAddEndPuncttrue
\mciteSetBstMidEndSepPunct{\mcitedefaultmidpunct}
{\mcitedefaultendpunct}{\mcitedefaultseppunct}\relax
\EndOfBibitem
\bibitem[Laturia and Vandenberghe(2017)Laturia, and
  Vandenberghe]{relax-epsilon}
Laturia,~A.; Vandenberghe,~W.~G. Dielectric properties of mono- and bilayers
  determined from first principles. 2017 International Conference on Simulation
  of Semiconductor Processes and Devices (SISPAD). 2017; pp 337--340\relax
\mciteBstWouldAddEndPuncttrue
\mciteSetBstMidEndSepPunct{\mcitedefaultmidpunct}
{\mcitedefaultendpunct}{\mcitedefaultseppunct}\relax
\EndOfBibitem
\bibitem[Sohier \latin{et~al.}(2016)Sohier, Calandra, and Mauri]{Sohier_2016}
Sohier,~T.; Calandra,~M.; Mauri,~F. Two-dimensional Fr{\"o}hlich interaction in
  transition-metal dichalcogenide monolayers: Theoretical modeling and
  first-principles calculations. \emph{Phys. Rev. B} \textbf{2016}, \emph{94},
  085415\relax
\mciteBstWouldAddEndPuncttrue
\mciteSetBstMidEndSepPunct{\mcitedefaultmidpunct}
{\mcitedefaultendpunct}{\mcitedefaultseppunct}\relax
\EndOfBibitem
\bibitem[Heyd \latin{et~al.}(2003)Heyd, Scuseria, and Ernzerhof]{Heyd_2003}
Heyd,~J.; Scuseria,~G.~E.; Ernzerhof,~M. Hybrid functionals based on a screened
  Coulomb potential. \emph{J. Chem. Phys.} \textbf{2003}, \emph{118},
  8207--8215\relax
\mciteBstWouldAddEndPuncttrue
\mciteSetBstMidEndSepPunct{\mcitedefaultmidpunct}
{\mcitedefaultendpunct}{\mcitedefaultseppunct}\relax
\EndOfBibitem
\bibitem[Heyd \latin{et~al.}(2006)Heyd, Scuseria, and Ernzerhof]{HSE_2006}
Heyd,~J.; Scuseria,~G.~E.; Ernzerhof,~M. Erratum: ``Hybrid functionals based on
  a screened Coulomb potential'' [J. Chem. Phys. 118, 8207 (2003)]. \emph{J.
  Chem. Phys.} \textbf{2006}, \emph{124}, 219906\relax
\mciteBstWouldAddEndPuncttrue
\mciteSetBstMidEndSepPunct{\mcitedefaultmidpunct}
{\mcitedefaultendpunct}{\mcitedefaultseppunct}\relax
\EndOfBibitem
\bibitem[Perdew \latin{et~al.}(1996)Perdew, Burke, and Ernzerhof]{Perdew_1996}
Perdew,~J.~P.; Burke,~K.; Ernzerhof,~M. Generalized Gradient Approximation Made
  Simple. \emph{Phys. Rev. Lett.} \textbf{1996}, \emph{77}, 3865--3868\relax
\mciteBstWouldAddEndPuncttrue
\mciteSetBstMidEndSepPunct{\mcitedefaultmidpunct}
{\mcitedefaultendpunct}{\mcitedefaultseppunct}\relax
\EndOfBibitem
\bibitem[Ernzerhof and Scuseria(1999)Ernzerhof, and Scuseria]{Ernzerhof99}
Ernzerhof,~M.; Scuseria,~G.~E. Assessment of the Perdew--Burke--Ernzerhof
  exchange-correlation functional. \emph{J. Chem. Phys.} \textbf{1999},
  \emph{110}, 5029--5036\relax
\mciteBstWouldAddEndPuncttrue
\mciteSetBstMidEndSepPunct{\mcitedefaultmidpunct}
{\mcitedefaultendpunct}{\mcitedefaultseppunct}\relax
\EndOfBibitem
\bibitem[Paier \latin{et~al.}(2005)Paier, Hirschl, Marsman, and
  Kresse]{Paier_2005_PBE}
Paier,~J.; Hirschl,~R.; Marsman,~M.; Kresse,~G. The Perdew--Burke--Ernzerhof
  exchange-correlation functional applied to the G2-1 test set using a
  plane-wave basis set. \emph{J. Chem. Phys.} \textbf{2005}, \emph{122},
  234102\relax
\mciteBstWouldAddEndPuncttrue
\mciteSetBstMidEndSepPunct{\mcitedefaultmidpunct}
{\mcitedefaultendpunct}{\mcitedefaultseppunct}\relax
\EndOfBibitem
\bibitem[Hedin(1965)]{Hedin_1965}
Hedin,~L. New Method for Calculating the One-Particle Green's Function with
  Application to the Electron-Gas Problem. \emph{Phys. Rev.} \textbf{1965},
  \emph{139}, A796--A823\relax
\mciteBstWouldAddEndPuncttrue
\mciteSetBstMidEndSepPunct{\mcitedefaultmidpunct}
{\mcitedefaultendpunct}{\mcitedefaultseppunct}\relax
\EndOfBibitem
\bibitem[Onida \latin{et~al.}(2002)Onida, Reining, and Rubio]{Onida_2002}
Onida,~G.; Reining,~L.; Rubio,~A. Electronic excitations: density-functional
  versus many-body Green's-function approaches. \emph{Rev. Mod. Phys.}
  \textbf{2002}, \emph{74}, 601--659\relax
\mciteBstWouldAddEndPuncttrue
\mciteSetBstMidEndSepPunct{\mcitedefaultmidpunct}
{\mcitedefaultendpunct}{\mcitedefaultseppunct}\relax
\EndOfBibitem
\bibitem[Matthes \latin{et~al.}(2016)Matthes, Pulci, and
  Bechstedt]{Matthes_2016}
Matthes,~L.; Pulci,~O.; Bechstedt,~F. Influence of out-of-plane response on
  optical properties of two-dimensional materials: First principles approach.
  \emph{Phys. Rev. B} \textbf{2016}, \emph{94}, 205408\relax
\mciteBstWouldAddEndPuncttrue
\mciteSetBstMidEndSepPunct{\mcitedefaultmidpunct}
{\mcitedefaultendpunct}{\mcitedefaultseppunct}\relax
\EndOfBibitem
\bibitem[Israelachvili(2011)]{Israelachvili_2011}
Israelachvili,~J.~N. Interactions Involving the Polarization of Molecules.
  \emph{Intermolecular and Surface Forces} \textbf{2011}, 91--106\relax
\mciteBstWouldAddEndPuncttrue
\mciteSetBstMidEndSepPunct{\mcitedefaultmidpunct}
{\mcitedefaultendpunct}{\mcitedefaultseppunct}\relax
\EndOfBibitem
\bibitem[Olsen \latin{et~al.}(2016)Olsen, Latini, Rasmussen, and
  Thygesen]{Olsen_2016_hydrogen}
Olsen,~T.; Latini,~S.; Rasmussen,~F.; Thygesen,~K.~S. Simple Screened Hydrogen
  Model of Excitons in Two-Dimensional Materials. \emph{Phys. Rev. Lett.}
  \textbf{2016}, \emph{116}, 056401\relax
\mciteBstWouldAddEndPuncttrue
\mciteSetBstMidEndSepPunct{\mcitedefaultmidpunct}
{\mcitedefaultendpunct}{\mcitedefaultseppunct}\relax
\EndOfBibitem
\bibitem[Jiang \latin{et~al.}(2017)Jiang, Liu, Li, and Duan]{Jiang_2017_Eg_Eb}
Jiang,~Z.; Liu,~Z.; Li,~Y.; Duan,~W. Scaling Universality between Band Gap and
  Exciton Binding Energy of Two-Dimensional Semiconductors. \emph{Phys. Rev.
  Lett.} \textbf{2017}, \emph{118}\relax
\mciteBstWouldAddEndPuncttrue
\mciteSetBstMidEndSepPunct{\mcitedefaultmidpunct}
{\mcitedefaultendpunct}{\mcitedefaultseppunct}\relax
\EndOfBibitem
\bibitem[Tobik and Dal~Corso(2004)Tobik, and Dal~Corso]{T_bik_2004}
Tobik,~J.; Dal~Corso,~A. Electric fields with ultrasoft pseudo-potentials:
  Applications to benzene and anthracene. \emph{J. Chem. Phys.} \textbf{2004},
  \emph{120}, 9934--9941\relax
\mciteBstWouldAddEndPuncttrue
\mciteSetBstMidEndSepPunct{\mcitedefaultmidpunct}
{\mcitedefaultendpunct}{\mcitedefaultseppunct}\relax
\EndOfBibitem
\bibitem[Wiser(1963)]{Wiser_1963}
Wiser,~N. Dielectric Constant with Local Field Effects Included. \emph{Phys.
  Rev.} \textbf{1963}, \emph{129}, 62–--69\relax
\mciteBstWouldAddEndPuncttrue
\mciteSetBstMidEndSepPunct{\mcitedefaultmidpunct}
{\mcitedefaultendpunct}{\mcitedefaultseppunct}\relax
\EndOfBibitem
\bibitem[Benedict \latin{et~al.}(1995)Benedict, Louie, and
  Cohen]{Benedict_1995}
Benedict,~L.~X.; Louie,~S.~G.; Cohen,~M.~L. Static polarizabilities of
  single-wall carbon nanotubes. \emph{Phys. Rev. B} \textbf{1995}, \emph{52},
  8541--8549\relax
\mciteBstWouldAddEndPuncttrue
\mciteSetBstMidEndSepPunct{\mcitedefaultmidpunct}
{\mcitedefaultendpunct}{\mcitedefaultseppunct}\relax
\EndOfBibitem
\bibitem[Markel(2016)]{Markel_2016}
Markel,~V.~A. Introduction to the Maxwell Garnett approximation: tutorial.
  \emph{J. Opt. Soc. Am. A} \textbf{2016}, \emph{33}, 1244\relax
\mciteBstWouldAddEndPuncttrue
\mciteSetBstMidEndSepPunct{\mcitedefaultmidpunct}
{\mcitedefaultendpunct}{\mcitedefaultseppunct}\relax
\EndOfBibitem
\bibitem[Meyer and Vanderbilt(2001)Meyer, and
  Vanderbilt]{Meyer_2001_dipole_slab}
Meyer,~B.; Vanderbilt,~D. Ab initiostudy of BaTiO$_3$ and PbTiO$_3$ surfaces in
  external electric fields. \emph{Phys. Rev. B} \textbf{2001}, \emph{63},
  205426\relax
\mciteBstWouldAddEndPuncttrue
\mciteSetBstMidEndSepPunct{\mcitedefaultmidpunct}
{\mcitedefaultendpunct}{\mcitedefaultseppunct}\relax
\EndOfBibitem
\bibitem[Wurstbauer \latin{et~al.}(2010)Wurstbauer, R{\"o}ling, Wurstbauer,
  Wegscheider, Vaupel, Thiesen, and Weiss]{graphene-epsilon10}
Wurstbauer,~U.; R{\"o}ling,~C.; Wurstbauer,~U.; Wegscheider,~W.; Vaupel,~M.;
  Thiesen,~P.~H.; Weiss,~D. Imaging ellipsometry of graphene. \emph{Appl. Phys.
  Lett.} \textbf{2010}, \emph{97}, 231901\relax
\mciteBstWouldAddEndPuncttrue
\mciteSetBstMidEndSepPunct{\mcitedefaultmidpunct}
{\mcitedefaultendpunct}{\mcitedefaultseppunct}\relax
\EndOfBibitem
\bibitem[Yim \latin{et~al.}(2014)Yim, O'Brien, McEvoy, Winters, Mirza, Lunney,
  and Duesberg]{Duesberg14}
Yim,~C.; O'Brien,~M.; McEvoy,~N.; Winters,~S.; Mirza,~I.; Lunney,~J.~G.;
  Duesberg,~G.~S. Investigation of the optical properties of MoS$_{2}$ thin
  films using spectroscopic ellipsometry. \emph{Appl. Phys. Lett.}
  \textbf{2014}, \emph{104}, 103114\relax
\mciteBstWouldAddEndPuncttrue
\mciteSetBstMidEndSepPunct{\mcitedefaultmidpunct}
{\mcitedefaultendpunct}{\mcitedefaultseppunct}\relax
\EndOfBibitem
\bibitem[Shen \latin{et~al.}(2013)Shen, Hsu, Li, and Liu]{Chiang13}
Shen,~C.-C.; Hsu,~Y.-T.; Li,~L.-J.; Liu,~H.-L. Charge Dynamics and Electronic
  Structures of Monolayer MoS 2 Films Grown by Chemical Vapor Deposition.
  \emph{Appl. Phys. Express} \textbf{2013}, \emph{6}, 125801\relax
\mciteBstWouldAddEndPuncttrue
\mciteSetBstMidEndSepPunct{\mcitedefaultmidpunct}
{\mcitedefaultendpunct}{\mcitedefaultseppunct}\relax
\EndOfBibitem
\bibitem[Li \latin{et~al.}(2014)Li, Birdwell, Amani, Burke, Ling, Lee, Liang,
  Peng, Richter, Kong, Gundlach, and Nguyen]{Kong14}
Li,~W.; Birdwell,~A.~G.; Amani,~M.; Burke,~R.~A.; Ling,~X.; Lee,~Y.-H.;
  Liang,~X.; Peng,~L.; Richter,~C.~A.; Kong,~J.; Gundlach,~D.~J.; Nguyen,~N.~V.
  Broadband optical properties of large-area monolayer CVD molybdenum
  disulfide. \emph{Phys. Rev. B} \textbf{2014}, \emph{90}, 195434\relax
\mciteBstWouldAddEndPuncttrue
\mciteSetBstMidEndSepPunct{\mcitedefaultmidpunct}
{\mcitedefaultendpunct}{\mcitedefaultseppunct}\relax
\EndOfBibitem
\bibitem[Li \latin{et~al.}(2014)Li, Chernikov, Zhang, Rigosi, Hill, van~der
  Zande, Chenet, Shih, Hone, and Heinz]{Li_2014}
Li,~Y.; Chernikov,~A.; Zhang,~X.; Rigosi,~A.; Hill,~H.~M.; van~der
  Zande,~A.~M.; Chenet,~D.~A.; Shih,~E.-M.; Hone,~J.; Heinz,~T.~F. Measurement
  of the optical dielectric function of monolayer transition-metal
  dichalcogenides:MoS$_{2}$,MoSe$_2$,WS$_{2}$, andWSe$_2$. \emph{Phys. Rev. B}
  \textbf{2014}, \emph{90}, 205422\relax
\mciteBstWouldAddEndPuncttrue
\mciteSetBstMidEndSepPunct{\mcitedefaultmidpunct}
{\mcitedefaultendpunct}{\mcitedefaultseppunct}\relax
\EndOfBibitem
\bibitem[Wilson and Yoffe(1969)Wilson, and Yoffe]{Yoffe-Wilson69}
Wilson,~J.; Yoffe,~A. The transition metal dichalcogenides discussion and
  interpretation of the observed optical, electrical and structural properties.
  \emph{Adv. Phys.} \textbf{1969}, \emph{18}, 193--335\relax
\mciteBstWouldAddEndPuncttrue
\mciteSetBstMidEndSepPunct{\mcitedefaultmidpunct}
{\mcitedefaultendpunct}{\mcitedefaultseppunct}\relax
\EndOfBibitem
\bibitem[Meckbach \latin{et~al.}(2018)Meckbach, Stroucken, and
  Koch]{Meckbach_2018}
Meckbach,~L.; Stroucken,~T.; Koch,~S.~W. Influence of the effective layer
  thickness on the ground-state and excitonic properties of transition-metal
  dichalcogenide systems. \emph{Phys. Rev. B} \textbf{2018}, \emph{97},
  035425\relax
\mciteBstWouldAddEndPuncttrue
\mciteSetBstMidEndSepPunct{\mcitedefaultmidpunct}
{\mcitedefaultendpunct}{\mcitedefaultseppunct}\relax
\EndOfBibitem
\bibitem[Trolle \latin{et~al.}(2017)Trolle, Pedersen, and
  V{\'e}niard]{Trolle_2017_eps_subst}
Trolle,~M.~L.; Pedersen,~T.~G.; V{\'e}niard,~V. Model dielectric function for
  2D semiconductors including substrate screening. \emph{Sci. Rep.}
  \textbf{2017}, \emph{7}, 39844\relax
\mciteBstWouldAddEndPuncttrue
\mciteSetBstMidEndSepPunct{\mcitedefaultmidpunct}
{\mcitedefaultendpunct}{\mcitedefaultseppunct}\relax
\EndOfBibitem
\bibitem[Aspnes(1982)]{Aspnes_1982}
Aspnes,~D. Optical properties of thin films. \emph{Thin Solid Films}
  \textbf{1982}, \emph{89}, 249--262\relax
\mciteBstWouldAddEndPuncttrue
\mciteSetBstMidEndSepPunct{\mcitedefaultmidpunct}
{\mcitedefaultendpunct}{\mcitedefaultseppunct}\relax
\EndOfBibitem
\bibitem[Laturia \latin{et~al.}(2018)Laturia, Van~de Put, and
  Vandenberghe]{Laturia_2018}
Laturia,~A.; Van~de Put,~M.~L.; Vandenberghe,~W.~G. Dielectric properties of
  hexagonal boron nitride and transition metal dichalcogenides: from monolayer
  to bulk. \emph{npj 2D Mater. Appl.} \textbf{2018}, \emph{2}, 6\relax
\mciteBstWouldAddEndPuncttrue
\mciteSetBstMidEndSepPunct{\mcitedefaultmidpunct}
{\mcitedefaultendpunct}{\mcitedefaultseppunct}\relax
\EndOfBibitem
\bibitem[Tancogne-Dejean \latin{et~al.}(2015)Tancogne-Dejean, Giorgetti, and
  V{\'e}niard]{Tancogne_Dejean_2015}
Tancogne-Dejean,~N.; Giorgetti,~C.; V{\'e}niard,~V. Optical properties of
  surfaces with supercellab initiocalculations: Local-field effects.
  \emph{Phys. Rev. B} \textbf{2015}, \emph{92}, 245308\relax
\mciteBstWouldAddEndPuncttrue
\mciteSetBstMidEndSepPunct{\mcitedefaultmidpunct}
{\mcitedefaultendpunct}{\mcitedefaultseppunct}\relax
\EndOfBibitem
\bibitem[Finkenrath(1988)]{Finkenrath_1988}
Finkenrath,~H. The moss rule and the influence of doping on the optical
  dielectric constant of semiconductors---II. \emph{Infrared Physics}
  \textbf{1988}, \emph{28}, 363--366\relax
\mciteBstWouldAddEndPuncttrue
\mciteSetBstMidEndSepPunct{\mcitedefaultmidpunct}
{\mcitedefaultendpunct}{\mcitedefaultseppunct}\relax
\EndOfBibitem
\bibitem[Ravindra and Srivastava(1980)Ravindra, and
  Srivastava]{Ravindra_1980_model}
Ravindra,~N.~M.; Srivastava,~V.~K. Electronic polarizability as a function of
  the penn gap in semiconductors. \emph{Infrared Phys.} \textbf{1980},
  \emph{20}, 67--69\relax
\mciteBstWouldAddEndPuncttrue
\mciteSetBstMidEndSepPunct{\mcitedefaultmidpunct}
{\mcitedefaultendpunct}{\mcitedefaultseppunct}\relax
\EndOfBibitem
\bibitem[Ravindra and Srivastava(1979)Ravindra, and
  Srivastava]{Ravindra_1979_eps_Eg}
Ravindra,~N.; Srivastava,~V. Variation of refractive index with energy gap in
  semiconductors. \emph{Infrared Phys.} \textbf{1979}, \emph{19},
  603--604\relax
\mciteBstWouldAddEndPuncttrue
\mciteSetBstMidEndSepPunct{\mcitedefaultmidpunct}
{\mcitedefaultendpunct}{\mcitedefaultseppunct}\relax
\EndOfBibitem
\bibitem[Haastrup \latin{et~al.}(2018)Haastrup, Strange, Pandey, Deilmann,
  Schmidt, Hinsche, Gjerding, Torelli, Larsen, Riis-Jensen, and
  et~al.]{Haastrup_2018}
Haastrup,~S.; Strange,~M.; Pandey,~M.; Deilmann,~T.; Schmidt,~P.~S.;
  Hinsche,~N.~F.; Gjerding,~M.~N.; Torelli,~D.; Larsen,~P.~M.;
  Riis-Jensen,~A.~C.; et~al., The Computational 2D Materials Database:
  high-throughput modeling and discovery of atomically thin crystals. \emph{2D
  Mater.} \textbf{2018}, \emph{5}, 042002\relax
\mciteBstWouldAddEndPuncttrue
\mciteSetBstMidEndSepPunct{\mcitedefaultmidpunct}
{\mcitedefaultendpunct}{\mcitedefaultseppunct}\relax
\EndOfBibitem
\bibitem[Mounet \latin{et~al.}(2018)Mounet, Gibertini, Schwaller, Campi,
  Merkys, Marrazzo, Sohier, Castelli, Cepellotti, Pizzi, and
  et~al.]{Mounet_2018}
Mounet,~N.; Gibertini,~M.; Schwaller,~P.; Campi,~D.; Merkys,~A.; Marrazzo,~A.;
  Sohier,~T.; Castelli,~I.~E.; Cepellotti,~A.; Pizzi,~G.; et~al.,
  Two-dimensional materials from high-throughput computational exfoliation of
  experimentally known compounds. \emph{Nat. Nanotechnol.} \textbf{2018},
  \emph{13}, 246--252\relax
\mciteBstWouldAddEndPuncttrue
\mciteSetBstMidEndSepPunct{\mcitedefaultmidpunct}
{\mcitedefaultendpunct}{\mcitedefaultseppunct}\relax
\EndOfBibitem
\bibitem[Kuc and Heine(2015)Kuc, and Heine]{Heine15}
Kuc,~A.; Heine,~T. The electronic structure calculations of two-dimensional
  transition-metal dichalcogenides in the presence of external electric and
  magnetic fields. \emph{Chem. Soc. Rev.} \textbf{2015}, \emph{44},
  2603--2614\relax
\mciteBstWouldAddEndPuncttrue
\mciteSetBstMidEndSepPunct{\mcitedefaultmidpunct}
{\mcitedefaultendpunct}{\mcitedefaultseppunct}\relax
\EndOfBibitem
\bibitem[Lee \latin{et~al.}(2017)Lee, Huang, Sumpter, and Yoon]{Lee_2017}
Lee,~J.; Huang,~J.; Sumpter,~B.~G.; Yoon,~M. Strain-engineered optoelectronic
  properties of 2D transition metal dichalcogenide lateral heterostructures.
  \emph{2D Materials} \textbf{2017}, \emph{4}, 021016\relax
\mciteBstWouldAddEndPuncttrue
\mciteSetBstMidEndSepPunct{\mcitedefaultmidpunct}
{\mcitedefaultendpunct}{\mcitedefaultseppunct}\relax
\EndOfBibitem
\bibitem[Shearer \latin{et~al.}(2016)Shearer, Slattery, Stapleton, Shapter, and
  Gibson]{Shearer_2016}
Shearer,~C.~J.; Slattery,~A.~D.; Stapleton,~A.~J.; Shapter,~J.~G.;
  Gibson,~C.~T. Accurate thickness measurement of graphene.
  \emph{Nanotechnology} \textbf{2016}, \emph{27}, 125704\relax
\mciteBstWouldAddEndPuncttrue
\mciteSetBstMidEndSepPunct{\mcitedefaultmidpunct}
{\mcitedefaultendpunct}{\mcitedefaultseppunct}\relax
\EndOfBibitem
\bibitem[Antoine \latin{et~al.}(1999)Antoine, Dugourd, Rayane, Benichou,
  Broyer, Chandezon, and Guet]{Antoine_1999}
Antoine,~R.; Dugourd,~P.; Rayane,~D.; Benichou,~E.; Broyer,~M.; Chandezon,~F.;
  Guet,~C. Direct measurement of the electric polarizability of isolated C60
  molecules. \emph{J. Chem. Phys.} \textbf{1999}, \emph{110}, 9771--9772\relax
\mciteBstWouldAddEndPuncttrue
\mciteSetBstMidEndSepPunct{\mcitedefaultmidpunct}
{\mcitedefaultendpunct}{\mcitedefaultseppunct}\relax
\EndOfBibitem
\bibitem[Cherniavskaya \latin{et~al.}(2003)Cherniavskaya, Chen, Weng, Yuditsky,
  and Brus]{Cherniavskaya_2003}
Cherniavskaya,~O.; Chen,~L.; Weng,~V.; Yuditsky,~L.; Brus,~L.~E. Quantitative
  Noncontact Electrostatic Force Imaging of Nanocrystal Polarizability.
  \emph{J. Phys. Chem. B} \textbf{2003}, \emph{107}, 1525--1531\relax
\mciteBstWouldAddEndPuncttrue
\mciteSetBstMidEndSepPunct{\mcitedefaultmidpunct}
{\mcitedefaultendpunct}{\mcitedefaultseppunct}\relax
\EndOfBibitem
\bibitem[Krauss and Brus(1999)Krauss, and Brus]{Krauss_1999_EFM}
Krauss,~T.~D.; Brus,~L.~E. Charge, Polarizability, and Photoionization of
  Single Semiconductor Nanocrystals. \emph{Phys. Rev. Lett.} \textbf{1999},
  \emph{83}, 4840--4843\relax
\mciteBstWouldAddEndPuncttrue
\mciteSetBstMidEndSepPunct{\mcitedefaultmidpunct}
{\mcitedefaultendpunct}{\mcitedefaultseppunct}\relax
\EndOfBibitem
\bibitem[Kumar \latin{et~al.}(2016)Kumar, Chauhan, Agarwal, and
  Bhowmick]{Kumar_2016_jpcc}
Kumar,~P.; Chauhan,~Y.~S.; Agarwal,~A.; Bhowmick,~S. Thickness and Stacking
  Dependent Polarizability and Dielectric Constant of Graphene--Hexagonal Boron
  Nitride Composite Stacks. \emph{J. Phys. Chem. C} \textbf{2016}, \emph{120},
  17620--17626\relax
\mciteBstWouldAddEndPuncttrue
\mciteSetBstMidEndSepPunct{\mcitedefaultmidpunct}
{\mcitedefaultendpunct}{\mcitedefaultseppunct}\relax
\EndOfBibitem
\bibitem[Andersen \latin{et~al.}(2015)Andersen, Latini, and
  Thygesen]{Andersen_2015_dielec_vdWH}
Andersen,~K.; Latini,~S.; Thygesen,~K.~S. Dielectric {Genome} of van der
  {Waals} {Heterostructures}. \emph{Nano Lett.} \textbf{2015}, \emph{15},
  4616--4621\relax
\mciteBstWouldAddEndPuncttrue
\mciteSetBstMidEndSepPunct{\mcitedefaultmidpunct}
{\mcitedefaultendpunct}{\mcitedefaultseppunct}\relax
\EndOfBibitem
\bibitem[Pulci \latin{et~al.}(2014)Pulci, Marsili, Garbuio, Gori, Kupchak, and
  Bechstedt]{Pulci_2014}
Pulci,~O.; Marsili,~M.; Garbuio,~V.; Gori,~P.; Kupchak,~I.; Bechstedt,~F.
  Excitons in two-dimensional sheets with honeycomb symmetry. \emph{Phys.
  Status Solidi B} \textbf{2014}, \emph{252}, 72--77\relax
\mciteBstWouldAddEndPuncttrue
\mciteSetBstMidEndSepPunct{\mcitedefaultmidpunct}
{\mcitedefaultendpunct}{\mcitedefaultseppunct}\relax
\EndOfBibitem
\bibitem[Banwell and McCash(1994)Banwell, and McCash]{Banwell_1994}
Banwell,~C.~N.; McCash,~E.~M. \emph{Fundamentals of molecular spectroscopy};
  McGraw-Hill New York, 1994; Vol. 851\relax
\mciteBstWouldAddEndPuncttrue
\mciteSetBstMidEndSepPunct{\mcitedefaultmidpunct}
{\mcitedefaultendpunct}{\mcitedefaultseppunct}\relax
\EndOfBibitem
\bibitem[Pedersen(2016)]{Pedersen_2016}
Pedersen,~T.~G. Exciton Stark shift and electroabsorption in monolayer
  transition-metal dichalcogenides. \emph{Phys. Rev. B} \textbf{2016},
  \emph{94}\relax
\mciteBstWouldAddEndPuncttrue
\mciteSetBstMidEndSepPunct{\mcitedefaultmidpunct}
{\mcitedefaultendpunct}{\mcitedefaultseppunct}\relax
\EndOfBibitem
\bibitem[Klein \latin{et~al.}(2016)Klein, Wierzbowski, Regler, Becker,
  Heimbach, M{\"u}ller, Kaniber, and Finley]{Klein_2016}
Klein,~J.; Wierzbowski,~J.; Regler,~A.; Becker,~J.; Heimbach,~F.;
  M{\"u}ller,~K.; Kaniber,~M.; Finley,~J.~J. Stark Effect Spectroscopy of Mono-
  and Few-Layer MoS$_{2}$. \emph{Nano Lett.} \textbf{2016}, \emph{16},
  1554--1559\relax
\mciteBstWouldAddEndPuncttrue
\mciteSetBstMidEndSepPunct{\mcitedefaultmidpunct}
{\mcitedefaultendpunct}{\mcitedefaultseppunct}\relax
\EndOfBibitem
\bibitem[Roch \latin{et~al.}(2018)Roch, Leisgang, Froehlicher, Makk, Watanabe,
  Taniguchi, Sch{\"o}nenberger, and Warburton]{Roch_2018}
Roch,~J.~G.; Leisgang,~N.; Froehlicher,~G.; Makk,~P.; Watanabe,~K.;
  Taniguchi,~T.; Sch{\"o}nenberger,~C.; Warburton,~R.~J. Quantum-Confined Stark
  Effect in a MoS$_{2}$ Monolayer van der Waals Heterostructure. \emph{Nano
  Lett.} \textbf{2018}, \emph{18}, 1070--1074\relax
\mciteBstWouldAddEndPuncttrue
\mciteSetBstMidEndSepPunct{\mcitedefaultmidpunct}
{\mcitedefaultendpunct}{\mcitedefaultseppunct}\relax
\EndOfBibitem
\bibitem[Verzhbitskiy \latin{et~al.}(2019)Verzhbitskiy, Vella, Watanabe,
  Taniguchi, and Eda]{Verzhbitskiy19}
Verzhbitskiy,~I.; Vella,~D.; Watanabe,~K.; Taniguchi,~T.; Eda,~G. Suppressed
  Out-of-Plane Polarizability of Free Excitons in Monolayer WSe$_2$. \emph{ACS
  Nano} \textbf{2019}, \emph{13}, 3218--3224\relax
\mciteBstWouldAddEndPuncttrue
\mciteSetBstMidEndSepPunct{\mcitedefaultmidpunct}
{\mcitedefaultendpunct}{\mcitedefaultseppunct}\relax
\EndOfBibitem
\bibitem[Li \latin{et~al.}(2014)Li, Santos, Xing, Cappelluti, Rold{\'a}n, Chen,
  Watanabe, and Taniguchi]{Liluhua_2014}
Li,~L.~H.; Santos,~E. J.~G.; Xing,~T.; Cappelluti,~E.; Rold{\'a}n,~R.;
  Chen,~Y.; Watanabe,~K.; Taniguchi,~T. Dielectric Screening in Atomically Thin
  Boron Nitride Nanosheets. \emph{Nano Lett.} \textbf{2014}, \emph{15},
  218--223\relax
\mciteBstWouldAddEndPuncttrue
\mciteSetBstMidEndSepPunct{\mcitedefaultmidpunct}
{\mcitedefaultendpunct}{\mcitedefaultseppunct}\relax
\EndOfBibitem
\bibitem[Tian \latin{et~al.}(2016)Tian, Rice, Santos, and Shih]{Tian_2016}
Tian,~T.; Rice,~P.; Santos,~E. J.~G.; Shih,~C.-J. Multiscale Analysis for
  Field-Effect Penetration through Two-Dimensional Materials. \emph{Nano Lett.}
  \textbf{2016}, \emph{16}, 5044--5052\relax
\mciteBstWouldAddEndPuncttrue
\mciteSetBstMidEndSepPunct{\mcitedefaultmidpunct}
{\mcitedefaultendpunct}{\mcitedefaultseppunct}\relax
\EndOfBibitem
\bibitem[Li \latin{et~al.}(2018)Li, Tian, Cai, Shih, and Santos]{Li_2018}
Li,~L.~H.; Tian,~T.; Cai,~Q.; Shih,~C.-J.; Santos,~E. J.~G. Asymmetric electric
  field screening in van der Waals heterostructures. \emph{Nat. Commun.}
  \textbf{2018}, \emph{9}, 1271\relax
\mciteBstWouldAddEndPuncttrue
\mciteSetBstMidEndSepPunct{\mcitedefaultmidpunct}
{\mcitedefaultendpunct}{\mcitedefaultseppunct}\relax
\EndOfBibitem
\bibitem[Tran \latin{et~al.}(2014)Tran, Soklaski, Liang, and Yang]{Tran_2014}
Tran,~V.; Soklaski,~R.; Liang,~Y.; Yang,~L. Layer-controlled band gap and
  anisotropic excitons in few-layer black phosphorus. \emph{Phys. Rev. B}
  \textbf{2014}, \emph{89}, 235319\relax
\mciteBstWouldAddEndPuncttrue
\mciteSetBstMidEndSepPunct{\mcitedefaultmidpunct}
{\mcitedefaultendpunct}{\mcitedefaultseppunct}\relax
\EndOfBibitem
\bibitem[Chernikov \latin{et~al.}(2014)Chernikov, Berkelbach, Hill, Rigosi, Li,
  Aslan, Reichman, Hybertsen, and Heinz]{Chernikov_2014_EB_MoS2_2D3D}
Chernikov,~A.; Berkelbach,~T.~C.; Hill,~H.~M.; Rigosi,~A.; Li,~Y.;
  Aslan,~O.~B.; Reichman,~D.~R.; Hybertsen,~M.~S.; Heinz,~T.~F. Exciton Binding
  Energy and Nonhydrogenic Rydberg Series in Monolayer WS$_{2}$. \emph{Phys.
  Rev. Lett.} \textbf{2014}, \emph{113}, 076802\relax
\mciteBstWouldAddEndPuncttrue
\mciteSetBstMidEndSepPunct{\mcitedefaultmidpunct}
{\mcitedefaultendpunct}{\mcitedefaultseppunct}\relax
\EndOfBibitem
\bibitem[Berkelbach \latin{et~al.}(2013)Berkelbach, Hybertsen, and
  Reichman]{Berkelbach_2013}
Berkelbach,~T.~C.; Hybertsen,~M.~S.; Reichman,~D.~R. Theory of neutral and
  charged excitons in monolayer transition metal dichalcogenides. \emph{Phys.
  Rev. B} \textbf{2013}, \emph{88}, 045318\relax
\mciteBstWouldAddEndPuncttrue
\mciteSetBstMidEndSepPunct{\mcitedefaultmidpunct}
{\mcitedefaultendpunct}{\mcitedefaultseppunct}\relax
\EndOfBibitem
\bibitem[Kresse and Hafner(1993)Kresse, and Hafner]{Kresse_1993}
Kresse,~G.; Hafner,~J. Ab initiomolecular dynamics for liquid metals.
  \emph{Phys. Rev. B} \textbf{1993}, \emph{47}, 558--561\relax
\mciteBstWouldAddEndPuncttrue
\mciteSetBstMidEndSepPunct{\mcitedefaultmidpunct}
{\mcitedefaultendpunct}{\mcitedefaultseppunct}\relax
\EndOfBibitem
\bibitem[Kresse and Furthm{\"u}ller(1996)Kresse, and
  Furthm{\"u}ller]{Kresse_1996_1}
Kresse,~G.; Furthm{\"u}ller,~J. Efficiency of ab-initio total energy
  calculations for metals and semiconductors using a plane-wave basis set.
  \emph{Comput. Mater. Sci.} \textbf{1996}, \emph{6}, 15--50\relax
\mciteBstWouldAddEndPuncttrue
\mciteSetBstMidEndSepPunct{\mcitedefaultmidpunct}
{\mcitedefaultendpunct}{\mcitedefaultseppunct}\relax
\EndOfBibitem
\bibitem[Kresse and Furthm{\"u}ller(1996)Kresse, and
  Furthm{\"u}ller]{Kresse_1996_2}
Kresse,~G.; Furthm{\"u}ller,~J. Efficient iterative schemes forab
  initiototal-energy calculations using a plane-wave basis set. \emph{Phys.
  Rev. B} \textbf{1996}, \emph{54}, 11169--11186\relax
\mciteBstWouldAddEndPuncttrue
\mciteSetBstMidEndSepPunct{\mcitedefaultmidpunct}
{\mcitedefaultendpunct}{\mcitedefaultseppunct}\relax
\EndOfBibitem
\bibitem[Kresse and Joubert(1999)Kresse, and
  Joubert]{Kresse_1999_pseudopotentials}
Kresse,~G.; Joubert,~D. From ultrasoft pseudopotentials to the projector
  augmented-wave method. \emph{Phys. Rev. B} \textbf{1999}, \emph{59},
  1758--1775\relax
\mciteBstWouldAddEndPuncttrue
\mciteSetBstMidEndSepPunct{\mcitedefaultmidpunct}
{\mcitedefaultendpunct}{\mcitedefaultseppunct}\relax
\EndOfBibitem
\bibitem[Lee \latin{et~al.}(2010)Lee, Murray, Kong, Lundqvist, and
  Langreth]{Lee_2010_vdFD2}
Lee,~K.; Murray,~E.~D.; Kong,~L.; Lundqvist,~B.~I.; Langreth,~D.~C.
  Higher-accuracy van der Waals density functional. \emph{Phys. Rev. B}
  \textbf{2010}, \emph{82}, 081101(R)\relax
\mciteBstWouldAddEndPuncttrue
\mciteSetBstMidEndSepPunct{\mcitedefaultmidpunct}
{\mcitedefaultendpunct}{\mcitedefaultseppunct}\relax
\EndOfBibitem
\end{mcitethebibliography}


\providecommand{\latin}[1]{#1}
\makeatletter
\providecommand{\doi}
  {\begingroup\let\do\@makeother\dospecials
  \catcode`\{=1 \catcode`\}=2 \doi@aux}
\providecommand{\doi@aux}[1]{\endgroup\texttt{#1}}
\makeatother
\providecommand*\mcitethebibliography{\thebibliography}
\csname @ifundefined\endcsname{endmcitethebibliography}
  {\let\endmcitethebibliography\endthebibliography}{}
\begin{mcitethebibliography}{46}
\providecommand*\natexlab[1]{#1}
\providecommand*\mciteSetBstSublistMode[1]{}
\providecommand*\mciteSetBstMaxWidthForm[2]{}
\providecommand*\mciteBstWouldAddEndPuncttrue
  {\def\EndOfBibitem{\unskip.}}
\providecommand*\mciteBstWouldAddEndPunctfalse
  {\let\EndOfBibitem\relax}
\providecommand*\mciteSetBstMidEndSepPunct[3]{}
\providecommand*\mciteSetBstSublistLabelBeginEnd[3]{}
\providecommand*\EndOfBibitem{}
\mciteSetBstSublistMode{f}
\mciteSetBstMaxWidthForm{subitem}{(\alph{mcitesubitemcount})}
\mciteSetBstSublistLabelBeginEnd
  {\mcitemaxwidthsubitemform\space}
  {\relax}
  {\relax}

\bibitem[Rozzi \latin{et~al.}(2006)Rozzi, Varsano, Marini, Gross, and
  Rubio]{Rozzi_2006}
Rozzi,~C.~A.; Varsano,~D.; Marini,~A.; Gross,~E. K.~U.; Rubio,~A. Exact Coulomb
  cutoff technique for supercell calculations. \emph{Phys. Rev. B}
  \textbf{2006}, \emph{73}, 205119\relax
\mciteBstWouldAddEndPuncttrue
\mciteSetBstMidEndSepPunct{\mcitedefaultmidpunct}
{\mcitedefaultendpunct}{\mcitedefaultseppunct}\relax
\EndOfBibitem
\bibitem[H\"{u}ser \latin{et~al.}(2013)H\"{u}ser, Olsen, and
  Thygesen]{Hueser_2013_2Dvs3D}
H\"{u}ser,~F.; Olsen,~T.; Thygesen,~K.~S. How dielectric screening in
  two-dimensional crystals affects the convergence of excited-state
  calculations: Monolayer MoS$_{2}$. \emph{Phys. Rev. B} \textbf{2013},
  \emph{88}, 245309\relax
\mciteBstWouldAddEndPuncttrue
\mciteSetBstMidEndSepPunct{\mcitedefaultmidpunct}
{\mcitedefaultendpunct}{\mcitedefaultseppunct}\relax
\EndOfBibitem
\bibitem[Matthes \latin{et~al.}(2016)Matthes, Pulci, and
  Bechstedt]{Matthes_2016}
Matthes,~L.; Pulci,~O.; Bechstedt,~F. Influence of out-of-plane response on
  optical properties of two-dimensional materials: First principles approach.
  \emph{Phys. Rev. B} \textbf{2016}, \emph{94}, 205408\relax
\mciteBstWouldAddEndPuncttrue
\mciteSetBstMidEndSepPunct{\mcitedefaultmidpunct}
{\mcitedefaultendpunct}{\mcitedefaultseppunct}\relax
\EndOfBibitem
\bibitem[Laturia \latin{et~al.}(2018)Laturia, Van~de Put, and
  Vandenberghe]{Laturia_2018}
Laturia,~A.; Van~de Put,~M.~L.; Vandenberghe,~W.~G. Dielectric properties of
  hexagonal boron nitride and transition metal dichalcogenides: from monolayer
  to bulk. \emph{npj 2D Mater. Appl.} \textbf{2018}, \emph{2}, 6\relax
\mciteBstWouldAddEndPuncttrue
\mciteSetBstMidEndSepPunct{\mcitedefaultmidpunct}
{\mcitedefaultendpunct}{\mcitedefaultseppunct}\relax
\EndOfBibitem
\bibitem[Markel(2016)]{Markel_2016}
Markel,~V.~A. Introduction to the Maxwell Garnett approximation: tutorial.
  \emph{J. Opt. Soc. Am. A} \textbf{2016}, \emph{33}, 1244\relax
\mciteBstWouldAddEndPuncttrue
\mciteSetBstMidEndSepPunct{\mcitedefaultmidpunct}
{\mcitedefaultendpunct}{\mcitedefaultseppunct}\relax
\EndOfBibitem
\bibitem[Choi \latin{et~al.}(2015)Choi, Cui, Lan, and Zhang]{Choi_linear_2015}
Choi,~J.-H.; Cui,~P.; Lan,~H.; Zhang,~Z. Linear {Scaling} of the {Exciton}
  {Binding} {Energy} versus the {Band} {Gap} of {Two}-{Dimensional}
  {Materials}. \emph{Phys. Rev. Lett.} \textbf{2015}, \emph{115}, 066403\relax
\mciteBstWouldAddEndPuncttrue
\mciteSetBstMidEndSepPunct{\mcitedefaultmidpunct}
{\mcitedefaultendpunct}{\mcitedefaultseppunct}\relax
\EndOfBibitem
\bibitem[Olsen \latin{et~al.}(2016)Olsen, Latini, Rasmussen, and
  Thygesen]{Olsen_2016_hydrogen}
Olsen,~T.; Latini,~S.; Rasmussen,~F.; Thygesen,~K.~S. Simple Screened Hydrogen
  Model of Excitons in Two-Dimensional Materials. \emph{Phys. Rev. Lett.}
  \textbf{2016}, \emph{116}, 056401\relax
\mciteBstWouldAddEndPuncttrue
\mciteSetBstMidEndSepPunct{\mcitedefaultmidpunct}
{\mcitedefaultendpunct}{\mcitedefaultseppunct}\relax
\EndOfBibitem
\bibitem[Jiang \latin{et~al.}(2017)Jiang, Liu, Li, and Duan]{Jiang_2017_Eg_Eb}
Jiang,~Z.; Liu,~Z.; Li,~Y.; Duan,~W. Scaling Universality between Band Gap and
  Exciton Binding Energy of Two-Dimensional Semiconductors. \emph{Phys. Rev.
  Lett.} \textbf{2017}, \emph{118}\relax
\mciteBstWouldAddEndPuncttrue
\mciteSetBstMidEndSepPunct{\mcitedefaultmidpunct}
{\mcitedefaultendpunct}{\mcitedefaultseppunct}\relax
\EndOfBibitem
\bibitem[Pulci \latin{et~al.}(2014)Pulci, Marsili, Garbuio, Gori, Kupchak, and
  Bechstedt]{Pulci_2014}
Pulci,~O.; Marsili,~M.; Garbuio,~V.; Gori,~P.; Kupchak,~I.; Bechstedt,~F.
  Excitons in two-dimensional sheets with honeycomb symmetry. \emph{Phys.
  Status Solidi B} \textbf{2014}, \emph{252}, 72--77\relax
\mciteBstWouldAddEndPuncttrue
\mciteSetBstMidEndSepPunct{\mcitedefaultmidpunct}
{\mcitedefaultendpunct}{\mcitedefaultseppunct}\relax
\EndOfBibitem
\bibitem[Davies(1952)]{Davies_1952}
Davies,~P. Polarizabilities of long chain conjugated molecules. \emph{Trans.
  Faraday Soc.} \textbf{1952}, \emph{48}, 789--795\relax
\mciteBstWouldAddEndPuncttrue
\mciteSetBstMidEndSepPunct{\mcitedefaultmidpunct}
{\mcitedefaultendpunct}{\mcitedefaultseppunct}\relax
\EndOfBibitem
\bibitem[Sabirov(2014)]{Sabirov_2014}
Sabirov,~D.~S. Polarizability as a landmark property for fullerene chemistry
  and materials science. \emph{RSC Adv.} \textbf{2014}, \emph{4},
  44996--45028\relax
\mciteBstWouldAddEndPuncttrue
\mciteSetBstMidEndSepPunct{\mcitedefaultmidpunct}
{\mcitedefaultendpunct}{\mcitedefaultseppunct}\relax
\EndOfBibitem
\bibitem[Benedict \latin{et~al.}(1995)Benedict, Louie, and
  Cohen]{Benedict_1995}
Benedict,~L.~X.; Louie,~S.~G.; Cohen,~M.~L. Static polarizabilities of
  single-wall carbon nanotubes. \emph{Phys. Rev. B} \textbf{1995}, \emph{52},
  8541--8549\relax
\mciteBstWouldAddEndPuncttrue
\mciteSetBstMidEndSepPunct{\mcitedefaultmidpunct}
{\mcitedefaultendpunct}{\mcitedefaultseppunct}\relax
\EndOfBibitem
\bibitem[Davies(1997)]{davies_physics_1997}
Davies,~J.~H. \emph{The {Physics} of {Low}-dimensional {Semiconductors}: {An}
  {Introduction}}; Cambridge University Press, 1997\relax
\mciteBstWouldAddEndPuncttrue
\mciteSetBstMidEndSepPunct{\mcitedefaultmidpunct}
{\mcitedefaultendpunct}{\mcitedefaultseppunct}\relax
\EndOfBibitem
\bibitem[Adler(1962)]{Adler_1962}
Adler,~S.~L. Quantum Theory of the Dielectric Constant in Real Solids.
  \emph{Phys. Rev.} \textbf{1962}, \emph{126}, 413--420\relax
\mciteBstWouldAddEndPuncttrue
\mciteSetBstMidEndSepPunct{\mcitedefaultmidpunct}
{\mcitedefaultendpunct}{\mcitedefaultseppunct}\relax
\EndOfBibitem
\bibitem[Hybertsen and Louie(1987)Hybertsen, and Louie]{Hybertsen_1987}
Hybertsen,~M.~S.; Louie,~S.~G. Ab initiostatic dielectric matrices from the
  density-functional approach. I. Formulation and application to semiconductors
  and insulators. \emph{Phys. Rev. B} \textbf{1987}, \emph{35},
  5585--5601\relax
\mciteBstWouldAddEndPuncttrue
\mciteSetBstMidEndSepPunct{\mcitedefaultmidpunct}
{\mcitedefaultendpunct}{\mcitedefaultseppunct}\relax
\EndOfBibitem
\bibitem[Ihn(2009)]{ihn_semiconductor_2009}
Ihn,~T. \emph{Semiconductor {Nanostructures}: {Quantum} states and electronic
  transport}; Oxford University Press, 2009\relax
\mciteBstWouldAddEndPuncttrue
\mciteSetBstMidEndSepPunct{\mcitedefaultmidpunct}
{\mcitedefaultendpunct}{\mcitedefaultseppunct}\relax
\EndOfBibitem
\bibitem[Fowler(1984)]{Fowler_1984}
Fowler,~P. Energy, polarizability and size of confined one-electron systems.
  \emph{Molecular Phys.} \textbf{1984}, \emph{53}, 865--889\relax
\mciteBstWouldAddEndPuncttrue
\mciteSetBstMidEndSepPunct{\mcitedefaultmidpunct}
{\mcitedefaultendpunct}{\mcitedefaultseppunct}\relax
\EndOfBibitem
\bibitem[Maize \latin{et~al.}(2011)Maize, Antonacci, and Marsiglio]{Maize_2011}
Maize,~M.~A.; Antonacci,~M.~A.; Marsiglio,~F. The static electric
  polarizability of a particle bound by a finite potential well. \emph{American
  Journal of Physics} \textbf{2011}, \emph{79}, 222--225\relax
\mciteBstWouldAddEndPuncttrue
\mciteSetBstMidEndSepPunct{\mcitedefaultmidpunct}
{\mcitedefaultendpunct}{\mcitedefaultseppunct}\relax
\EndOfBibitem
\bibitem[Heyd \latin{et~al.}(2005)Heyd, Peralta, Scuseria, and
  Martin]{Heyd_2005}
Heyd,~J.; Peralta,~J.~E.; Scuseria,~G.~E.; Martin,~R.~L. Energy band gaps and
  lattice parameters evaluated with the Heyd-Scuseria-Ernzerhof screened hybrid
  functional. \emph{J. Chem. Phys.} \textbf{2005}, \emph{123}, 174101\relax
\mciteBstWouldAddEndPuncttrue
\mciteSetBstMidEndSepPunct{\mcitedefaultmidpunct}
{\mcitedefaultendpunct}{\mcitedefaultseppunct}\relax
\EndOfBibitem
\bibitem[Kumar \latin{et~al.}(2016)Kumar, Bhadoria, Kumar, Bhowmick, Chauhan,
  and Agarwal]{Kumar_2016_PRB}
Kumar,~P.; Bhadoria,~B.~S.; Kumar,~S.; Bhowmick,~S.; Chauhan,~Y.~S.;
  Agarwal,~A. Thickness and electric-field-dependent polarizability and
  dielectric constant in phosphorene. \emph{Phys.Rev. B} \textbf{2016},
  \emph{93}, 195428\relax
\mciteBstWouldAddEndPuncttrue
\mciteSetBstMidEndSepPunct{\mcitedefaultmidpunct}
{\mcitedefaultendpunct}{\mcitedefaultseppunct}\relax
\EndOfBibitem
\bibitem[Kumar \latin{et~al.}(2016)Kumar, Chauhan, Agarwal, and
  Bhowmick]{Kumar_2016_jpcc}
Kumar,~P.; Chauhan,~Y.~S.; Agarwal,~A.; Bhowmick,~S. Thickness and Stacking
  Dependent Polarizability and Dielectric Constant of Graphene--Hexagonal Boron
  Nitride Composite Stacks. \emph{J. Phys. Chem. C} \textbf{2016}, \emph{120},
  17620--17626\relax
\mciteBstWouldAddEndPuncttrue
\mciteSetBstMidEndSepPunct{\mcitedefaultmidpunct}
{\mcitedefaultendpunct}{\mcitedefaultseppunct}\relax
\EndOfBibitem
\bibitem[Haastrup \latin{et~al.}(2018)Haastrup, Strange, Pandey, Deilmann,
  Schmidt, Hinsche, Gjerding, Torelli, Larsen, Riis-Jensen, and
  et~al.]{Haastrup_2018}
Haastrup,~S.; Strange,~M.; Pandey,~M.; Deilmann,~T.; Schmidt,~P.~S.;
  Hinsche,~N.~F.; Gjerding,~M.~N.; Torelli,~D.; Larsen,~P.~M.;
  Riis-Jensen,~A.~C.; et~al., The Computational 2D Materials Database:
  high-throughput modeling and discovery of atomically thin crystals. \emph{2D
  Mater.} \textbf{2018}, \emph{5}, 042002\relax
\mciteBstWouldAddEndPuncttrue
\mciteSetBstMidEndSepPunct{\mcitedefaultmidpunct}
{\mcitedefaultendpunct}{\mcitedefaultseppunct}\relax
\EndOfBibitem
\bibitem[Arnaud \latin{et~al.}(2006)Arnaud, Leb{\`e}gue, Rabiller, and
  Alouani]{Arnaud_2006_exc_hBN}
Arnaud,~B.; Leb{\`e}gue,~S.; Rabiller,~P.; Alouani,~M. Huge {Excitonic}
  {Effects} in {Layered} {Hexagonal} {Boron} {Nitride}. \emph{Phys. Rev. Lett.}
  \textbf{2006}, \emph{96}, 026402\relax
\mciteBstWouldAddEndPuncttrue
\mciteSetBstMidEndSepPunct{\mcitedefaultmidpunct}
{\mcitedefaultendpunct}{\mcitedefaultseppunct}\relax
\EndOfBibitem
\bibitem[Ramasubramaniam(2012)]{Ramasubramaniam_2012}
Ramasubramaniam,~A. Large excitonic effects in monolayers of molybdenum and
  tungsten dichalcogenides. \emph{Physical Review B} \textbf{2012}, \emph{86},
  115409\relax
\mciteBstWouldAddEndPuncttrue
\mciteSetBstMidEndSepPunct{\mcitedefaultmidpunct}
{\mcitedefaultendpunct}{\mcitedefaultseppunct}\relax
\EndOfBibitem
\bibitem[Chernikov \latin{et~al.}(2014)Chernikov, Berkelbach, Hill, Rigosi, Li,
  Aslan, Reichman, Hybertsen, and Heinz]{Chernikov_2014_EB_MoS2_2D3D}
Chernikov,~A.; Berkelbach,~T.~C.; Hill,~H.~M.; Rigosi,~A.; Li,~Y.;
  Aslan,~O.~B.; Reichman,~D.~R.; Hybertsen,~M.~S.; Heinz,~T.~F. Exciton Binding
  Energy and Nonhydrogenic Rydberg Series in Monolayer WS$_{2}$. \emph{Phys.
  Rev. Lett.} \textbf{2014}, \emph{113}, 076802\relax
\mciteBstWouldAddEndPuncttrue
\mciteSetBstMidEndSepPunct{\mcitedefaultmidpunct}
{\mcitedefaultendpunct}{\mcitedefaultseppunct}\relax
\EndOfBibitem
\bibitem[Gould and Bucko(2016)Gould, and Bucko]{Gould_2016_jctc}
Gould,~T.; Bucko,~T. C6 Coefficients and Dipole Polarizabilities for All Atoms
  and Many Ions in Rows 1--6 of the Periodic Table. \emph{J. Chem. Theory and
  Comput.} \textbf{2016}, \emph{12}, 3603--3613\relax
\mciteBstWouldAddEndPuncttrue
\mciteSetBstMidEndSepPunct{\mcitedefaultmidpunct}
{\mcitedefaultendpunct}{\mcitedefaultseppunct}\relax
\EndOfBibitem
\bibitem[Finkenrath(1988)]{Finkenrath_1988}
Finkenrath,~H. The moss rule and the influence of doping on the optical
  dielectric constant of semiconductors---II. \emph{Infrared Physics}
  \textbf{1988}, \emph{28}, 363--366\relax
\mciteBstWouldAddEndPuncttrue
\mciteSetBstMidEndSepPunct{\mcitedefaultmidpunct}
{\mcitedefaultendpunct}{\mcitedefaultseppunct}\relax
\EndOfBibitem
\bibitem[Ketterson(2016)]{ketterson_physics_2016}
Ketterson,~J.~B. \emph{The {Physics} of {Solids}}; Oxford University Press,
  2016\relax
\mciteBstWouldAddEndPuncttrue
\mciteSetBstMidEndSepPunct{\mcitedefaultmidpunct}
{\mcitedefaultendpunct}{\mcitedefaultseppunct}\relax
\EndOfBibitem
\bibitem[Nazarov(2015)]{Nazarov_2015_2D_3D}
Nazarov,~V.~U. Electronic excitations in quasi-2D crystals: what theoretical
  quantities are relevant to experiment? \emph{New J. Phys.} \textbf{2015},
  \emph{17}, 073018\relax
\mciteBstWouldAddEndPuncttrue
\mciteSetBstMidEndSepPunct{\mcitedefaultmidpunct}
{\mcitedefaultendpunct}{\mcitedefaultseppunct}\relax
\EndOfBibitem
\bibitem[Bechstedt \latin{et~al.}(2012)Bechstedt, Matthes, Gori, and
  Pulci]{Bechstedt_2012}
Bechstedt,~F.; Matthes,~L.; Gori,~P.; Pulci,~O. Infrared absorbance of silicene
  and germanene. \emph{Appl. Phys. Lett.} \textbf{2012}, \emph{100},
  261906\relax
\mciteBstWouldAddEndPuncttrue
\mciteSetBstMidEndSepPunct{\mcitedefaultmidpunct}
{\mcitedefaultendpunct}{\mcitedefaultseppunct}\relax
\EndOfBibitem
\bibitem[Ning \latin{et~al.}(2015)Ning, Lu, Li, Chen, Dou, Wang, Rehman, Cao,
  and Jin]{Ning_2015}
Ning,~M.-Q.; Lu,~M.-M.; Li,~J.-B.; Chen,~Z.; Dou,~Y.-K.; Wang,~C.-Z.;
  Rehman,~F.; Cao,~M.-S.; Jin,~H.-B. Two-dimensional nanosheets of MoS$_{2}$: a
  promising material with high dielectric properties and microwave absorption
  performance. \emph{Nanoscale} \textbf{2015}, \emph{7}, 15734--15740\relax
\mciteBstWouldAddEndPuncttrue
\mciteSetBstMidEndSepPunct{\mcitedefaultmidpunct}
{\mcitedefaultendpunct}{\mcitedefaultseppunct}\relax
\EndOfBibitem
\bibitem[Li \latin{et~al.}(2014)Li, Chernikov, Zhang, Rigosi, Hill, van~der
  Zande, Chenet, Shih, Hone, and Heinz]{Li_2014}
Li,~Y.; Chernikov,~A.; Zhang,~X.; Rigosi,~A.; Hill,~H.~M.; van~der
  Zande,~A.~M.; Chenet,~D.~A.; Shih,~E.-M.; Hone,~J.; Heinz,~T.~F. Measurement
  of the optical dielectric function of monolayer transition-metal
  dichalcogenides:MoS$_{2}$,MoSe$_2$,WS$_{2}$, andWSe$_2$. \emph{Phys. Rev. B}
  \textbf{2014}, \emph{90}, 205422\relax
\mciteBstWouldAddEndPuncttrue
\mciteSetBstMidEndSepPunct{\mcitedefaultmidpunct}
{\mcitedefaultendpunct}{\mcitedefaultseppunct}\relax
\EndOfBibitem
\bibitem[Yao \latin{et~al.}(2014)Yao, Koski, Luo, Cha, Hu, Kong, Narasimhan,
  Huo, and Cui]{Yao_2014}
Yao,~J.; Koski,~K.~J.; Luo,~W.; Cha,~J.~J.; Hu,~L.; Kong,~D.;
  Narasimhan,~V.~K.; Huo,~K.; Cui,~Y. Optical transmission enhacement through
  chemically tuned two-dimensional bismuth chalcogenide nanoplates. \emph{Nat.
  Commun.} \textbf{2014}, \emph{5}, 5670\relax
\mciteBstWouldAddEndPuncttrue
\mciteSetBstMidEndSepPunct{\mcitedefaultmidpunct}
{\mcitedefaultendpunct}{\mcitedefaultseppunct}\relax
\EndOfBibitem
\bibitem[Wu \latin{et~al.}(2015)Wu, Pak, Liu, Zhou, Wu, Zhu, Lin, Han, Ren,
  Peng, and et~al.]{Wu_2015}
Wu,~D.; Pak,~A.~J.; Liu,~Y.; Zhou,~Y.; Wu,~X.; Zhu,~Y.; Lin,~M.; Han,~Y.;
  Ren,~Y.; Peng,~H.; et~al., Thickness-Dependent Dielectric Constant of
  Few-Layer In2Se3 Nanoflakes. \emph{Nano Letters} \textbf{2015}, \emph{15},
  8136--8140\relax
\mciteBstWouldAddEndPuncttrue
\mciteSetBstMidEndSepPunct{\mcitedefaultmidpunct}
{\mcitedefaultendpunct}{\mcitedefaultseppunct}\relax
\EndOfBibitem
\bibitem[Gobre and Tkatchenko(2013)Gobre, and Tkatchenko]{Gobre_2013}
Gobre,~V.~V.; Tkatchenko,~A. Scaling laws for van der Waals interactions in
  nanostructured materials. \emph{Nat. Commun.} \textbf{2013}, \emph{4},
  2341\relax
\mciteBstWouldAddEndPuncttrue
\mciteSetBstMidEndSepPunct{\mcitedefaultmidpunct}
{\mcitedefaultendpunct}{\mcitedefaultseppunct}\relax
\EndOfBibitem
\bibitem[Sze and Ng(2006)Sze, and Ng]{sze_appendix_2006}
Sze,~S.; Ng,~K.~K. \emph{Physics of {Semiconductor} {Devices}}; John Wiley \&
  Sons, Inc., 2006; pp 789--789\relax
\mciteBstWouldAddEndPuncttrue
\mciteSetBstMidEndSepPunct{\mcitedefaultmidpunct}
{\mcitedefaultendpunct}{\mcitedefaultseppunct}\relax
\EndOfBibitem
\bibitem[Miller(1990)]{Miller_1990}
Miller,~K.~J. Calculation of the molecular polarizability tensor. \emph{J. Am.
  Chem. Soc.} \textbf{1990}, \emph{112}, 8543--8551\relax
\mciteBstWouldAddEndPuncttrue
\mciteSetBstMidEndSepPunct{\mcitedefaultmidpunct}
{\mcitedefaultendpunct}{\mcitedefaultseppunct}\relax
\EndOfBibitem
\bibitem[Ramprasad and Shi(2006)Ramprasad, and Shi]{Ramprasad_2006}
Ramprasad,~R.; Shi,~N. Polarizability of phthalocyanine based molecular
  systems: A first-principles electronic structure study. \emph{Appl. Phys.
  Lett.} \textbf{2006}, \emph{88}, 222903\relax
\mciteBstWouldAddEndPuncttrue
\mciteSetBstMidEndSepPunct{\mcitedefaultmidpunct}
{\mcitedefaultendpunct}{\mcitedefaultseppunct}\relax
\EndOfBibitem
\bibitem[Matsuda \latin{et~al.}(2010)Matsuda, Tahir-Kheli, and
  Goddard]{Matsuda_2010}
Matsuda,~Y.; Tahir-Kheli,~J.; Goddard,~W.~A. Definitive Band Gaps for
  Single-Wall Carbon Nanotubes. \emph{J. Phys. Chem. Lett.} \textbf{2010},
  \emph{1}, 2946--2950\relax
\mciteBstWouldAddEndPuncttrue
\mciteSetBstMidEndSepPunct{\mcitedefaultmidpunct}
{\mcitedefaultendpunct}{\mcitedefaultseppunct}\relax
\EndOfBibitem
\bibitem[Jensen \latin{et~al.}(2000)Jensen, Schmidt, Mikkelsen, and
  {\AA}strand]{Jensen_2000}
Jensen,~L.; Schmidt,~O.~H.; Mikkelsen,~K.~V.; {\AA}strand,~P.-O. Static and
  Frequency-Dependent Polarizability Tensors for Carbon Nanotubes. \emph{J.
  Phys. Chem. B} \textbf{2000}, \emph{104}, 10462--10466\relax
\mciteBstWouldAddEndPuncttrue
\mciteSetBstMidEndSepPunct{\mcitedefaultmidpunct}
{\mcitedefaultendpunct}{\mcitedefaultseppunct}\relax
\EndOfBibitem
\bibitem[Brothers \latin{et~al.}(2008)Brothers, Izmaylov, Scuseria, and
  Kudin]{Brothers_2008}
Brothers,~E.~N.; Izmaylov,~A.~F.; Scuseria,~G.~E.; Kudin,~K.~N. Analytically
  Calculated Polarizability of Carbon Nanotubes: Single Wall, Coaxial, and
  Bundled Systems. \emph{J. Phys. Chem. C} \textbf{2008}, \emph{112},
  1396--1400\relax
\mciteBstWouldAddEndPuncttrue
\mciteSetBstMidEndSepPunct{\mcitedefaultmidpunct}
{\mcitedefaultendpunct}{\mcitedefaultseppunct}\relax
\EndOfBibitem
\bibitem[Salzner \latin{et~al.}(1998)Salzner, Pickup, Poirier, and
  Lagowski]{Salzner_1998}
Salzner,~U.; Pickup,~P.~G.; Poirier,~R.~A.; Lagowski,~J.~B. Accurate Method for
  Obtaining Band Gaps in Conducting Polymers Using a DFT/Hybrid Approach.
  \emph{J. Phys. Chem. A} \textbf{1998}, \emph{102}, 2572--2578\relax
\mciteBstWouldAddEndPuncttrue
\mciteSetBstMidEndSepPunct{\mcitedefaultmidpunct}
{\mcitedefaultendpunct}{\mcitedefaultseppunct}\relax
\EndOfBibitem
\bibitem[Hinchliffe \latin{et~al.}(2005)Hinchliffe, Nikolaidi, and
  Soscun~Machado]{Hinchliffe_2005}
Hinchliffe,~A.; Nikolaidi,~B.; Soscun~Machado,~H. Density functional studies of
  the dipole polarizabilities of the linear polyacenes benzene through
  nonacene. \emph{Open Chem.} \textbf{2005}, \emph{3}, 361--369\relax
\mciteBstWouldAddEndPuncttrue
\mciteSetBstMidEndSepPunct{\mcitedefaultmidpunct}
{\mcitedefaultendpunct}{\mcitedefaultseppunct}\relax
\EndOfBibitem
\bibitem[Lin and Nori(1994)Lin, and Nori]{Lin_1994}
Lin,~Y.-L.; Nori,~F. Electronic structure of single- and multiple-shell carbon
  fullerenes. \emph{Phys. Rev. B} \textbf{1994}, \emph{49}, 5020--5023\relax
\mciteBstWouldAddEndPuncttrue
\mciteSetBstMidEndSepPunct{\mcitedefaultmidpunct}
{\mcitedefaultendpunct}{\mcitedefaultseppunct}\relax
\EndOfBibitem
\bibitem[Martin \latin{et~al.}(2008)Martin, Sild, Maran, and
  Karelson]{Martin_2008}
Martin,~D.; Sild,~S.; Maran,~U.; Karelson,~M. QSPR Modeling of the
  Polarizability of Polyaromatic Hydrocarbons and Fullerenes. \emph{J. Phys.
  Chem. C} \textbf{2008}, \emph{112}, 4785--4790\relax
\mciteBstWouldAddEndPuncttrue
\mciteSetBstMidEndSepPunct{\mcitedefaultmidpunct}
{\mcitedefaultendpunct}{\mcitedefaultseppunct}\relax
\EndOfBibitem
\end{mcitethebibliography}
